# Constraining the climate and ocean pH of the early Earth with a geological carbon cycle model

Joshua Krissansen-Totton[a,b,1], Giada N. Arney[b,c,d], and David C. Catling[a,b]

[a]Department of Earth and Space Sciences, University of Washington, Seattle, WA 98195; [b]Virtual Planetary Laboratory Team, NASA Astrobiology Institute, Seattle, WA 98195; [c]Planetary Systems Laboratory, NASA Goddard Space Flight Center, Greenbelt, MD 20771; and [d]Sellers Exoplanet Environments Collaboration, NASA Goddard Space Flight Center, Greenbelt, MD 20771

**The early Earth's environment is controversial. Climatic estimates range from hot to glacial, and inferred marine pH spans strongly alkaline to acidic. Better understanding of early climate and ocean chemistry would improve our knowledge of the origin of life and its coevolution with the environment. Here, we use a geological carbon cycle model with ocean chemistry to calculate self-consistent histories of climate and ocean pH. Our carbon cycle model includes an empirically justified temperature and pH dependence of seafloor weathering, allowing the relative importance of continental and seafloor weathering to be evaluated. We find that the Archean climate was likely temperate (0–50 °C) due to the combined negative feedbacks of continental and seafloor weathering. Ocean pH evolves monotonically from $6.6^{+0.6}_{-0.4}$ (2σ) at 4.0 Ga to $7.0^{+0.7}_{-0.5}$ (2σ) at the Archean–Proterozoic boundary, and to $7.9^{+0.1}_{-0.2}$ (2σ) at the Proterozoic–Phanerozoic boundary. This evolution is driven by the secular decline of $pCO_2$, which in turn is a consequence of increasing solar luminosity, but is moderated by carbonate alkalinity delivered from continental and seafloor weathering. Archean seafloor weathering may have been a comparable carbon sink to continental weathering, but is less dominant than previously assumed, and would not have induced global glaciation. We show how these conclusions are robust to a wide range of scenarios for continental growth, internal heat flow evolution and outgassing history, greenhouse gas abundances, and changes in the biotic enhancement of weathering.**

Constraining the climate and ocean chemistry of the early Earth is crucial for understanding the emergence of life, the subsequent coevolution of life and the environment, and as a point of reference for evaluating the habitability of terrestrial exoplanets. However, the surface temperature of the early Earth is debated. Oxygen isotopes in chert have low $\delta^{18}O$ values in the Archean (1). If this isotope record reflects the temperature-dependent equilibrium fractionation of $^{18}O$ and $^{16}O$ between silica and seawater, then this would imply mean surface temperatures around 70 ± 15 °C at 3.3 Ga (2). A hot early Earth is also supported by possible evidence for a low viscosity Archean ocean (3), and the thermostability of reconstructed ancestral proteins (4), including those purportedly reflective of Archean photic zone temperatures (5). Silicon isotopes in cherts have also been interpreted to infer 60–80 °C Archean seawater temperatures (6).

Alternatively, the trend in $\delta^{18}O$ over Earth history has been interpreted as a change in the oxygen isotope composition of seawater (7), or hydrothermal alteration of the seafloor (8). Isotopic analyses using deuterium (9) and phosphates (10) report Archean surface temperatures <40 °C. Archean glacial deposits (ref. 11 and references therein) also suggest an early Earth with ice caps, or at least transient cool periods. A geological carbon cycle model of Sleep and Zahnle (12) predicts Archean and Hadean temperatures below 0 °C due to efficient seafloor weathering. An analysis combining general circulation model (GCM) outputs with a carbon cycle model predicts more moderate temperatures at 3.8 Ga (13). Resolving these conflicting interpretations would provide a better understanding of the conditions for the origin and early evolution of life.

Ocean pH is another important environmental parameter because it partitions carbon between the atmosphere and ocean and is thus linked to climate. Additionally, many biosynthetic pathways hypothesized to be important for the origin of life are strongly pH dependent (14–16), and so constraining the pH of the early ocean would inform their viability. Furthermore, bacterial biomineralization is favorable at higher environmental pH values because this allows cells to more easily attract cations through deprotonation (17). Arguably, low environmental pH values would be an obstacle to the evolution of advanced life due to biomineralization inhibition (18). Finally, many $pO_2$ proxies are pH dependent (19–21), and so understanding the history of pH would enable better quantification of the history of $pO_2$.

However, just as with climate, debate surrounds empirical constraints on Archean ocean pH. Empirical constraints are scant and conflicting. Based on the scarcity of gypsum pseudomorphs before 1.8 Ga, Grotzinger and Kasting (22) argued that the Archean ocean pH was likely between 5.7 and 8.6. However, others note the presence of Archean gypsum as early as 3.5 Ga (23); its scarcity could be explained by low sulfate (24). Blättler et al. (24) interpreted Archean Ca isotopes to reflect high Ca/alkalinity ratios, which in turn would rule out high pH and high $pCO_2$ values. Friend et al. (25) argued for qualitatively circumneutral to weakly alkaline Archean ocean pH based on rare Earth element anomalies.

## Significance

The climate and ocean pH of the early Earth are important for understanding the origin and early evolution of life. However, estimates of early climate range from below freezing to over 70 °C, and ocean pH estimates span from strongly acidic to alkaline. To better constrain environmental conditions, we applied a self-consistent geological carbon cycle model to the last 4 billion years. The model predicts a temperate (0–50 °C) climate and circumneutral ocean pH throughout the Precambrian due to stabilizing feedbacks from continental and seafloor weathering. These environmental conditions under which life emerged and diversified were akin to the modern Earth. Similar stabilizing feedbacks on climate and ocean pH may operate on earthlike exoplanets, implying life elsewhere could emerge in comparable environments.

Author contributions: J.K.-T. and D.C.C. designed research; J.K.-T. performed research; J.K.-T., G.N.A., and D.C.C. analyzed data; G.N.A. performed climate model calculations; and J.K.-T. and D.C.C. wrote the paper.

The authors declare no conflict of interest.

This article is a PNAS Direct Submission.

Data deposition: The Python source code is available on GitHub (https://github.com/joshuakt/early-earth-carbon-cycle).

[1]To whom correspondence should be addressed. Email: joshkt@uw.edu.

Theoretical arguments for the evolution of ocean pH also disagree. By analogy with modern alkaline lakes, Kempe and Degens (26) argued for a pH 9–11 "soda ocean" on the early Earth, but mass balance challenges such an idea (27). Additionally, high pH oceans (>9.0) would shift the $NH_3$–$NH_4^+$ aqueous equilibrium toward $NH_3$, which would volatilize and fractionate nitrogen isotopes in a way that is not observed in marine sediments (28). The conventional view of the evolution of ocean pH is that the secular decline of $pCO_2$ over Earth history has driven an increase in ocean pH from acidic to modern slightly alkaline. For example, Halevy and Bachan (29) modeled ocean chemistry over Earth history with prescribed $pCO_2$ and climate histories, and reported a monotonic pH evolution broadly consistent with this view. However, it has also been argued that seafloor weathering buffered ocean pH to near-modern values throughout Earth history (12).

On long timescales, both climate and ocean pH are controlled by the geological carbon cycle. The conventional view of the carbon cycle is that carbon outgassing into the atmosphere–ocean system is balanced by continental silicate weathering and subsequent marine carbonate formation (30, 31). The weathering of silicates is temperature and $pCO_2$ dependent, which provides a natural thermostat to buffer climate against changes in stellar luminosity and outgassing. This mechanism is widely believed to explain the relative stability of Earth's climate despite a ~30% increase in solar luminosity since 4.0 Ga (31).

A possible complimentary negative feedback to continental weathering is provided by seafloor weathering. Such weathering occurs when the seawater circulating in off-axis hydrothermal systems reacts with the surrounding basalt and releases cations, which then precipitate as carbonates in the pore space (32, 33). If the rate of basalt dissolution and pore-space carbonate precipitation depends on the carbon content of the atmosphere–ocean system via $pCO_2$, temperature, or pH, then seafloor weathering could provide an additional negative feedback (12).

The existence of a negative feedback to balance the carbon cycle on million-year timescales is undisputed. Without it, atmospheric $CO_2$ would be depleted, leading to a runaway icehouse, or would accumulate to excessive levels (34). However, the relative importance of continental and seafloor weathering in providing this negative feedback, and the overall effectiveness of these climate-stabilizing and pH-buffering feedbacks on the early Earth are unknown.

In this study, we apply a geological carbon cycle model with ocean chemistry to the entirety of Earth history. The inclusion of ocean carbon chemistry enables us to model the evolution of ocean pH and realistically capture the pH-dependent and temperature-dependent kinetics of seafloor weathering. This is a significant improvement on previous geological carbon cycle models (e.g., refs. 12 and 35) that omit ocean chemistry and instead adopt an arbitrary power-law dependence on $pCO_2$ for seafloor weathering which, as we show, overestimates $CO_2$ drawdown on the early Earth. By coupling seafloor weathering to Earth's climate and the geological carbon cycle, we calculate self-consistent histories of Earth's climate and pH evolution, and evaluate the relative importance of continental and seafloor weathering through time. The pH evolution we calculate is therefore more robust than that of Halevy and Bachan (29) because, unlike their model, we do not prescribe $pCO_2$ and temperature histories.

Our approach remains agnostic on unresolved issues, such as the history of continental growth, internal heat flow, and the biological enhancement of weathering, because we include a broad range of values for these parameters. Our conclusions are therefore robust to uncertainties in Earth system evolution. We find that a hot early Earth is very unlikely, and pH should, on average, have monotonically increased since 4.0 Ga, buffered somewhat by continental and seafloor weathering.

## Methods

The geological carbon cycle model builds on that described in Krissansen-Totton and Catling (36). Here, we summarize its key features, and additional details are provided in the *SI Appendix*. The Python source code is available on GitHub at github.com/joshuakt/early-earth-carbon-cycle.

We model the time evolution of the carbon cycle using two separate boxes representing the atmosphere–ocean system and the pore space in the seafloor (Fig. 1 and *SI Appendix A*). We track carbon and carbonate alkalinity fluxes into and between these boxes, and assume that the bulk ocean is in equilibrium with the atmosphere.

Many of the parameters in our model are uncertain, and so we adopt a range of values (*SI Appendix*, Table S1) based on spread in the literature rather than point estimates. Each parameter range was sampled uniformly, and the forward model was run 10,000 times to build distributions for model outputs such as $pCO_2$, pH, and temperature. Model outputs are compared with proxy data for $pCO_2$, temperature, and carbonate precipitation (*SI Appendix D*).

Continental silicate weathering is described by the following function:

$$F_{sil} = f_{bio} f_{land} F_{sil}^{mod} \left( \frac{pCO_2}{pCO_2^{mod}} \right)^{\alpha} \exp(\Delta T_S / T_e) \qquad [1]$$

Here, $f_{bio}$ is the biological enhancement of weathering (see below), $f_{land}$ is the continental land fraction relative to modern, $F_{sil}^{mod}$ is the modern continental silicate weathering flux (Tmol $y^{-1}$), $\Delta T_S = T_S - T_S^{mod}$ is the difference in global mean surface temperature, $T_S$, relative to preindustrial modern, $T_S^{mod}$. The exponent $\alpha$ is an empirical constant that determines the dependence of weathering on the partial pressure of carbon dioxide relative to modern, $pCO_2/pCO_2^{mod}$. An *e*-folding temperature, $T_e$, defines the temperature dependence of weathering. A similar expression for carbonate weathering is described in *SI Appendix A*.

The land fraction, $f_{land}$, and biological modifier, $f_{bio}$, account for the growth of continents and the biological enhancement of continental weathering, respectively. We adopt a broad range of continental growth curves that encompasses literature estimates (Fig. 2*A* and *SI Appendix A*). For our nominal model, we assume Archean land fraction was anywhere between 10% and 75% of modern land fraction (Fig. 2*A*), but we also consider a no-land Archean endmember (Fig. 2*B*).

To account for the possible biological enhancement of weathering in the Phanerozoic due to vascular land plants, lichens, bryophytes, and ectomycorrhizal fungi, we adopt a broad range of histories for the biological enhancement of weathering, $f_{bio}$ (Fig. 2*C*). The lower end of this range is consistent with estimates of biotic enhancement of weathering from the literature (37–39).

The dissolution of basalt in the seafloor is dependent on the spreading rate, pore-space pH, and pore-space temperature (*SI Appendix A*). This formulation is based on the validated parameterization in ref. 36. Pore-space temperatures are a function of climate and geothermal heat flow. Empirical data and fully coupled global climate models reveal a linear relationship between deep ocean temperature and surface climate (36). Equations relating pore-space temperature, deep ocean temperature, and sediment thickness are provided in *SI Appendix A*.

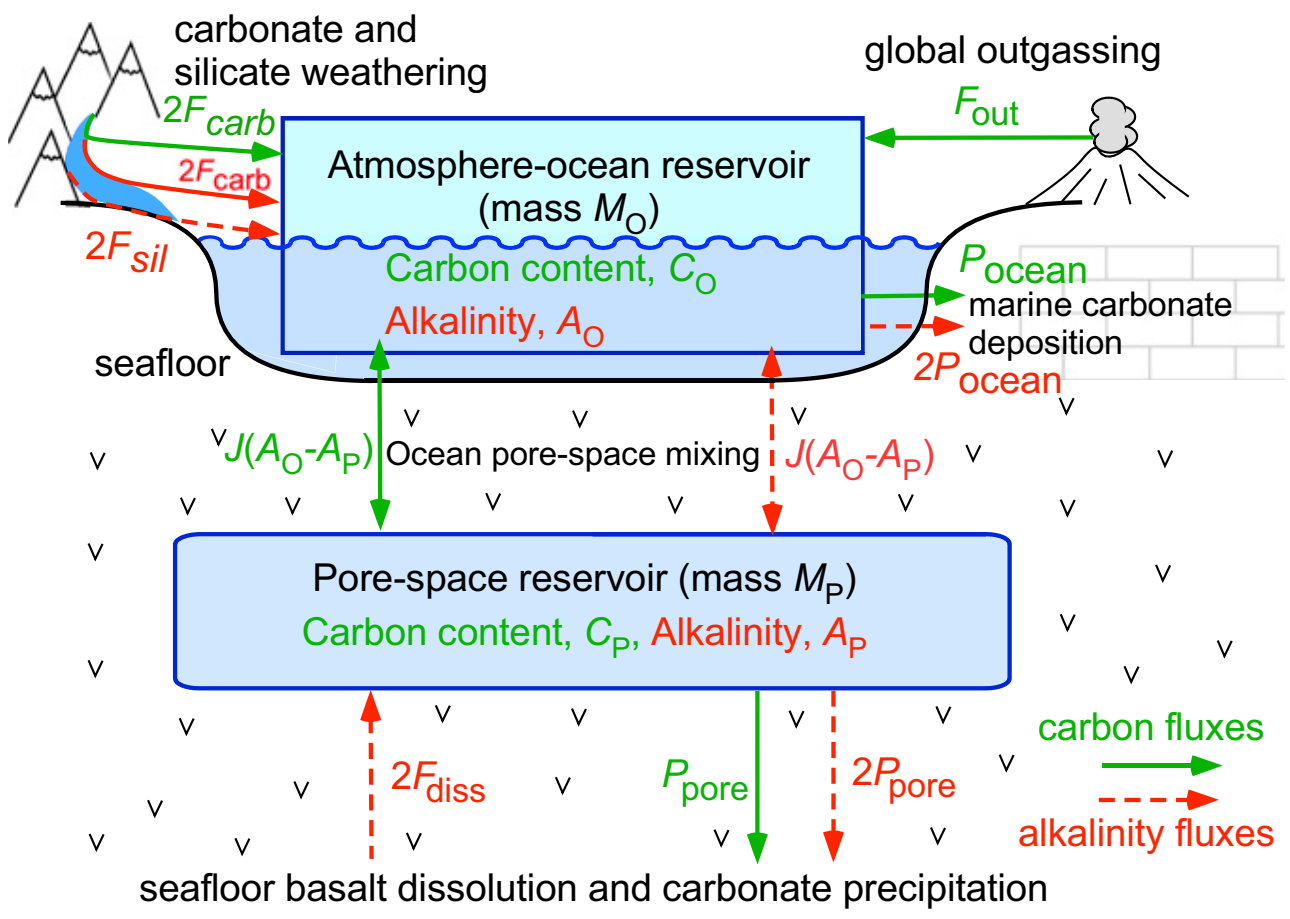

**Fig. 1.** Schematic of carbon cycle model used in this study. Carbon fluxes (Tmol C $y^{-1}$) are denoted by solid green arrows, and alkalinity fluxes (Tmol eq $y^{-1}$) are denoted by red dashed arrows. The fluxes into/out of the atmosphere–ocean system are outgassing, $F_{out}$, silicate weathering, $F_{sil}$, carbonate weathering, $F_{carb}$, and marine carbonate precipitation, $P_{ocean}$. The fluxes into/out of the pore space are basalt dissolution, $F_{diss}$, and pore-space carbonate precipitation, $P_{pore}$. Alkalinity fluxes are multiplied by 2 because the uptake or release of one mole of carbon as carbonate is balanced by a cation with a 2+ charge (typically $Ca^{2+}$). A constant mixing flux, $J$ (kg $y^{-1}$), exchanges carbon and alkalinity between the atmosphere–ocean system and pore space.

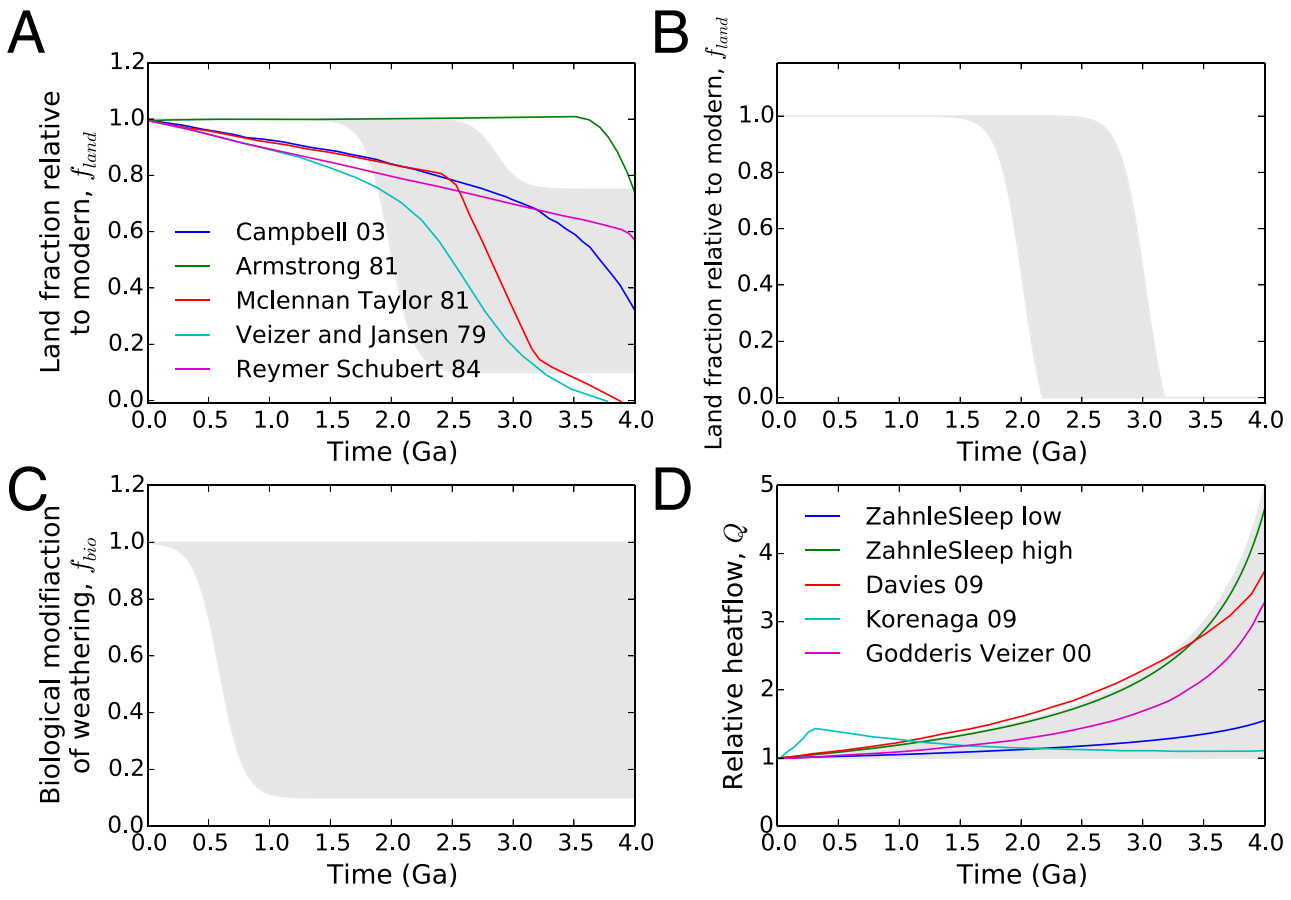

**Fig. 2.** Gray shaded regions are ranges assumed for selected model input parameters. (*A*) Range of continental growth curves assumed in our nominal model, $f_{land}$. Various literature estimates are plotted alongside the model growth curve (*SI Appendix A*). (*B*) Range of continental growth curves for an endmember of no Archean land; (*C*) range for biological enhancement of weathering histories, $f_{bio}$; and (*D*) range of internal heat flow histories, *Q*, compared with literature estimates (*SI Appendix A*).

Carbon leaves the atmosphere–ocean system through carbonate precipitation in the ocean and pore space of the oceanic crust. At each time step, the carbon abundances and alkalinities are used to calculate the carbon speciation, atmospheric $pCO_2$, and saturation state assuming chemical equilibrium. Saturation states are then used to calculate carbonate precipitation fluxes (*SI Appendix A*). We allow calcium (Ca) abundance to evolve with alkalinity, effectively assuming no processes are affecting Ca abundances other than carbonate and silicate weathering, seafloor dissolution, and carbonate precipitation. The consequences of this simplification are explored in the sensitivity analysis in *SI Appendix C*. We do not track organic carbon burial because organic burial only constitutes 10–30% of total carbon burial for the vast majority of Earth history (40), and so the inorganic carbon cycle is the primary control.

The treatment of tectonic and interior processes is important for specifying outgassing and subduction flux histories. We avoid tracking crustal and mantle reservoirs because explicitly parameterizing how outgassing fluxes relate to crustal production and reservoirs assumes modern-style plate tectonics has operated throughout Earth history (e.g., ref. 12) and might not be valid. Evidence exists for Archean subduction in eclogitic diamonds (41) and sulfur mass-independent fractionation in ocean island basalts ostensibly derived from recycled Archean crust (42). However, other tectonic modes have been proposed for the early Earth such as heat-pipe volcanism (43), delamination and shallow convection (44), or a stagnant lid regime (45).

Our generalized parameterizations for heat flow, spreading rates, and outgassing histories are described in *SI Appendix A*. Fig. 2*D* shows our assumed range of internal heat flow histories compared with estimates from the literature. Spreading rate is connected to crustal production via a power law, which spans endmember cases (*SI Appendix A*). These parameterizations provide an extremely broad range of heat flow, outgassing, and crustal production histories, and do not assume a fixed coupling between these variables.

We used a 1D radiative convective model (46) to create a grid of mean surface temperatures as a function of solar luminosity and $pCO_2$. The grid of temperature outputs was fitted with a 2D polynomial (*SI Appendix E*). We initially neglect other greenhouse gases besides $CO_2$ and $H_2O$, albedo changes, and assumed a constant total pressure over Earth history. However, later we consider these influences, such as including methane ($CH_4$) in the Precambrian. The evolution of solar luminosity is conventionally parameterized (47).

Our model has been demonstrated for the last 100 Ma against abundant proxy data (36) and it can broadly reproduce Sleep and Zahnle (12) if we replace our kinetic formulation of seafloor weathering with their simpler $CO_2$-dependent expression (*SI Appendix B*). Agreement with ref. 12 confirms that the omission of crustal and mantle reservoirs does not affect our conclusions.

## Results

Fig. 3 shows the evolution of the geological carbon cycle over Earth history according to our nominal model. Here, we have used our kinetic parameterization of seafloor weathering rather than the arbitrary $pCO_2$ power law adopted in previous studies (*SI Appendixes A* and *B*). We have also assumed a range of continental growth curves (Fig. 2*A*), a range of Phanerozoic biological weathering enhancements (Fig. 2*C*), and a range of temperature dependencies of weathering from ref. 36. Proxies for surface temperature, atmospheric $pCO_2$, and seafloor weathering flux are plotted alongside model outputs for comparison (*SI Appendix D*). For all results, both 95% confidence intervals and median values are plotted for key carbon cycle outputs. Median values are calculated at each time step, and consequently their evolution does not necessarily resemble the most probable time evolution of carbon cycle variables. In practice, however, most individual model realizations tend to track the median, at least qualitatively (*SI Appendix A*).

We observe that modeled temperatures are relatively constant throughout Earth history, with Archean temperatures ranging from 271 to 314 K. The combination of continental and seafloor weathering efficiently buffers climate against changes in luminosity, outgassing, and biological evolution. This temperature history is broadly consistent with glacial constraints and recent isotope proxies (Fig. 3*D*). The continental weathering buffer dominates over the seafloor weathering buffer for most of Earth history, but in the Archean the two carbon sinks are comparable (*SI Appendix*, Fig. S1). Indeed, if seafloor weathering were artificially held constant, then continental weathering alone may be unable to efficiently buffer the climate of the early Earth—the temperature distribution at 4.0 Ga extends to 370 K, and the atmospheric $pCO_2$ distribution extends to 7 bar (*SI Appendix*, Fig. S3).

In our nominal model, the median Archean surface temperature is slightly higher than modern surface temperatures. If solar evolution were the only driver of the carbon cycle, then Archean temperatures would necessarily be cooler than modern temperatures; weathering feedbacks can mitigate this cooling but not produce warming. Warmer Archean climates are possible because elevated internal heat flow, lower continental land fraction, and lessened biological enhancement of weathering all act to warm to Precambrian climate. These three factors produce a comparable warming effect (*SI Appendix*, Fig. S17*A* and *Appendix C*), although the magnitude of each is highly uncertain and so temperate Archean temperatures cannot be uniquely attributed to any one variable.

Continental and seafloor weathering also buffer ocean pH against changes in luminosity and outgassing. Ocean pH increases monotonically over Earth history from 6.3–7.2 at 4.0 Ga, to 6.5–7.7 at 2.5 Ga, and to the modern value of 8.2. The broad range of parameterizations does not tightly constrain the history of atmospheric $pCO_2$, but the model $pCO_2$ outputs encompass paleosol proxies (Fig. 3*B*).

The results described above assume an Archean landmass fraction between 0.1 and 0.75 times the modern land fraction. Next, we consider the endmember scenario of zero Archean landmass. This is unrealistic because abundant evidence exists for Archean land (48). However, a zero land fraction case could represent a scenario where continental fraction is sufficiently small that continental silicate weathering becomes supply limited (e.g., ref. 49).

Fig. 4 shows model outputs for the zero land fraction case. When continental weathering drops to zero, seafloor weathering increases dramatically to balance the carbon cycle (Fig. 4*F*). This is largely a consequence of the temperature-dependent feedback of seafloor weathering. The climate warms by 10–15 K (Fig. 4*D*) before the temperature-dependent seafloor weathering flux is sufficiently large to balance the carbon cycle. Even in this extreme case, the median Archean temperature is ~305 K, and the upper end of the temperature distribution at 4 Ga only extends to ~328 K, excluding a "hot" Archean of 60–80 °C. Archean $pCO_2$ (and pH) are slightly higher (lower) in the zero land case, but the seafloor weathering feedback is still an effective buffer.

Finally, we investigated whether the inclusion of methane as a Precambrian greenhouse gas would substantially change our results.

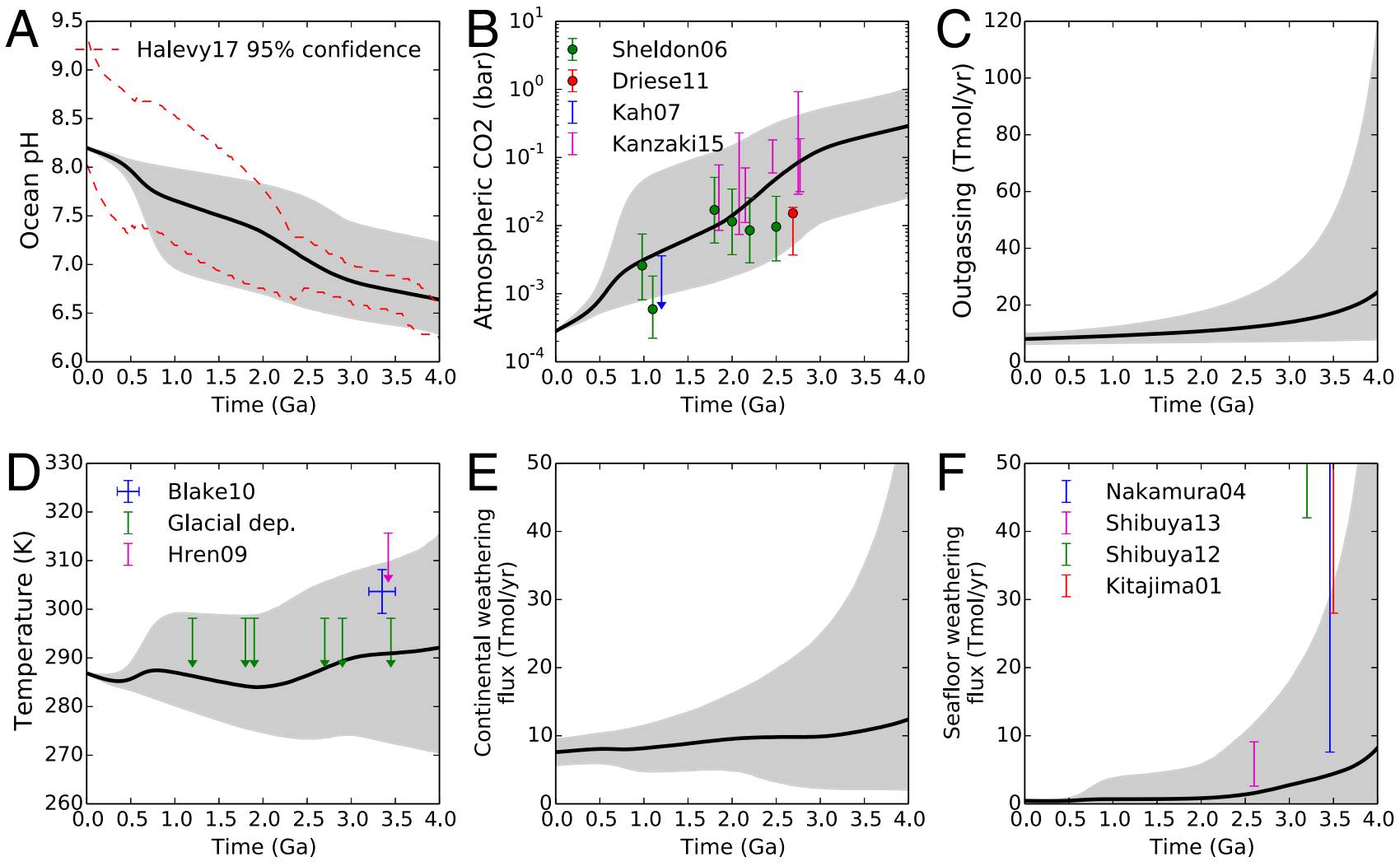

**Fig. 3.** Nominal model outputs. Gray shaded regions represent 95% confidence intervals, and black lines are the median outputs. (*A*) Ocean pH with the 95% confidence interval from Halevy and Bachan (29) plotted with red dashed lines for comparison. Our model predicts a monotonic evolution of pH from slightly acidic values at 4.0 Ga to slightly alkaline modern values. (*B*) Atmospheric $pCO_2$ plotted alongside proxies from the literature. (*C*) Global outgassing flux. (*D*) Mean surface temperature plotted alongside glacial and geochemical proxies from the literature. Our model predicts surface temperatures have been temperate throughout Earth history. (*E*) Continental silicate weathering flux. (*F*) Seafloor weathering flux plotted alongside flux estimates from Archean altered seafloor basalt. dep, deposit.

Fig. 5 shows model outputs where we have assumed 100 ppm Proterozoic methane and 1% Archean methane levels (*SI Appendix E*). The temperature changes are smaller than what might be expected if only methane levels were changing. This is because $pCO_2$ drops in response to the imposed temperature increase—$pCO_2$ must drop otherwise weathering sinks would exceed source fluxes. The $pCO_2$ distribution at 4.0 Ga is shifted downward relative to the nominal case with no other greenhouse cases, and ocean pH increases in response to this $pCO_2$ drop. Note that for parts of parameter space where $CO_2/CH_4 \gtrsim 0.2$ (50), our temperatures should be considered upper limits because a photochemical haze would form, cooling the climate (*SI Appendix E*).

Thus, even with considerable warming from an additional greenhouse gas, the median temperature at 4.0 Ga is below 300 K, and the temperature distribution extends to 320 K, again excluding a hot Archean. *SI Appendix*, Fig. S7 shows the results for the most extreme case of no Archean land and high methane abundances. Even in this extreme scenario, the seafloor weathering flux successfully buffers the climate to a median 4.0 Ga value of ∼310 K. Archean pH values are closer to circumneutral when methane is included due to lower $pCO_2$, but there is still a monotonic evolution in pH over Earth history.

## Discussion

Previously, Sleep and Zahnle (12) modeled the evolution of the geological carbon cycle over Earth history and reported the Hadean and Archean mean surface temperatures below 0 °C, unless atmospheric methane abundances were very high. In contrast, we find that Archean temperatures were likely temperate, regardless of methane abundances. This disparity can be ascribed to the differing treatments of seafloor weathering. Sleep and Zahnle (12) did not include ocean chemistry in their model (they effectively fix pH), and were thus forced to parameterize seafloor weathering using a power-law $pCO_2$ dependence with a fitted exponent. This parameterization overestimates the role of Archean seafloor weathering. Experiments with basalt dissolution reveal a weak pH dependence and a moderate temperature dependence, but no direct $pCO_2$ dependence (see discussion in ref. 36).

The change from a $CO_2$-dependent parameterization to a temperature-dependent parameterization means the seafloor weathering feedback better stabilizes climate against increasing luminosity. For a purely temperature-dependent weathering feedback, decreasing luminosity does not change climate, as the weathering flux must remain constant to maintain carbon cycle balance. Instead, $CO_2$ adjusts upwards to maintain the same temperature at lower insolation. In contrast, for a purely $CO_2$-dependent weathering feedback, a decrease in solar luminosity will result in a temperature decrease (see also *SI Appendix B*). In short, the pH-dependent and temperature-dependent seafloor weathering parameterization we apply stabilizes climate and prevents a globally glaciated early Earth. This result is broadly consistent with a single time point at 3.8 Ga that calculated equilibrium surface temperatures using a GCM and geological carbon cycle model (13).

The only way to produce Archean climates below 0 °C in our model is to assume the Archean outgassing flux was 1–5× lower than the modern flux (*SI Appendix*, Fig. S12). However, dramatically lowered Archean outgassing fluxes contradict known outgassing proxies and probably require both a stagnant lid tectonic regime and a mantle more reduced than zircon data suggest, which lowers the portion of outgassed $CO_2$ (*SI Appendix C*). Moreover, even when outgassing is low, frozen climates are not guaranteed (*SI Appendix*, Fig. S12).

Our model gives a monotonic evolution of ocean pH from 6.3–7.7 in the Archean (95% confidence), to 6.5–8.1 (95% confidence) in the Proterozoic, and increasing to 8.2 in the modern surface ocean. This history is broadly consistent with that of Halevy and Bachan (29) (Figs. 3 and 4). Halevy and Bachan (29) tracked Na, Cl, Mg, and K exchanges with continental and oceanic crust, and related these fluxes to the thermal evolution of the Earth. Minor constituents such as HS, $NH_3$, $Fe^{2+}$, and $SO_4^{2-}$ were also considered. However, they prescribe many features of the carbon cycle rather than apply a self-consistent model as we have done here. Specifically, they imposed $pCO_2$ to ensure near-modern temperatures throughout Earth history. Consequently, the explicit temperature dependence of both seafloor and continental weathering were omitted. Additionally, subduction and outgassing were assumed to be directly proportional, a limited range of heat flow histories were adopted, and continental silicate weathering was described using an overall $pCO_2$ power-law dependence with no allowance for changing land fraction or biogenic enhancement weathering. Thus, the uncertainty envelopes for the early Earth ocean pH are underestimated in ref. 29 as can be seen by their uncertainty diminishing further back in time. Good agreement with the results of ref. 29 confirms that the details of ocean chemistry are of secondary importance to pH evolution, and that the monotonic evolution of pH is instead driven by solar luminosity evolution, buffered by enhanced continental and seafloor weathering under high $pCO_2$ conditions. The two models also agree because the carbon cycle buffers to near-modern temperatures, allowing the constant temperature assumption of ref. 29, but there is no way of knowing the effectiveness of the buffer without a self-consistent model of the carbon cycle.

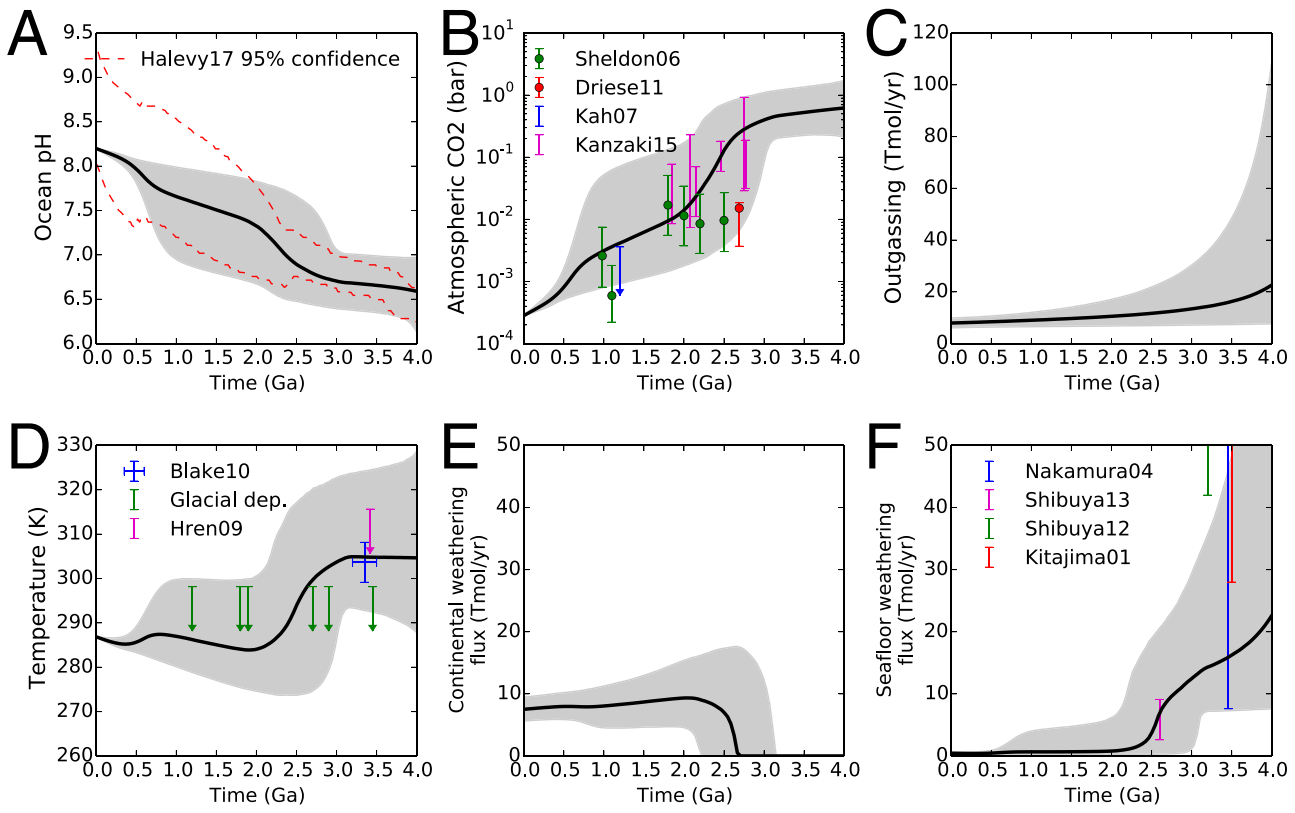

**Fig. 4.** No Archean land endmember scenario. Panels *A–F*, lines, and, shadings are the same as in Fig. 3. (*E*) Continental weathering drops to zero in the Archean, but (*F*) seafloor weathering increases due to its temperature dependence to balance the carbon cycle. This causes an increase in surface temperature in the Archean, (*D*) but conditions are still temperate throughout Earth history. The evolution of (*A*) ocean pH and (*B*) $pCO_2$ are similar to the nominal model. dep, deposit.

One caveat for our results is that the ocean chemistry is incomplete. Specifically, Ca abundances in the ocean and pore space are controlled entirely by alkalinity fluxes from continental and seafloor weathering. In reality, Ca abundances are modulated by other processes such as the hydrothermal exchange of Ca and Mg in the seafloor, dolomitization, and clay formation (51). To explore whether neglecting these processes would affect our results we conducted sensitivity tests with a large ensemble of Ca evolutions (*SI Appendix C*). High Archean Ca abundances might be expected to produce more acidic oceans because carbonate abundances are lower for the same saturation state. However, this effect is buffered by decreases in Ca and $CO_3^{2-}$ activity coefficients (complexing), and so model outputs look very similar to our nominal model for a broad range of Ca abundance trajectories (*SI Appendix*, Fig. S9).

In our nominal climate model, we did not include the effects of changing atmospheric pressure or albedo changes. These effects are likely to be modest compared with the other sources of uncertainty in our model. Lower surface albedo from a reduced Archean land fraction can contribute at most 5 W/m$^2$ of radiative forcing (52), which would cause only a few degrees of warming. Halving Archean total pressure—as has been suggested by paleopressure proxies (53)—would cool the Earth by ~5 K because of the loss of pressure broadening, thereby offsetting the lower land fraction (54). Changes in cloud cover could, in principle, induce larger warming, but the required conditions for >10 K warming are highly speculative (52). In any case, the effects of pressure changes and albedo changes are unlikely to affect our conclusions because the temperature changes they induce will be compensated by $pCO_2$ variations to balance the carbon cycle. Sensitivity analyses where massive amounts of Archean warming are imposed (+30 K) still result in temperate surface temperatures because of this $pCO_2$ compensation (*SI Appendix*, Fig. S10).

Although our model outputs are broadly consistent with paleosol proxies and glacial constraints, some disagreement occurs with selected seafloor weathering proxies. Proxies for seafloor carbonate precipitation were estimated by using the average carbonate abundances in Archean oceanic crust, scaled by the model spreading rate at that time multiplied by an assumed carbonatization depth (*SI Appendix D*). Our modeled seafloor carbonate precipitation fluxes agree with that of Nakamura and Kato (55) and Shibuya et al. (56), but undershoot crustal carbonate abundances reported by Shibuya et al. (57) and Kitajima et al. (58). It is difficult to construct a balanced carbon cycle model with seafloor weathering fluxes in excess of 100 Tmol C/y as these latter two studies imply, and so the discrepancy may be because those oceanic crust samples are not representative of global carbonatization flux, or because some of the carbonate is secondary. The only way to approach the carbonate abundances reported by Shibuya et al. (57) and Kitajima et al. (58) is to impose very high Archean outgassing (e.g., up to 60× the modern flux; *SI Appendix*, Fig. S11), but even then the fit is marginal. If high Archean crustal carbonate estimates were truly primary, then Archean outgassing would have been very high, and so Earth's internal heat flow would have decreased dramatically over Earth's history, contrary to Korenaga (59).

We conclude that current best knowledge of Earth's geologic carbon cycle precludes a hot Archean. Our results are insensitive to assumptions about ocean chemistry, internal evolution, and weathering parameterizations, so a hot early Earth would require some fundamental error in current understanding of the carbon cycle. Increasing the biotic enhancement of weathering by several orders of magnitude as proposed by Schwartzman (60) does not produce a hot Archean because this is mathematically equivalent to zeroing out the continental weathering flux (Fig. 4). In this case the temperature-dependent seafloor weathering feedback buffers the climate of the Earth to moderate temperatures (*SI Appendix*, Fig. S14). Dramatic temperature increases (or decreases) due to albedo changes also do not change our conclusions due to the buffering effect of the carbon cycle (see above). If both continental and seafloor weathering become supply limited (e.g., refs. 49 and 61), then temperatures could easily exceed 50 °C. However, in this case the carbon cycle would be out of balance, leading to excessive $pCO_2$ accumulation within a few hundred million years unless buffered by some other, unknown feedback.

## Conclusions

The early Earth was probably temperate. Continental and seafloor weathering buffer Archean surface temperatures to 0–50 °C. This result holds for a broad range of assumptions about the evolution of internal heat flow, crustal production, spreading rates, and the biotic enhancement of continental weathering.

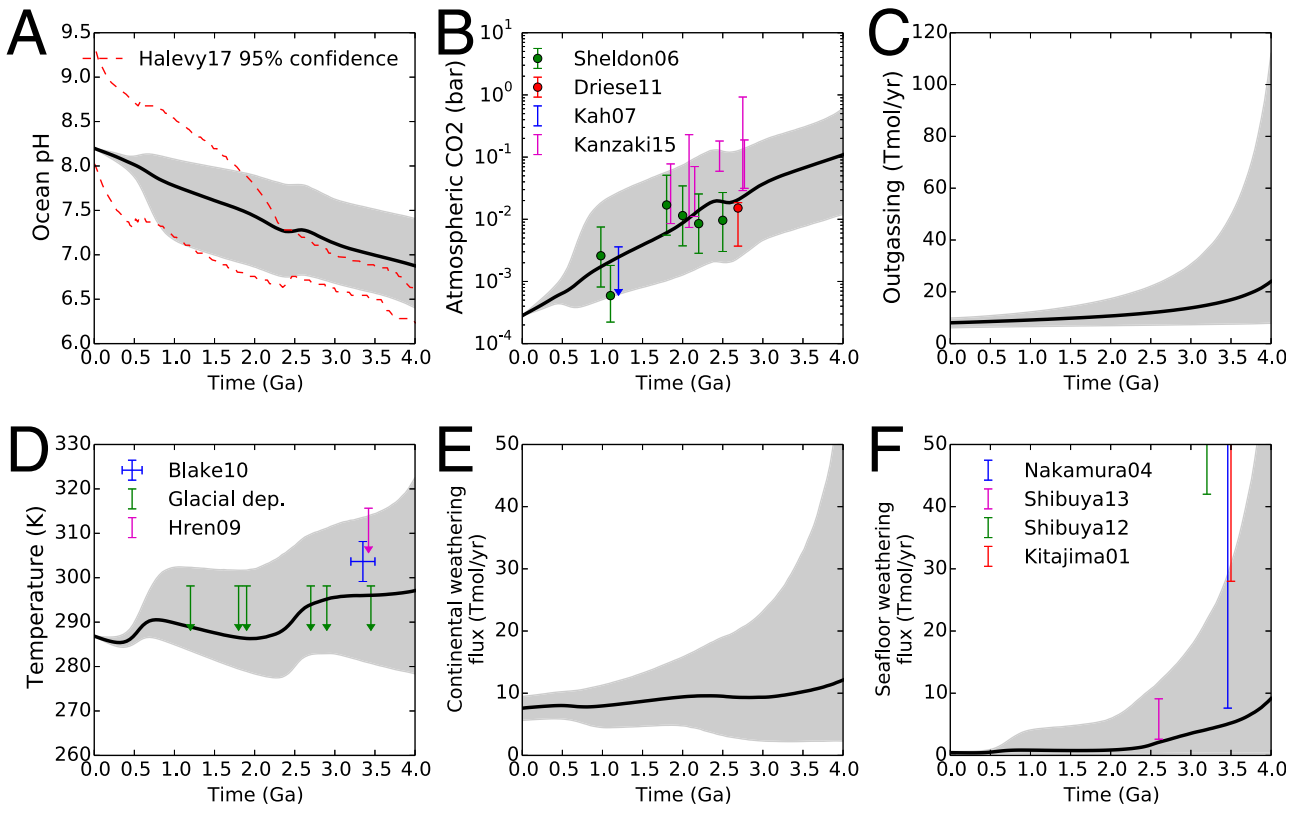

**Fig. 5.** Default continental growth range with imposed 100 ppm methane in the Proterozoic and 1% methane in the Archean. Panels *A–F*, lines, and, shadings are the same as in Fig. 3. (*D*) Temperature increases sharply in the Archean due to methane, but by less than what would be expected if $pCO_2$ were unchanged. In practice, there is a compensating decrease in atmospheric $pCO_2$, (*B*) which must occur to balance the carbon cycle. Otherwise temperatures would be too high and weathering sinks would exceed outgassing sources. Because $pCO_2$ is lower, Archean pH values are closer to circumneutral (*A*). dep, deposit.

Even in extreme scenarios with negligible subaerial Archean land and high methane abundances, a hot Archean (>50 °C) is unlikely. Sub-0 °C climates are also unlikely unless the Archean outgassing flux was unrealistically lower than the modern flux.

The seafloor weathering feedback is important, but less dominant than previously assumed. Consequently, the early Earth would not have been in a snowball state due to $pCO_2$ drawdown from seafloor weathering. In principle, little to no methane is required to maintain a habitable surface climate, although methane should be expected in the anoxic Archean atmosphere once methanogenesis evolved (ref. 62, chap. 11).

Ignoring transient excursions, the pH of Earth's ocean has evolved monotonically from $6.6^{+0.6}_{-0.4}$ at 4.0 Ga (2σ) to $7.0^{+0.7}_{-0.5}$ at 2.5 Ga (2σ), and 8.2 in the modern ocean. This evolution is robust to assumptions about ocean chemistry, internal heat flow, and other carbon cycle parameterizations. Consequently, similar feedbacks may control ocean pH and climate on other Earthlike planets with basaltic seafloors and silicate continents, suggesting that life elsewhere could emerge in comparable environments to those on our early planet.

**ACKNOWLEDGMENTS.** We thank Roger Buick, Michael Way, Anthony Del Genio, and Mark Chandler, and the two anonymous reviewers for helpful discussions and insightful contributions. This work was supported by NASA Exobiology Program Grant NNX15AL23G awarded to D.C.C., the Simons Collaboration on the Origin of Life Award 511570, and by the NASA Astrobiology Institute's Virtual Planetary Laboratory, Grant NNA13AA93A. J.K.-T. is supported by NASA Headquarters under the NASA Earth and Space Science Fellowship program, Grant NNX15AR63H.

1. Knauth LP (2005) Temperature and salinity history of the Precambrian ocean: Implications for the course of microbial evolution. *Palaeogeogr Palaeoclimatol Palaeoecol* 219:53–69.
2. Knauth LP, Lowe DR (2003) High Archean climatic temperature inferred from oxygen isotope geochemistry of cherts in the 3.5 Ga Swaziland Supergroup, South Africa. *Geol Soc Am Bull* 115:566–580.
3. Fralick P, Carter JE (2011) Neoarchean deep marine paleotemperature: Evidence from turbidite successions. *Precambrian Res* 191:78–84.
4. Gaucher EA, Govindarajan S, Ganesh OK (2008) Palaeotemperature trend for Precambrian life inferred from resurrected proteins. *Nature* 451:704–707.
5. Garcia AK, Schopf JW, Yokobori SI, Akanuma S, Yamagishi A (2017) Reconstructed ancestral enzymes suggest long-term cooling of Earth's photic zone since the Archean. *Proc Natl Acad Sci USA* 114:4619–4624.
6. Robert F, Chaussidon M (2006) A palaeotemperature curve for the Precambrian oceans based on silicon isotopes in cherts. *Nature* 443:969–972.
7. Kasting JF, et al. (2006) Paleoclimates, ocean depth, and the oxygen isotopic composition of seawater. *Earth Planet Sci Lett* 252:82–93.
8. van den Boorn SH, van Bergen MJ, Nijman W, Vroon PZ (2007) Dual role of seawater and hydrothermal fluids in early Archean chert formation: Evidence from silicon isotopes. *Geology* 35:939–942.
9. Hren MT, Tice MM, Chamberlain CP (2009) Oxygen and hydrogen isotope evidence for a temperate climate 3.42 billion years ago. *Nature* 462:205–208.
10. Blake RE, Chang SJ, Lepland A (2010) Phosphate oxygen isotopic evidence for a temperate and biologically active Archaean ocean. *Nature* 464:1029–1032.
11. de Wit MJ, Furnes H (2016) 3.5-Ga hydrothermal fields and diamictites in the Barberton Greenstone Belt-Paleoarchean crust in cold environments. *Sci Adv* 2:e1500368.
12. Sleep NH, Zahnle K (2001) Carbon dioxide cycling and implications for climate on ancient Earth. *J Geophys Res Planets* 106:1373–1399.
13. Charnay B, Hir GL, Fluteau F, Forget F, Catling DC (2017) A warm or a cold early Earth? New insights from a 3-D climate-carbon model. *Earth Planet Sci Lett* 474:97–109.
14. Keller MA, Kampjut D, Harrison SA, Ralser M (2017) Sulfate radicals enable a non-enzymatic Krebs cycle precursor. *Nat Ecol Evol* 1:83.
15. Dora Tang TY, et al. (2014) Fatty acid membrane assembly on coacervate microdroplets as a step towards a hybrid protocell model. *Nat Chem* 6:527–533.
16. Powner MW, Sutherland JD, Szostak JW (2010) Chemoselective multicomponent one-pot assembly of purine precursors in water. *J Am Chem Soc* 132:16677–16688.
17. Konhauser K, Riding R (2012) Bacterial biomineralization. *Fundamentals of Geobiology* (Wiley–Blackwell, Hoboken, NJ), pp 105–130.
18. Knoll AH (2003) Biomineralization and evolutionary history. *Rev Mineral Geochem* 54:329–356.
19. Liu X, et al. (2016) Tracing Earth's O2 evolution using Zn/Fe ratios in marine carbonates. *Geochem Persp Let* 2:24–34.
20. Garvin J, Buick R, Anbar AD, Arnold GL, Kaufman AJ (2009) Isotopic evidence for an aerobic nitrogen cycle in the latest Archean. *Science* 323:1045–1048.
21. Planavsky NJ, et al. (2014) Earth history. Low mid-Proterozoic atmospheric oxygen levels and the delayed rise of animals. *Science* 346:635–638.
22. Grotzinger JP, Kasting JF (1993) New constraints on Precambrian ocean composition. *J Geol* 101:235–243.
23. Buick R, Dunlop J (1990) Evaporitic sediments of early Archaean age from the Warrawoona Group, North Pole, Western Australia. *Sedimentology* 37:247–277.
24. Blättler C, et al. (2017) Constraints on ocean carbonate chemistry and pCO2 in the Archaean and Palaeoproterozoic. *Nat Geosci* 10:41–45.
25. Friend CR, Nutman AP, Bennett VC, Norman M (2008) Seawater-like trace element signatures (REE+ Y) of Eoarchaean chemical sedimentary rocks from southern West Greenland, and their corruption during high-grade metamorphism. *Contrib Mineral Petrol* 155:229–246.
26. Kempe S, Degens ET (1985) An early soda ocean? *Chem Geol* 53:95–108.
27. Sleep NH, Zahnle K, Neuhoff PS (2001) Initiation of clement surface conditions on the earliest Earth. *Proc Natl Acad Sci USA* 98:3666–3672.
28. Stüeken E, Buick R, Schauer A (2015) Nitrogen isotope evidence for alkaline lakes on late Archean continents. *Earth Planet Sci Lett* 411:1–10.
29. Halevy I, Bachan A (2017) The geologic history of seawater pH. *Science* 355:1069–1071.
30. Berner RA (2004) *The Phanerozoic Carbon Cycle: CO2 and O2* (Oxford Univ Press, New York).
31. Walker JC, Hays P, Kasting JF (1981) A negative feedback mechanism for the long-term stabilization of Earth's surface temperature. *J Geophys Res Oceans* 86:9776–9782.
32. Brady PV, Gíslason SR (1997) Seafloor weathering controls on atmospheric CO 2 and global climate. *Geochim Cosmochim Acta* 61:965–973.
33. Coogan LA, Dosso SE (2015) Alteration of ocean crust provides a strong temperature dependent feedback on the geological carbon cycle and is a primary driver of the Sr-isotopic composition of seawater. *Earth Planet Sci Lett* 415:38–46.
34. Berner RA, Caldeira K (1997) The need for mass balance and feedback in the geochemical carbon cycle. *Geology* 25:955–956.
35. Franck S, Kossacki KJ, von Bloh W, Bounama C (2002) Long-term evolution of the global carbon cycle: Historic minimum of global surface temperature at present. *Tellus B Chem Phys Meterol* 54:325–343.
36. Krissansen-Totton J, Catling DC (2017) Constraining climate sensitivity and continental versus seafloor weathering using an inverse geological carbon cycle model. *Nat Commun* 8:15423.
37. Taylor L, Banwart S, Leake J, Beerling DJ (2011) Modeling the evolutionary rise of ectomycorrhiza on sub-surface weathering environments and the geochemical carbon cycle. *Am J Sci* 311:369–403.
38. Moulton KL, West J, Berner RA (2000) Solute flux and mineral mass balance approaches to the quantification of plant effects on silicate weathering. *Am J Sci* 300:539–570.
39. Arthur M, Fahey T (1993) Controls on soil solution chemistry in a subalpine forest in north-central Colorado. *Soil Sci Soc Am J* 57:1122–1130.
40. Krissansen-Totton J, Buick R, Catling DC (2015) A statistical analysis of the carbon isotope record from the Archean to Phanerozoic and implications for the rise of oxygen. *Am J Sci* 315:275–316.
41. Shirey SB, Richardson SH (2011) Start of the Wilson cycle at 3 Ga shown by diamonds from subcontinental mantle. *Science* 333:434–436.
42. Delavault H, Chauvel C, Thomassot E, Devey CW, Dazas B (2016) Sulfur and lead isotopic evidence of relic Archean sediments in the Pitcairn mantle plume. *Proc Natl Acad Sci USA* 113:12952–12956.
43. Moore WB, Webb AAG (2013) Heat-pipe Earth. *Nature* 501:501–505.
44. Foley SF, Buhre S, Jacob DE (2003) Evolution of the Archaean crust by delamination and shallow subduction. *Nature* 421:249–252.
45. Debaille V, et al. (2013) Stagnant-lid tectonics in early Earth revealed by 142 Nd variations in late Archean rocks. *Earth Planet Sci Lett* 373:83–92.
46. Kopparapu RK, et al. (2013) Habitable zones around main-sequence stars: New estimates. *Astrophys J* 765:131.
47. Gough D (1981) Solar interior structure and luminosity variations. *Sol Phys* 74:21–34.
48. Sleep NH (2010) The Hadean-Archaean environment. *Cold Spring Harb Perspect Biol* 2:a002527.
49. Foley BJ (2015) The role of plate tectonic-climate coupling and exposed land area in the development of habitable climates on rocky planets. *Astrophys J* 812:36.
50. Trainer MG, et al. (2006) Organic haze on Titan and the early Earth. *Proc Natl Acad Sci USA* 103:18035–18042.
51. Higgins JA, Schrag DP (2015) The Mg isotopic composition of Cenozoic seawater–Evidence for a link between Mg-clays, seawater Mg/Ca, and climate. *Earth Planet Sci Lett* 416:73–81.
52. Goldblatt C, Zahnle K (2010) Clouds and the faint young sun paradox. *Clim Past Discuss* 6:1163–1207.
53. Som SM, et al. (2016) Earth's air pressure 2.7 billion years ago constrained to less than half of modern levels. *Nat Geosci* 9:448–451.
54. Goldblatt C, et al. (2009) Nitrogen-enhanced greenhouse warming on early Earth. *Nat Geosci* 2:891–896.
55. Nakamura K, Kato Y (2004) Carbonatization of oceanic crust by the seafloor hydrothermal activity and its significance as a CO 2 sink in the early Archean. *Geochim Cosmochim Acta* 68:4595–4618.
56. Shibuya T, et al. (2013) Decrease of seawater CO 2 concentration in the late Archean: An implication from 2.6 Ga seafloor hydrothermal alteration. *Precambrian Res* 236:59–64.
57. Shibuya T, et al. (2012) Depth variation of carbon and oxygen isotopes of calcites in Archean altered upperoceanic crust: Implications for the CO 2 flux from ocean to oceanic crust in the Archean. *Earth Planet Sci Lett* 321:64–73.
58. Kitajima K, Maruyama S, Utsunomiya S, Liou J (2001) Seafloor hydrothermal alteration at an Archaean mid-ocean ridge. *J Metamorph Geol* 19:583–599.
59. Korenaga J (2008) Plate tectonics, flood basalts and the evolution of Earth's oceans. *Terra Nova* 20:419–439.
60. Schwartzman D (2002) *Life, Temperature, and the Earth: The Self-Organizing Biosphere* (Columbia Univ Press, New York).
61. Abbot DS, Cowan NB, Ciesla FJ (2012) Indication of insensitivity of planetary weathering behavior and habitable zone to surface land fraction. *Astrophys J* 756:178.
62. Catling DC, Kasting JF (2017) *Atmospheric Evolution on Inhabited and Lifeless Worlds* (Cambridge Univ Press, Cambridge, UK).

## Supplementary Material

### Appendix A: Additional Model Description

System of equations and weathering formulations

The system of equations describing the carbon cycle model are as follows (1):

$$
\begin{aligned}
\frac{dC_{\mathrm{O}}}{dt} &= \frac{-J\left(C_{\mathrm{O}}-C_{\mathrm{P}}\right)}{M_{\mathrm{O}}}+\frac{F_{\mathrm{out}}}{M_{\mathrm{O}}}+\frac{F_{\mathrm{carb}}}{M_{\mathrm{O}}}-\frac{P_{\mathrm{ocean}}}{M_{\mathrm{O}}} \\
\frac{dA_{\mathrm{O}}}{dt} &= \frac{-J\left(A_{\mathrm{O}}-A_{\mathrm{P}}\right)}{M_{\mathrm{O}}}+2\frac{F_{\mathrm{sil}}}{M_{\mathrm{O}}}+2\frac{F_{\mathrm{carb}}}{M_{\mathrm{O}}}-2\frac{P_{\mathrm{ocean}}}{M_{\mathrm{O}}} \\
\frac{dC_{\mathrm{P}}}{dt} &= \frac{J\left(C_{\mathrm{O}}-C_{\mathrm{P}}\right)}{M_{\mathrm{P}}}-\frac{P_{\mathrm{pore}}}{M_{\mathrm{P}}} \\
\frac{dA_{\mathrm{P}}}{dt} &= \frac{J\left(A_{\mathrm{O}}-A_{\mathrm{P}}\right)}{M_{\mathrm{P}}}+2\frac{F_{\mathrm{diss}}}{M_{\mathrm{P}}}-2\frac{P_{\mathrm{pore}}}{M_{\mathrm{P}}}
\end{aligned}
\qquad \text{(S1)}
$$

Here, $C_{\mathrm{O}}$ and $C_{\mathrm{P}}$ are the concentrations of carbon (Tmol C kg$^{-1}$) in the atmosphere-ocean and pore space, respectively. The carbon concentration in the pore-space is equivalent to the Dissolved Inorganic Carbon (DIC) abundance, $C_{\mathrm{P}}=\mathrm{DIC}_{\mathrm{P}}$, whereas carbon in the atmosphere-ocean reservoir is equal to marine dissolved inorganic carbon plus atmospheric carbon, $C_{\mathrm{O}}=\mathrm{DIC}_{\mathrm{O}}+\mathrm{pCO}_2\times s$, where $s$ is a scaling factor equal to the ratio of total number of moles per bar in the atmosphere divided by the mass of the ocean, $s=\left(1.8\times10^{20}\ \text{moles/bar}\right)/M_{\mathrm{O}}$, and $\mathrm{pCO}_2$ is in bar. Similarly, $A_{\mathrm{O}}=\mathrm{ALK}_{\mathrm{O}}$ and $A_{\mathrm{P}}=\mathrm{ALK}_{\mathrm{P}}$ are the carbonate alkalinities in the atmosphere-ocean and pore space, respectively (Tmol eq kg$^{-1}$). The global outgassing flux (Tmol C yr$^{-1}$) is specified by $F_{\mathrm{out}}$, whereas the rates of continental silicate weathering and carbonate weathering are $F_{\mathrm{sil}}$ and $F_{\mathrm{carb}}$, respectively (Tmol C yr$^{-1}$). Seafloor weathering from basalt dissolution (Tmol eq yr$^{-1}$) is $F_{\mathrm{diss}}$, and the precipitation flux of carbonates (Tmol C yr$^{-1}$) in the ocean and pore space are given by $P_{\mathrm{ocean}}$ and $P_{\mathrm{pore}}$, respectively. The mass of the ocean and the pore space are given by $M_{\mathrm{O}}=1.35\times10^{21}$ kg and $M_{\mathrm{P}}=1.35\times10^{19}$ kg, respectively (2). We do not track crustal and mantle reservoirs of carbon because of the difficulties associated with coupling crustal and mantle reservoirs described in the main text, and also because it enables us to reverse the direction of integration as the atmosphere-ocean system is always in quasi steady state (1).

Continental carbonate weathering is described by a similar function to continental silicate weathering (see main text), but we allow for a different $\mathrm{pCO}_2$ dependence:

$$
F_{\mathrm{carb}}=f_{bio}f_{land}F_{\mathrm{carb}}^{\mathrm{mod}}\left(\frac{\mathrm{pCO}_2}{\mathrm{pCO}_2^{\mathrm{mod}}}\right)^{\xi}\exp\left(\Delta T_{\mathrm{S}}/T_e\right) \qquad \text{(S2)}
$$

Here, $\xi$ is the exponent defining the $pCO_2$ dependence of carbonate weathering, and $F_{\text{carb}}^{\text{mod}}$ is the modern carbonate weathering flux (Tmol C $yr^{-1}$).

The dissolution of basalt in the seafloor, $F_{diss}$, is described using the following parameterization:

$$F_{\text{diss}} = k_{\text{diss}} r_{\text{spread-rate}} \exp\left(-E_{\text{bas}}/RT_{\text{pore}}\right)\left(\frac{\left[\text{H}^+\right]_{\text{P}}}{\left[\text{H}^+\right]_{\text{P}}^{\text{mod}}}\right)^{\gamma} \tag{S3}$$

Here $k_{\text{diss}}$ is a proportionality constant chosen to match the modern flux, $r_{\text{spread-rate}}$ is the spreading rate (see main text), $R$ is the universal gas constant, $\left[\text{H}^+\right]_{\text{P}}$ is the hydrogen ion molality in the pore-space, and $\left[\text{H}^+\right]_{\text{P}}^{\text{mod}}$ is the modern molality. The exponent $\gamma$ ranges from 0 to 0.5 based on kinetic experiments of the dissolution of basalt (see discussion in (1)). The temperature of the pore space is $T_{\text{pore}}$ (K), and the temperature dependence of basalt dissolution is described by an effective activation energy, $E_{\text{bas}} = 60-100$ kJ/mol (1).

Relationship between pore water and ocean temperature

By Fourier's Law, the temperature of the pore space will also depend on the heat flow from the interior relative to modern, $Q$, and the ocean sediment thickness, $S_{\text{thick}}$, as:

$$T_{\text{Pore}} = T_D + QS_{\text{thick}}/K \tag{S4}$$

Here, $K$ is conductivity of sediments. Throughout the Cenozoic and Mesozoic, the pore space is approximately 9 K warmer than the deep ocean (3). Modern sediment thickness is ~700m (4), and modern relative heat flow is $Q=1$. We thus fit the effective conductivity to be $K = 700/9 = 77.8$ m/K. Relative heat flow at earlier times is defined below. We assumed a range of sediment thickness evolutions given by:

$$S_{\text{thick}} = 700(1-(1-f_{\text{sed}})t/4) \tag{S5}$$

Here, $f_{\text{sed}}$ is the relative depth of sediments at 4.0 Ga, assumed to range 0.2 to 1. This range reflects the possible effects of lower Archean land fraction on marine sedimentation depth.

Land fraction and biological enhancement of weathering

Relative land fraction, $f_{land}$, is specified by the equation:

$$f_{land} = max\left\{0, 1-\frac{1}{1/(1-L_{\text{Archean}})+\exp(-10(t-t_{\text{grow}}))}\right\} \tag{S6}$$

Here, $L_{\text{Archean}}$ is the land fraction in the Archean relative to modern, $t$ is the time (Ga), and $t_{\text{grow}}$ =2-3 Ga is the time the continents grew from their Archean value to modern values. We nominally assume $L_{\text{Archean}}$ =0.1-0.75, but later set $L_{\text{Archean}}$ =-0.2 to represent a zero Archean land mass scenario. Equation (S6) describes a smooth step function that transitions from 1 to $L_{\text{Archean}}$ at $t_{\text{grow}}$ (Fig. 2a). The maximum value is necessary to prevent negative land fractions when modeling scenarios where the land fraction drops to zero at $t_{\text{grow}}$ (Fig. 2b). Our assumed function for the evolution of land fraction is compared to various literature estimates in Fig. 2a (5-9).

Note that $f_{land}$ in equation (S6) refers to emergent continent, whereas continental grow curves in Fig. 2a purportedly track the volume of continental crust. Early continents may have been submerged due to the early Earth's thermal state (10), and so the growth of continental crustal volume may or not map to emergent continental fraction. Note however, that there is rare Earth element evidence for extensive continental weathering at 2.7 Ga (11). In any case, the broad range of growth histories we consider, including zero emerged Archean continents, accounts for the possibility of submerged continents.

The biological modification of weathering, $f_{bio}$ , is specified by the following function:

$$f_{bio} = 1 - \frac{1}{1/(1 - B_{\text{Precambrian}}) + \exp(-10(t - t_{bio}))} \quad \text{(S7)}$$

Here, $t_{bio}$ =0.6 Ga is the time at which a stepwise increase in weatherability occurs, and $B_{\text{Precambrian}}$ is the relative Precambrian weatherability which ranges from 0.1-1. Equation (S7) also describes a smoothly varying step function from 1 to $B_{\text{Precambrian}}$ at 4.0 Ga.

Internal heat flow evolution, tectonics, and outgassing parameterizations

Internal heat flow relative to modern is specified by:

$$Q = (1 - t/4.5)^{-n_{out}} \quad \text{(S8)}$$

We allow the outgassing exponent, $n_{\text{out}}$, to vary from 0 to 0.73. Fig. 2d shows the range of heat flow histories assumed in this model compared to various estimates of Earth's heat flow from the literature (12-15), including both the conventional view of declining heat flow (12) and relatively constant heat flow (14).

We relate global outgassing to heat flow using a power law:

$$F_{\text{out}} = F_{\text{out}}^{\text{mod}} Q^{m} \quad \text{(S9)}$$

Here, $F_{\text{out}}^{\text{mod}}$ is the modern outgassing flux, and the exponent, $m$, ranges from 1 to 2. Given that our assumed range of heat flow histories allow relative heat flow at 4.0 Ga to vary from 1-5 times modern (eq. (S8)), this implies Archean outgassing could be anywhere between 1 to 25 times modern levels, an extremely broad range driven by current uncertainty in the history of $Q$ (Fig. 2d). We relate heat flow, crustal production, $r_{\text{crust-prod}}$, and spreading rate, $r_{\text{spread-rate}}$, using the following set of equations:

$$\begin{aligned} r_{\text{crust-prod}} &= Q \\ r_{\text{spread-rate}} &= (r_{\text{crust-prod}})^{\beta} = Q^{\beta} \end{aligned} \tag{S10}$$

Here, $\beta$ ranges from 0 to 2. An exponent of zero represents the endmember scenario whereby spreading rates are insensitive to crustal production. This could be because the tectonic mode of the early Earth is different to the modern, or because the maximum depth at which water-rock reactions occur is unchanged despite greater crustal thickness. An exponent of 2 represents the other endmember of a strong relationship between heat flow and the volume of oceanic crust available for seawater-rock interactions (16, p. 597).

Equilibrium chemistry of the ocean and carbonate precipitation

Carbonate alkalinity (ALK) and dissolved inorganic carbon (DIC) have the following standard definitions in our model:

$$\begin{aligned} \text{DIC} &= \left[\text{CO}_3^{2-}\right] + \left[\text{HCO}_3^-\right] + \left[\text{CO}_2\text{aq}\right] \\ \text{ALK} &= 2\left[\text{CO}_3^{2-}\right] + \left[\text{HCO}_3^-\right] \end{aligned} \tag{S11}$$

Given carbon and alkalinity in the atmosphere-ocean $\left(C_\text{O}, \text{ALK}_\text{O}\right)$ or the pore-space $\left(C_\text{P}, \text{ALK}_\text{P}\right)$, we can calculate ocean chemistry using the following set of equations:

$$\left[\text{CO}_2\text{aq}\right] = \text{pCO}_2 \times \text{H}_{\text{CO}_2} \tag{S12}$$

$$\left[\text{HCO}_3^-\right] = \frac{\left[\text{CO}_2\text{aq}\right] \times \text{K}_1^*}{\left[\text{H}^+\right]} \tag{S13}$$

$$\left[\text{CO}_3^{2-}\right] = \frac{\left[\text{HCO}_3^-\right] \times \text{K}_2^*}{\left[\text{H}^+\right]} \tag{S14}$$

$$\frac{\text{ALK}}{\text{K}_1^*\text{K}_2^*}\left(1 + \frac{s}{\text{H}_{\text{CO}_2}}\right)\left[\text{H}^+\right]^2 + \frac{\left(\text{ALK-}C\right)}{\text{K}_2^*}\left[\text{H}^+\right] + \left(\text{ALK-}2C\right) = 0 \tag{S15}$$

$$pH = \text{-}\log_{10}\left(\left[\text{H}^+\right]\right) \tag{S16}$$

Here, $\text{H}_{\text{CO}_2}$ is the Henry's law constant for $CO_2$, $\left[\text{CO}_2\text{aq}\right]$ is the sum of the concentrations of free $CO_2$ and $H_2CO_3$, and $\text{K}_1^*$ and $\text{K}_2^*$ are the first and second apparent dissociation constants of carbonic acid, respectively. Temperature-dependent expressions for these constants along with a derivation of equation (S15) can be found in Krissansen-Totton and Catling (1). The scaling factor, $s$, is also described in (1). At each time-step, this set of equations is solved separately for the ocean and the pore-space by substitution the generic carbon concentration and alkalinity $\left(C, \text{ALK}\right)$ for $\left(C_\text{O}, \text{ALK}_\text{O}\right)$ and $\left(C_\text{P}, \text{ALK}_\text{P}\right)$, respectively.

In the nominal model, calcium abundances evolve dynamically with alkalinity:

$$\left[\mathrm{Ca}^{2+}\right]=0.5(\left[\mathrm{ALK}\right]-\left[\mathrm{ALK}\right]_{\mathrm{initial}})+\left[\mathrm{Ca}^{2+}\right]_{\mathrm{initial}} \tag{S17}$$

This effectively assumes the only process affecting calcium abundances is alkalinity delivery from continental and seafloor weathering. However, later we relax this assumption to allow for a very broad range of calcium molality evolutions (see Appendix B).

At each time step, the saturation state of the ocean and the pore-space are calculated from the calcium and carbonate molalities:

$$\Omega_{\mathrm{O}}=\frac{\left[\mathrm{Ca}^{2+}\right]\left[\mathrm{CO}_3^{2-}\right]_{\mathrm{O}}}{\mathrm{K}_{\mathrm{sp}}} \quad \text{and} \quad \Omega_{\mathrm{P}}=\frac{\left[\mathrm{Ca}^{2+}\right]\left[\mathrm{CO}_3^{2-}\right]_{\mathrm{P}}}{\mathrm{K}_{\mathrm{sp}}} \tag{S18}$$

Here, $\mathrm{K}_{\mathrm{sp}}=\mathrm{K}_{\mathrm{sp}}\left(\mathrm{T}\right)$ is the temperature-dependent solubility product, the equation for which is described in Krissansen-Totton and Catling (1).

Fluxes of carbonate precipitation in the ocean and pore space of the oceanic crust are parameterized as follows:

$$\begin{aligned} P_{\mathrm{ocean}} &= k_{\mathrm{ocean}} f_{land}\left(\Omega_{\mathrm{O}}-1\right)^n \\ P_{\mathrm{pore}} &= k_{\mathrm{pore}}\left(\Omega_{\mathrm{P}}-1\right)^n \end{aligned} \tag{S19}$$

Here, the constants $k_{\mathrm{ocean}}$ and $k_{\mathrm{pore}}$ are chosen to fit the modern fluxes (see Table S2). The saturation state of the ocean, $\Omega_{\mathrm{O}}$, and the pore space, $\Omega_{\mathrm{P}}$, are calculated from ocean equilibrium chemistry as described above.

Justification for linear relationship between surface and deep ocean temperatures

In Fig. 7 of Krissansen-Totton and Catling (1) we presented GCM outputs and proxy data validating the linear relationship between surface temperatures and deep ocean temperatures. Fig. S18 shows an updated version of that figure where we extended the range to higher temperatures. There is a scarcity of fully coupled GCM outputs at these high temperatures, but we have added several data points simulating Earth at high insolation with different rotation rates (17), Cenomanian Cretaceous with 4x modern $CO_2$ (18) and modern Earth with 4x $CO_2$. These simulations produce equilibrium temperatures consistent with the trend line, validating our model parameterization. The high insolation and Cretaceous simulations assume a bathtub ocean that only extends to ~1300m, and although this is largely below the thermocline in Earth's oceans, further GCM simulations are needed to determine the precise relationship between deep ocean and surface temperatures. In light of the uncertainties in this relationship, we adopt a very broad range of gradients (grey region in Fig. S18).

Based on Fig. S18, the equation relating deep ocean temperature, $T_{\mathrm{D}}$ (K), to mean surface temperatures, $T_{\mathrm{S}}$ (K) is given by:

$$T_{\mathrm{D}}=\max\{\min\{a_{\mathrm{grad}}T_{\mathrm{S}}+b_{\mathrm{int}},T_{\mathrm{S}}\},271.15\} \tag{S20}$$

We assume a broad gradient range $a_{\mathrm{grad}}$=0.8 to 1.4, whereas the intercept, $b_{\mathrm{int}}$, is chosen to ensure consistency with modern surface and deep ocean temperatures (1). The minima and maxima ensure deep ocean temperatures do not exceed surface temperatures, or pass below the freezing point of salt water (ensures the red shaded region in Fig. S18 is prohibited).

The references for the GCMs plotted in Fig. S18 are: Li*, et al.* (19), Stouffer and Manabe (20), Danabasoglu and Gent (21), Chandler*, et al.* (18), and Way*, et al.* (17). These studies were chosen because they used fully coupled atmosphere-ocean GCMs with complete ocean circulation. Paleocene-Eocene Thermal Maximum (PETM) temperatures were sourced from Jones*, et al.* (22), Last Glacial Maximum (LGM) from Clark*, et al.* (23), and Cretaceous proxies from a composite dataset described in Krissansen-Totton and Catling (1). See Krissansen-Totton and Catling (1) for further explanation of GCM and proxy data.

Interpretation of median model outputs

Both 95% confidence intervals and median model outputs are reported in our results. Broadly speaking, individual model realizations undergo similar qualitative evolutions to the median curve. For example, Fig. S4a shows the surface temperature evolution for our nominal model in addition to 100 individual realizations drawn at random from the 10,000 used to generate confidence intervals. However, median curves cannot be used to infer the relative magnitude of swings in carbon cycle variables over Earth history. The individual realizations often exhibit much larger (or smaller) swings in variables than the median. For example, Fig. S4b shows all individual realizations with a 4.0 Ga temperature within 2 K of the nominal model median. Consequently, the evolution of the 95% confidence interval is the more important output because it shows the full range of probable outcomes. The median simply provides a convenient way to compare the average of envelopes.

**Appendix B: Model validation and comparison with previous studies**
For comparative purposes, we reproduced the model outputs of Sleep and Zahnle (13) who also modeled the evolution of the geological carbon cycle over Earth history. To replicate their results we replace our kinetic formulation of seafloor weathering (equation (S3)) with their $pCO_2$ and spreading rate dependence:

$$F_{\text{diss}} = k_{\text{diss}} r_{\text{spread-rate}} \left( \frac{\text{pCO}_2}{\text{pCO}_2^{\text{mod}}} \right)^{\mu} \tag{S21}$$

Additionally, we assume no changes in land fraction ( $f_{\text{land}} = 1$ ), no biological enhancement of weathering ( $f_{\text{bio}} = 1$ ), and change our initial conditions to match their model assumptions. All other model parameters are unchanged. Our model outputs are plotted alongside those of Sleep and Zahnle (13) in Fig. S2 (dashed line). In general, our model envelope encompasses their outputs. The only exception to this is atmospheric $pCO_2$, which is lower in our model. This is unsurprising because our model uses a different climate model with larger average climate sensitivity over the range of atmospheric $pCO_2$ explored in Fig. S2. This higher climate sensitivity is better in line with GCM and paleoclimate estimates (1, 24), and so $pCO_2$ is lower for the same changes in temperature. The agreement between the models confirms that the omission of crustal and mantle reservoirs does not affect our conclusions.

Similarly, our model can reproduce the results of Franck*, et al.* (25), who also applied a geological carbon cycle model to predict the evolution of atmospheric $pCO_2$ and temperature over Earth history. When Franck, *et al.* (25) adopted a strong $pCO_2$ dependence for seafloor weathering (essentially equation (S21) with μ=1), then their model predicts Archean temperatures well below freezing. When seafloor weathering kinetics are neglected, then their model predicts high Archean temperatures, which is also consistent with our model (see Fig. S3).

As noted in the main text, the main reason our nominal model results differ from those of Sleep and Zahnle (13) is because they assumed seafloor weathering is an exclusively $CO_2$ dependent feedback, whereas we allow for the temperature-dependence of weathering and effectively adopt a very weak $CO_2$ dependence via a pore-space pH dependence. This can be seen by plotting the relationship between atmospheric $pCO_2$ and pore space pH in our model (Fig. S5a). Note that there is no precise relationship because of the degeneracies in the carbonate chemistry system; in fact, the scatter would be even wider if we included sensitivity test outputs (e.g. varying Calcium). Nonetheless, we can derive an approximate $pCO_2$-pH relationship for illustrative purposes via linear regression:

$$\log_{10}(pCO_2) = -1.34 \times \mathrm{pH} + 8.44 \qquad \text{(S22)}$$

Using the definition of pH (equation (S16)), this can be recast in terms of hydrogen molality:

$$\left[\mathrm{H}^+\right]_\mathrm{P} = 10^{-8.44} \times pCO_2^{(1/1.34)} \qquad \text{(S23)}$$

In our nominal model, seafloor dissolution is related to hydrogen molality of the pore space by an exponent, $\gamma$ (equation (S3)), which varies from 0 to 0.5. This implies that the overall $pCO_2$ dependence of seafloor weathering in our model is $pCO_2^{(\gamma/1.34)}$, where the exponent $(\gamma/1.34)$ now ranges 0 to 0.37. This is a weak direct $pCO_2$ dependence compared to that of Sleep and Zahnle (13) (Fig. S5b). Furthermore, the gradients of individual model realizations are typically steeper than the best fit straight line, which would imply an even weaker $pCO_2$ dependence. Sleep and Zahnle (13) adopted values of 0.4 and 1.0 in their seafloor weathering exponent, and they did not include a direct temperature dependence. In contrast, our seafloor weathering parameterization (equation (S3)) has a strong temperature-dependence.

Why does shifting from a $pCO_2$ dependent weathering feedback to a temperature dependent weathering feedback result in a more temperate Archean climate? Fig S6a compares ensembles of temperature histories for two illustrative endmember cases: (i) no $CO_2$ dependence for both seafloor and continental weathering (red), and (ii), both the continental and seafloor weathering $CO_2$ dependencies at the maximum extent of our chosen parameter ranges (green). In these calculations we also keep continental fraction, the biological enhancement of weathering, and internal heat flow constant such that the only important driver of the system is increasing solar luminosity (although there are other minor effects such as from changes in sediment thickness). Temperature is constant in case (i) because the only way to ensure the constant outgassing flux balances the weathering flux is for temperature to remain fixed, and for $CO_2$ to decrease with time to compensate for the increase in luminosity. In contrast, in case (ii), $CO_2$ remains relatively steady with time to ensure a constant weathering flux, and temperature therefore tracks increasing luminosity.

Fig. S6b illustrates this effect by contrasting the seafloor weathering parameterization used in this study, and that of Sleep and Zahnle (13). In this figure, we contrast temperature outputs from our nominal model (Fig. 3), and those of the exact same model except that we have adopted Sleep and Zahnle's (13) seafloor weathering function. This comparison differs from Fig. S2 because here we preserve the land fraction, biological enhancement of weathering and initial conditions of our nominal model; we are not trying to reproduce (13), but rather demonstrate how our nominal model changes dramatically merely by replacing the seafloor weathering function. Fig. S6b shows the same effect as in Fig. S6a. Recall the $pCO_2$ exponent in (13) is 0.4 or 1.0 (in this figure an average of 0.7 was used), whereas in our model the effective exponent, via pH dependence, is 0-0.37. By weakening the $CO_2$ dependent feedback of seafloor weathering and strengthening the temperature dependence, the climate is buffered to near-modern temperatures and a frozen Archean is precluded.

**Appendix C: Sensitivity tests**

(i) Pore-space temperature unaffected by changes in internal heat flow.
Using the Fourier's Law to relate internal heat flow to pore-space temperatures is somewhat speculative, and so here we consider the case where pore-space temperature is determined entirely by deep ocean temperature, with a constant temperature difference (9 K) between the deep ocean temperature and the pore space temperature. This approach was validated in Krissansen-Totton and Catling (1) for the Cenozoic and Mesozoic carbon cycle. Fig. S15 shows model outputs for this scenario. The difference between this case and the nominal model are very minor. The main difference is that the Archean seafloor weathering flux is slightly lower than the nominal model because of the pore-space is at a lower temperature.

(ii) Broad range of calcium abundance histories.
Here we modify the calcium abundances in our model (equation (S17)) as follows:

$$\left[\mathrm{Ca}^{2+}\right] = 0.5(\mathrm{ALK} - \mathrm{ALK}_{\mathrm{initial}}) + \left[\mathrm{Ca}^{2+}\right]_{\mathrm{initial}} + \left[\mathrm{Ca}^{2+}\right]_{\mathrm{max}} \exp(t-4) \qquad \text{(S24)}$$

Here, the parameter $\left[\mathrm{Ca}^{2+}\right]_{\mathrm{max}}$ =0-0.5 mol/kg with a uniform distribution, and $t$ is the time in Ga. Model outputs for this broad range of calcium histories are shown in Fig. S8. This range spans plausible calcium histories from the literature. For instance Foriel*, et al.* (26) measured Ca abundances in 3.5 Ga fluid inclusions, and found the seawater source had 0.48 mol/kg Ca. This provides an upper limit on Archean Ca abundances because calcium may have been concentrated during evaporation. The calcium envelope derived from the model of Halevy and Bachan (27) has a maximum value of 0.3 mol/kg, also within our 0-0.5 mol/kg range. Note that De Ronde*, et al.* (28) reported Ca abundances in Archean seawater, but the relevant deposits were later shown to be of Quaternary age (29).

(iii) Broad range of calcium abundance histories with aqueous complexing.
In our nominal model and sensitivity test (ii), we ignored changes in Ca and carbonate activity coefficients. However, for high Ca abundance runs (0-0.5 mol/kg Ca in the Archean), the effects of aqueous complexing cannot be ignored and so Fig. S8 is likely unrealistic. Activity coefficients for calcium and carbonate ions were calculated using the Pitzer equations with the commercial software package Aspen Plus. The methodology for implementing these calculations is described in Krissansen-Totton*, et al.* (30). Starting with modern Earth ocean composition (Table 6 in Krissansen-Totton*, et al.* (30), we incrementally increased the Ca abundance and re-calculated the activity coefficients for calcium, $\gamma_{Ca^{2+}}$, and carbonate, $\gamma_{CO_3^{2-}}$. Sodium and chlorine ion abundances were adjusted at each new Ca value to preserve charge balance. Varying atmospheric $pCO_2$ from 350 ppm to 0.5 bar did not dramatically change the activity coefficients of interest. Fig. S20 shows the product of the activities coefficients, relative to their value on the modern Earth, as calculated using Aspen Plus. We fitted these data with a 3rd order polynomial (Fig. S20):

$$\frac{\gamma_{Ca^{2+}}\gamma_{CO_3^{2-}}}{\gamma_{Ca^{2+}}^{modern}\gamma_{CO_3^{2-}}^{modern}} = -7.89\left[Ca^{2+}\right]^3 + 10.87\left[Ca^{2+}\right]^2 - 5.2246\left[Ca^{2+}\right] + 1.0551 \qquad (S25)$$

Here, $\left[Ca^{2+}\right]$ is the calcium molality, and the 'modern' superscripts denote activity coefficient values in the modern ocean. This equation was then used to modify the saturation state equations (equation (S18)) in our carbon cycle model:

$$\Omega_O = \left(\frac{\gamma_{Ca^{2+}}\gamma_{CO_3^{2-}}}{\gamma_{Ca^{2+}}^{modern}\gamma_{CO_3^{2-}}^{modern}}\right)_O \frac{\left[Ca^{2+}\right]\left[CO_3^{2-}\right]_O}{K_{sp}} \quad \text{and} \quad \Omega_P = \left(\frac{\gamma_{Ca^{2+}}\gamma_{CO_3^{2-}}}{\gamma_{Ca^{2+}}^{modern}\gamma_{CO_3^{2-}}^{modern}}\right)_P \frac{\left[Ca^{2+}\right]\left[CO_3^{2-}\right]_P}{K_{sp}} \qquad (S26)$$

Fig. S9 shows model outputs when both these modified saturation state equations and variable Ca histories (case ii) are adopted. As reported in the main text, the acidifying effect of high Ca abundances and the pH buffering effect of lowered activity coefficients cancel out somewhat, and the resultant model outputs are very similar to the nominal model.

(iv) Extreme biotic enhancement of weathering
In the nominal model we assume biotic enhancement of weathering may have increased continental weathering by up to one order of magnitude over Earth history. However, large biological modifications have been proposed (e.g. Schwartzman (31)). Here, we ran our calculations with $B_{\text{Precambrian}}$ ranging from 0.001 to 0.1 (uniformly sampled in logspace). This range allows for biological effects to increasing continental weathering rates by up to three orders of magnitude over Earth history. The model outputs are shown in Fig. S14. Even in this extreme scenario, Archean temperatures remain below 50°C because of the buffering effect of seafloor weathering. Note also that this scenario is unlikely because it conflicts with Proterozoic $pCO_2$ proxies.

(v) No seafloor weathering
In this sensitivity test we hold seafloor weathering constant at its modern value. Because the modern flux is a small fraction of the continental weathering flux, this is roughly equivalent to setting seafloor weathering flux to zero. Fig. S3 shows the model outputs from this scenario. Extreme Archean temperatures (approaching 100°C) and $pCO_2$ (several bar) are possible in this scenario, illustrating the importance of seafloor weathering as a negative feedback early in Earth history. Note however, that temperate climates still fall within the Archean surface temperature envelope, and so the uncertainties in our model parameters are too large to precisely quantify the relative importance of continental and seafloor buffers in the Archean.

(vi) Extreme Archean outgassing
Xenon isotope tracers in Archean quartz suggest mantle outgassing at 3.3 Ga was 9.5±4.5 times greater than the modern flux (32), and would presumably have been even higher at 4.0 Ga. This does not imply the total Archean outgassing flux was 9.5 times the modern flux since a large fraction of modern outgassing is sourced from metamorphic gases and recycled, subducted carbon, both of which are unlikely to scale by the same factor. However, high outgassing scenarios still ought to be considered. In this particular scenario we set $n_{out}$ =0.73, and allow $m$ to vary from 1-3. This allows Archean outgassing values that are up to 60x the modern flux. Even in this extreme scenario, Archean surface temperatures were most likely temperate, with only a small probability of temperatures exceeding 50°C (Fig. S11). Ocean pH evolution is comparable to the nominal model. This scenario shows an unrealistic outgassing flux several orders of magnitude larger than the modern is required to produce a hot early Earth.

We also consider the extreme scenario where Archean outgassing is lower than the modern. Some thermal history models predict heat flow in the Archean comparable to or even slightly lower than the modern (14). Specifically, if the Archean Earth were in a stagnant lid regime (e.g. 33) then total outgassing may have been lower due to the absence of arc and mid ocean ridge volcanism, and because melting is suppressed by a thick crust. For example, Tosi*, et al.* (34) estimated that for a mantle with IW+1 (Iron-Wustite) oxygen fugacity, a stagnant lid Earth would outgas only 15 bar of $CO_2$ over 2 Ga, or 1.4 Tmol/yr. This is approximately 20% of the modern flux. For this low-outgassing sensitivity test, we parametrized outgassing and heatflow as follows:

$$F_{\text{out}} = F_{\text{out}}^{\text{mod}} Q \equiv F_{\text{out}}^{\text{mod}} \left[ 1 - \left(1 - f_{\text{out}}^{\text{Arch}}\right) \left(t / 4.5\right)^{2} \right] \quad \text{(S27)}$$

Here, $f_{\text{out}}^{\text{Arch}}$ is the outgassing flux at 4.0 Ga relative to the modern flux, which is assumed to range from 0.2 to 1.0 for this sensitivity test.

Fig. S12 shows model outputs for the low outgassing scenario described above. Less than half of the model runs result in a frozen Archean. However, an Archean outgassing flux five times less than the modern is probably unrealistic. In the stagnant lid calculations described above, a mantle redox state of IW+1 was assumed, whereas zircon data suggest Earth's mantle redox has remained close to Quartz-Fayalite-Magnetite (IW+3.5) since the Hadean (e.g. 35). Increasing mantle redox on stagnant lid planets predicts dramatically higher $CO_2$ fluxes (34). Additionally, low Archean outgassing contradicts the xenon outgassing proxies (32). Finally, geochemical and paleomagnetic evidence favors episodic subduction, and therefore greater outgassing, during the Archean over a permanent stagnant lid regime (36, 37). If atmospheric methane is added (Fig. S13), then frozen Archean climates can most likely be avoided, even with low outgassing.

(vii) Extreme albedo changes

In this scenario we have artificially added 30 K of warming in the Archean to simulate possible warming from dramatically lowering Earth's albedo. Surface temperatures increase by less than 30 K because there is a compensating decrease in $pCO_2$ to balance the carbon cycle (Fig. S10). Archean temperatures remain temperate, and ocean pH values are more basic because of the lower $pCO_2$. However, as noted in the main text, this scenario is speculative because 30 K warming requires an unrealistically large albedo change.

(vii) Stronger continental weathering thermostat

In Krissansen-Totton and Catling (1) we perform an inverse analysis with our geological carbon cycle model and conclude the effective temperature of continental silicate weathering, $T_e$, is very likely greater than 10°C, and hence we adopt the range $T_e$=10-40°C in our nominal model. However, extrapolating this value to all of Earth history may not be appropriate, and so we also consider a more conventional range for the temperature dependence of continental weathering, $T_e$=5-15°C (38-41). Fig. S16 shows out model outputs assuming these lower effective temperatures. None of our key conclusions are changed; median Archean temperatures are slightly cooler than the nominal model, and the pH envelope is slightly narrower.

(viii) Contributions to warm Archean temperatures
As noted in the results section, if solar luminosity were the only variable driving the carbon cycle, then Archean surface temperatures would necessarily be lower than modern surface temperatures. This is not reflected in our nominal model outputs because reduced continental land fraction, lowered biological enhancement of weathering, and elevated heat flow all act to warm the early Earth.

In this sensitivity test we repeated our nominal model calculations four times. Firstly, we held continental land fraction, the biological enhancement of weathering, and internal heat flow constant throughout Earth history. This shows the climate evolution if only solar luminosity is increasing, buffered by weathering feedbacks. Next, we repeated our calculations restoring the land fraction parameter to its nominal range, but we maintained constant values biological enhancement and heat flow. Similarly, we repeated the calculations restoring nominal values for the biological enhancement of weathering and holding the continental fraction heat flow constant, and finally heat flow was varied whilst the other two variables were held constant. These three cases isolate the warming effect from each variable such that they can be quantitatively compared.

Fig. S17a shows the median surface temperature evolutions for all four cases described above, in addition to the nominal model (Fig. 3). In the absence of any modifiers (luminosity increase only), surface temperature increases monotonically over Earth history. However, decreased land fraction, reduced biological enhancement of weathering, and higher heat flow all contribute comparable amounts of warming, on average. Unsurprisingly, elevated heat flow has the largest warming effect in the early Archean when heat flow was highest, whereas the warming from lessened biological enhancement of weathering is a step change at the Proterozoic-Phanerozoic boundary. Because each variable makes similar magnitude contributions to Archean warming, albeit with large uncertainties, we cannot attribute a temperate Archean climate to any one factor.

Nonetheless, we can say definitively that lower Precambrian land fraction is not necessary for temperate Archean climates. Fig S17b compares nominal model temperature evolution to a scenario where land fraction is constant throughout Earth history. The constant land fraction case is somewhat cooler, but above-zero Archean temperatures remain highly probable.

**Appendix D: Proxy data**
The $pCO_2$, temperature, and seafloor weathering proxies plotted alongside model outputs in Fig. 3-5 and many supplementary figures were sourced from a range of literature estimates. In order of increasing age, the glacial temperature constraints (green arrows) are Geboy*, et al.* (42), Kuipers*, et al.* (43), Williams (44), Ojakangas*, et al.* (45), Young*, et al.* (46), and de Wit and Furnes (47). Most of these constraints consist of diamictites or dropstones, and although paleolatitudes for many of these glacial deposits are uncertain or unknown, but we argue the existence of glacial deposits anywhere on Earth bounds global average temperatures to <25°C. The glaciation of Antarctica occurred approximately 35 Ma when global temperatures crossed below 20°C (48), and so an upper limit of 25°C is a more conservative estimate allowing for variations due to paleogeography. Of course, it is unclear whether the Precambrian glaciations reported in the literature represent transient glaciation or the long-term average temperature, hence the need for carbon cycle modeling. The other temperature constrains plotted on Fig. 3-5 are derived from $\delta^{18}O$ and deuterium isotopes in 3.42 Ga cherts (49), and $\delta^{18}O$ and organic phosphate (50).

Most of the atmospheric $pCO_2$ constraints plotted in Fig. 3-5 are from paleosols. The proxies from Sheldon (51) and Driese*, et al.* (52) are derived from a mass balance model of weathering, whereas the higher estimates of Kanzaki and Murakami (53) are derived from explicit modeling of the aqueous chemistry of soils. The Precambrian $pCO_2$ upper bound from Kah and Riding (54) is based on calcification of cyanobacterial sheaths.

The seafloor weathering constraints plotted on Fig. 3-5 are derived from carbonate abundances in hydrothermally altered Precambrian oceanic crust. These carbonate abundances are converted to fluxes by assuming a broad range of crustal production rates (see below). Nakamura and Kato (55) measured carbonate content in 3.46 Ga altered seafloor basalt, and found that if the Archean crustal production rate was 5x greater than today, and the depth of alteration was 500 m, then the seafloor weathering sink would be 38 Tmol C/yr. Therefore, for modern crustal production rates, the flux would be 7.6 Tmol C/yr. Using our parameterizations for spreading rate with their associated ranges (equation (S10)), the maximum possible seafloor weathering flux is $7.6\times10^{12}\times Q^2 = 7.6\times10^{12}\times[(1-3.46/4.5)^{-0.73}]^2 = 65$ Tmol C/yr. Thus based on the Nakamura and Kato (55) data, the 3.46 Ga seafloor weathering flux could be anywhere from 7.6-65 Tmol C/yr, and this is the interval plotted in Fig. 3-5. Similarly, the carbonate abundances measured in 2.6 Ga altered seafloor by Shibuya*, et al.* (56) implies a seafloor weathering sink of 2.6 Tmol C/yr assuming modern spreading rate. If our model maximum spreading rate at 2.6 Ga the flux is 9.1 Tmol C/yr, and so the range 2.6-9.1 Tmol C/yr is plotted on Fig. 3-5. Similar calculations using the carbonate abundances reported in Shibuya*, et al.* (57) and Kitajima*, et al.* (58) yield flux ranges of 42-260 Tmol C/yr, and 28-250 Tmol C/yr, respectively.

**Appendix E: Climate model description and polynomial fit**

The 1D radiative convective climate model used here was originally developed in Kasting and Ackerman (59) and was recently applied in studies on topics related to the habitable zones of different stars (60), the atmospheres of Earth-like planets around various stellar types (61-63), clouds in exoplanet atmospheres (64, 65), exoplanet false positive biosignatures (66), possible climates for Proxima Centauri b (67), and the climate of the early Earth (68, 69). Its suite of radiatively active gases include $O_3$, $CO_2$, $H_2O$, $O_2$, $CH_4$, and $C_2H_6$ whose mixing ratios can be set from an input file, or passed to it from a photochemical model. The model uses a δ-2 stream multiple scattering algorithm to calculate net absorbed solar radiation in each layer (70) with 38 spectral intervals spanning from 0.2 to 4.2 μm. Net outgoing IR radiation is computed with separate correlated-k coefficients using 55 spectral intervals spanning 0-15,000 $cm^{-1}$. This 1D model has been found to agree with 3D GCM model results within about 2-5 K for Archean earth simulations (69) (also see below for an additional GCM comparison).

We assume an Earth-mass planet with a surface albedo of 0.32, which includes the radiative effects of water clouds (60). We apply a Manabe/Wetherald relative humidity model for the troposphere (71) and specify a surface relative humidity of 0.8. Our relative humidity parameterization is described in more detail in Pavlov, *et al.* (72). The model atmosphere is divided into 101 layers up to 7x10$^{-5}$ bar. The model is considered converged when the incoming solar flux is balanced with the outgoing reflected and radiated fluxes.

The model output grid is plotted in Fig. S19 alongside the 4$^{th}$ order polynomial fit to these outputs. The polynomial fit to the model outputs is given by:

$$T_S = 3145.89 - 894.196x - 14231.92\,y - 3791.699\,x^2 + 18527.77x^2y - 33560.74x^2y^2 + 26297.61y^2$$
$$-7674.76xy^2 + 4461.16xy - 1569.25x^3 + 11329.25x^3y^3 - 21270.62y^3 - 14022.32x^3y^2$$
$$+26860.83x^2y^3 + 7684.55x^3y + 5722.64xy^3 - 178.264x^4 - 396.147x^4y^4 + 6399.091y^4$$
$$+875.423x^4y - 1605.364x^4y^2 + 1304.944x^4y^3 - 1569.3007y^4x - 8012.67y^4x^2 - 3415.93y^4x^3$$
(S28)

Here, $x = \log_{10}(pCO_2)$ where pCO2 is in bar, and $y = L_{relative}$ .

Our 1D radiative-convective climate model outputs are comparable to more sophisticated GCM simulations of Archean climate. Fig S21 compares our climate model, or rather the polynomial fit to our model, to Archean GCM simulations from Charnay, *et al.* (73), and the two temperature plots match quite closely, although the GCM has higher climate sensitivity, possibly due to cloud feedbacks.

To investigate the effect of methane, we calculated two additional climate model grids assuming 100 ppm methane, representing a high Proterozoic estimate (74), and 1% methane, representing a high Archean estimate (75). These model outputs were fitted with the following polynomials:

$$T_S(\text{Proterozoic}) = -28.0318 - 355.758x + 1134.4049y - 196.968357570535x^2 + 893.9465x^2y$$
$$-1178.9796x^2y^2 - 1253.544y^2 - 2246.677xy^2 + 1716.5297xy - 27.95989x^3 + 68.42046x^3y^3$$
$$+501.20787y^3 - 163.8806x^3y^2 + 488.804962942262x^2y^3 + 124.12437x^3y + 917.75019xy^3$$
(S29)

$$T_S(\text{Archean}) = -7.5499 - 156.53786x + 1095.694y - 42.3166x^2 + 333.328x^2y - 536.2506x^2y^2$$
$$-1240.45y^2 - 1541.568xy^2 + 1040.713xy - 1.46501x^3 + 24.54807x^3y^3$$
$$+513.5871y^3 - 48.7778x^3y^2 + 251.7894x^2y^3 + 26.3929x^3y + 686.9375xy^3$$
(S30)

Here, $x = \log_{10}(pCO_2)$ where $pCO_2$ is in bar, and $y = L_{relative}$ as before. Fig. S22 and S23 show the climate model outputs plus polynomial fits for the low and high methane cases, respectively. Finally, Fig. S24 shows a comparison of the three parameterizations (no methane, 100 ppm methane, and 1% methane) for $L_{relative} = 0.75$.

The climate model used to produce the polynomial fits described above did not include photochemistry. To test the impact of photochemistry, we coupled the climate model to a photochemical model using the methods described in Arney*, et al.* (69). The photochemical model is based on a 1-D model originally in Kasting*, et al.* (76) and significantly modernized as described in (77). This version of the model can simulate a range of atmospheric oxidation states ranging from extremely anoxic (pO2 = $10^{-14}$) to atmospheres with $O_2$ as the dominant atmospheric constituent (66, 78, 79). When run in coupled mode, the photochemical model can pass profiles of radiatively active gases to the climate model, and the climate model can pass the temperature and water vapor profile to the photochemical model. The models iterate until convergence is reached. For the atmospheres shown here that include photochemistry, we use our early Earth template described in Arney*, et al.* (69), including the standard gas mixing ratios/boundary conditions and reactions provided in the supplemental online materials of Arney*, et al.* (69). The photochemical model is converged when redox is conserved and when a re-run of the model with the previous run's final state as its initial conditions converges quickly (i.e. in < 50 timesteps). Fig. S24 shows a comparison between the coupled photochemical-climate model and the climate model fit. They match within a few degrees, and spot checks in other regions of parameter space match similarly well.

Note that for our photochemical coupling test, we did not examine parts of parameter space where photochemical haze may form from high levels of methane. For parts of parameter space where haze can form, i.e. for $CH_4/CO_2 > 0.2$ (80), our temperatures can be considered upper limits as haze formation cools the climate (68, 69). Photochemical haze would decrease the surface temperature of the planet, but the feedbacks explored here would then tend to increase atmospheric $pCO_2$, which would both raise surface temperatures and enable hazes to clear by decreasing the $CH_4/CO_2$ ratio.

The climate and photochemical models are publicly available and accessible from https://github.com/VirtualPlanetaryLaboratory/atmos.

Supplementary Tables

Supplementary Table 1 is described in the main text. Supplementary Table 2 shows all the initial values assumed in our model or ranges for variables that are uncertain. All other initial values are fully determined by the variables in this table.

**Supplementary Figures**

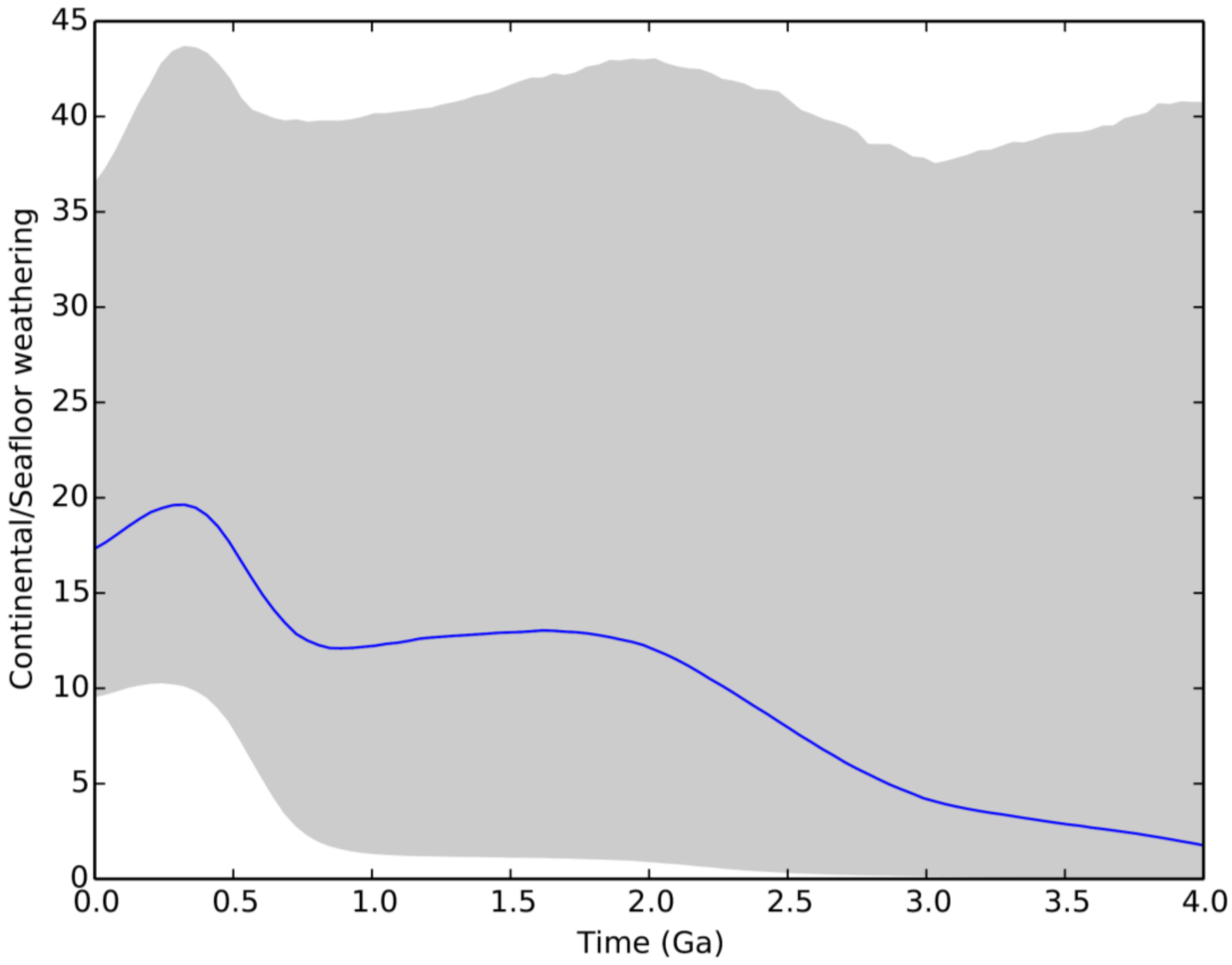

Fig. S1: Ratio of continental silicate weathering to seafloor weathering flux for the nominal model. Solid blue line is the median model output, and grey shaded region is the 95% confidence interval. On the modern Earth, the continental silicate weathering is much larger than the seafloor weathering sink, but at 4.0 Ga there is a high probability that the two fluxes were comparable.

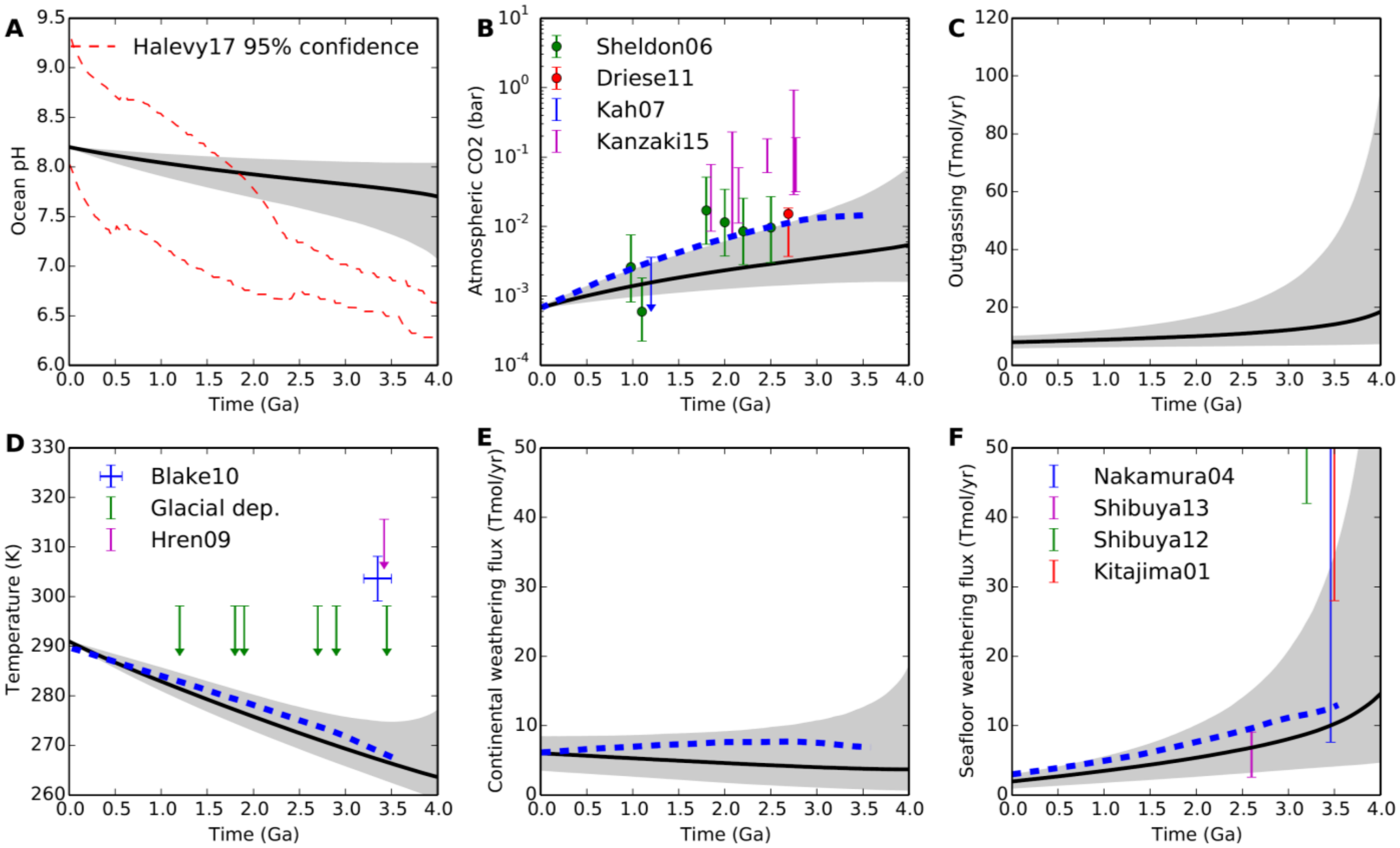

Fig. S2: Validation against Sleep and Zahnle (13). Dashed blue lines are model outputs from Sleep and Zahnle (13), and grey shaded regions are the 95% confidence intervals from our model where we have simplified our parameterizations to match their assumptions (e.g. no change in continental land fraction, no biogenic enhancement of weathering, and seafloor weathering dependent on $pCO_2$ with an arbitrary power law rather than pH and temperature-dependent kinetics). Black lines are the median outputs from our model. The two models are in broad agreement, which suggests the omission of surface reservoirs in our model is reasonable.

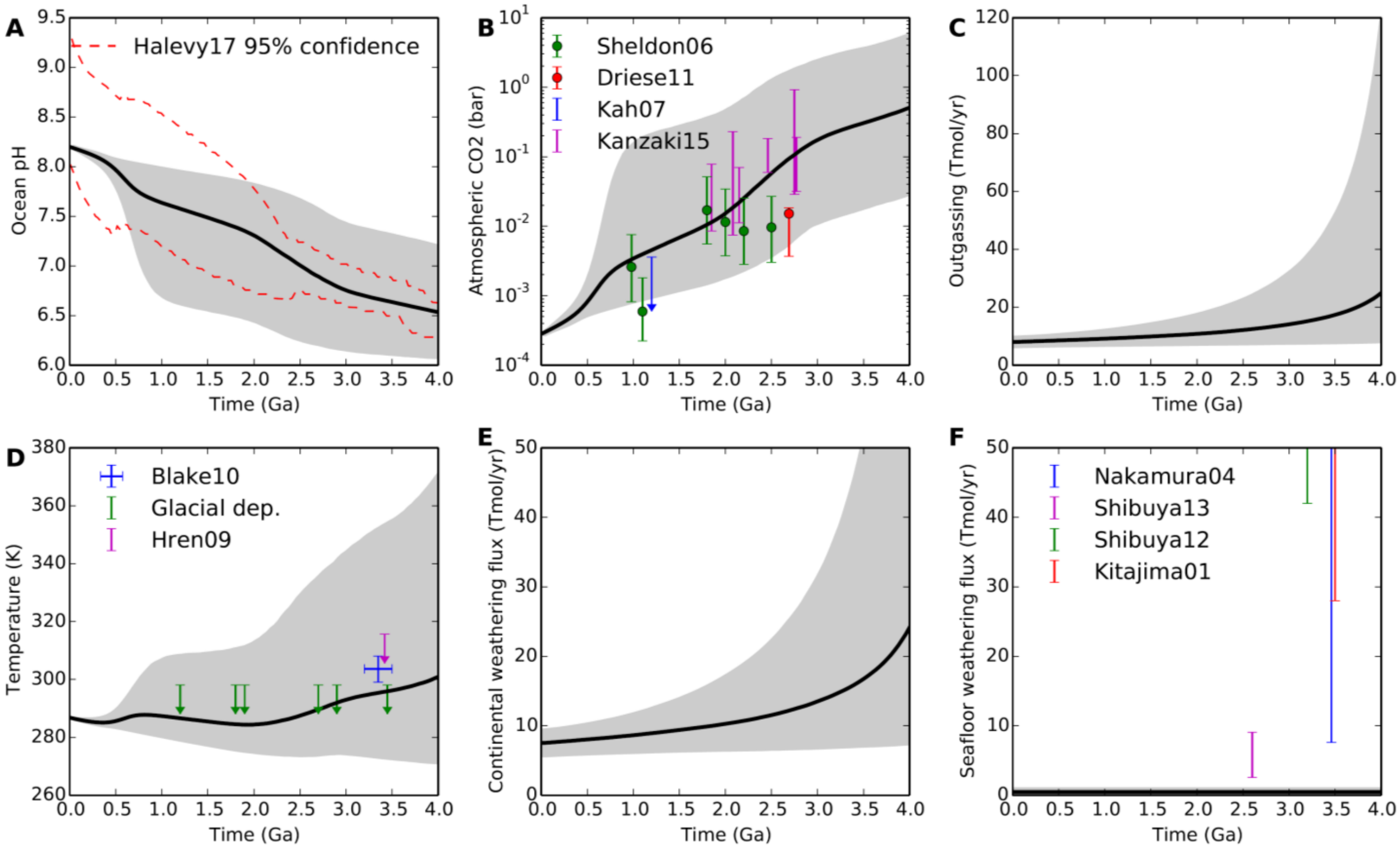

Fig. S3: Nominal model outputs except that the seafloor weathering flux has been held constant. Subfigures (a)-(f), lines, and, shadings are the same as in Fig. 3. (d) Very high temperatures (~100°C) are possible without the seafloor weathering buffer. (b) Archean $pCO_2$ was potentially several bar.

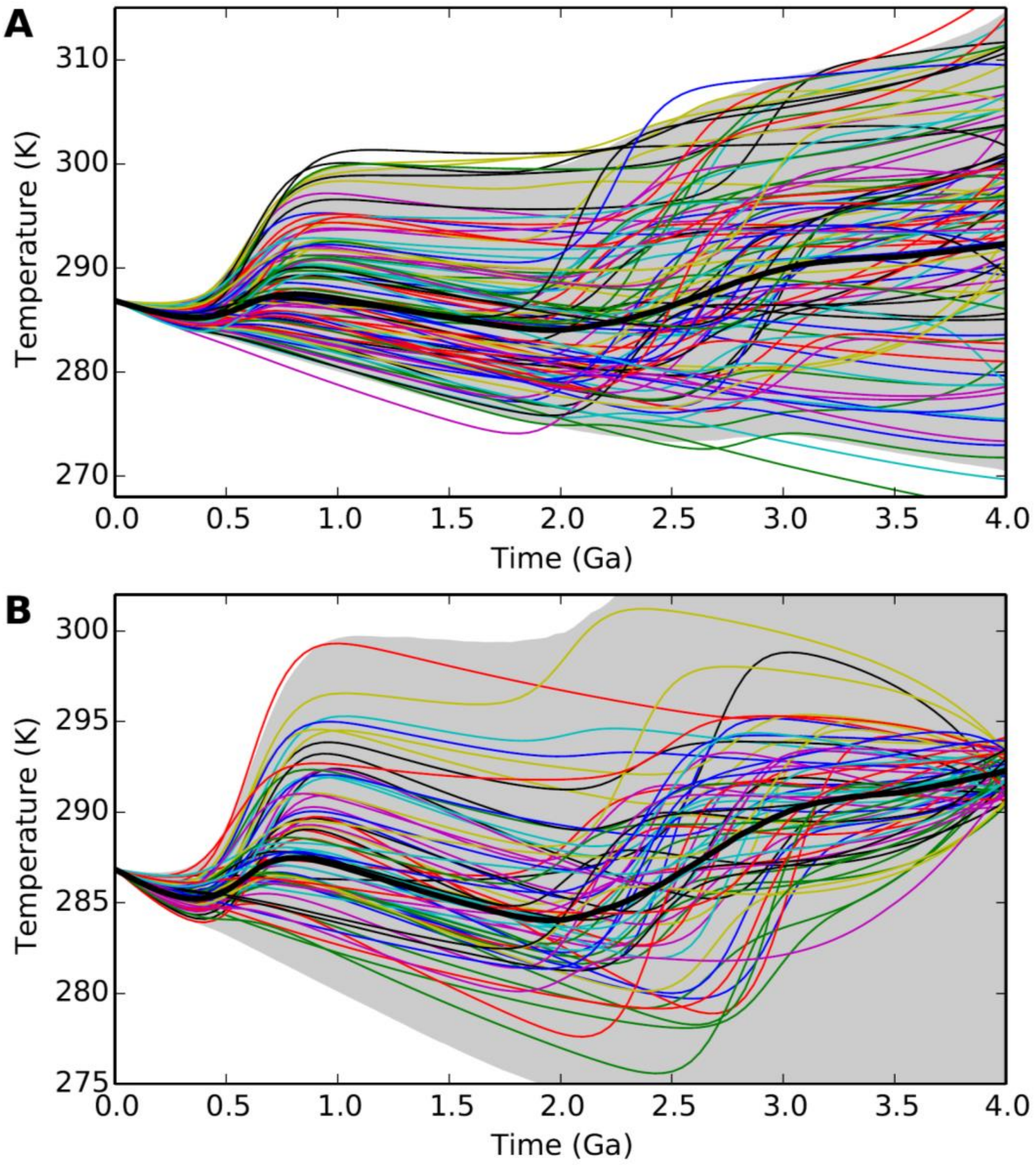

Fig. S4: (A) Grey envelope shows 95% confidence interval of surface temperature from nominal model, whereas colored lines show 100 individual realizations (drawn from the total distribution of 10,000 model realizations). The median curve (black) shares similar variations to individual model realizations. (B) Same as (A) but plotting individual model realizations with a 4.0 Ga temperature within 2 K of the median model output. Temperature fluctuations in individual realizations may vary in magnitude compared to the median output.

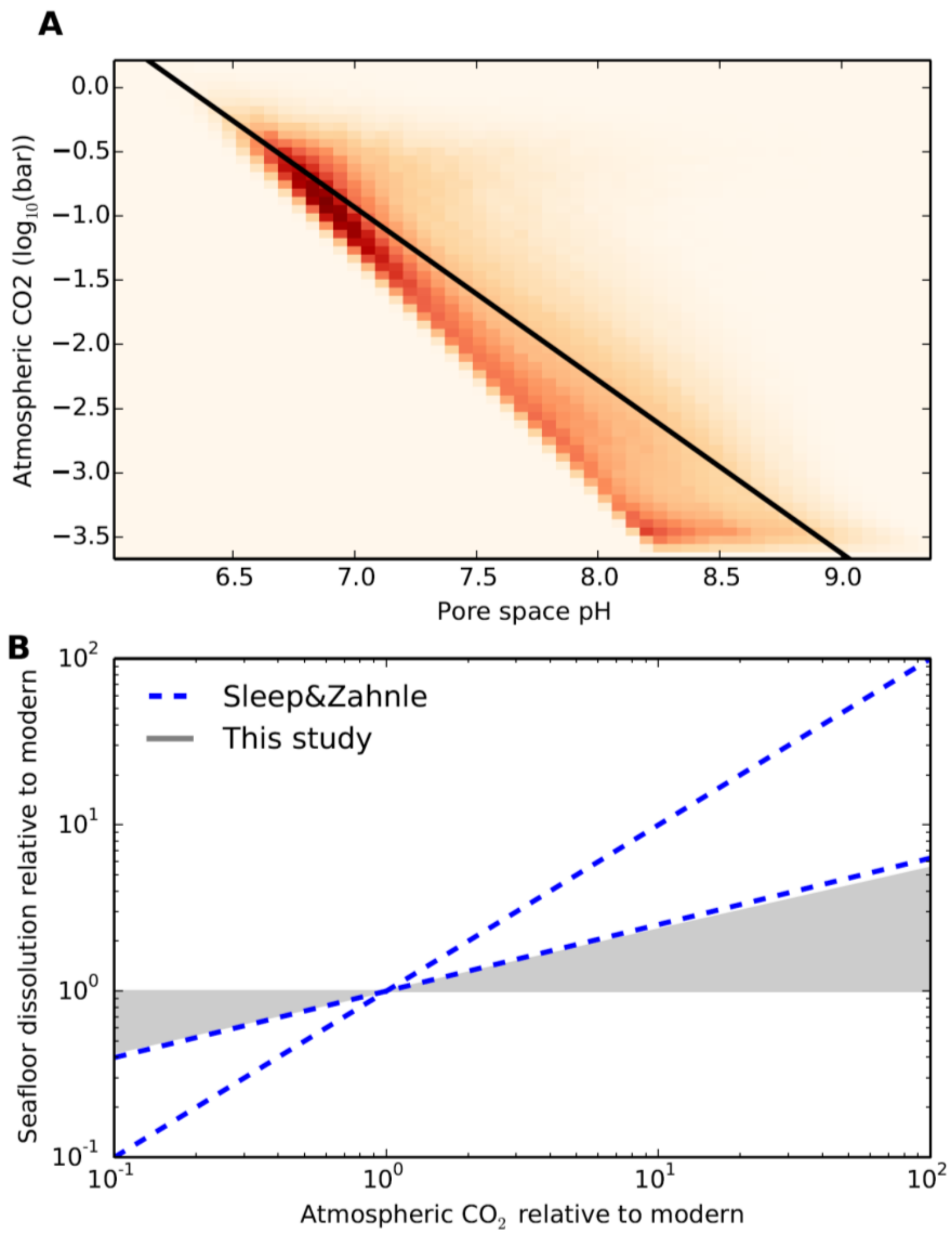

Fig. S5: (A) Relationship between atmospheric $CO_2$ and pore space pH in our nominal model. The red shading is a 2D histogram showing the density of atmospheric $CO_2$ and pore space pH combinations using all 10,000 nominal model runs. The blue black line is the best-fit linear regression to these outputs. The regression provides an approximate $pCO_2$-pH relationship, which we use in (B) to demonstrate that the overall $pCO_2$ dependence in our seafloor weathering parameterization is weaker than has been previously assumed. Here, the blue dashed lines show the weak and strong exponential dependencies of seafloor weathering on $pCO_2$ assumed in Sleep and Zahnle (2001). The grey shaded region shows the effective $pCO_2$ dependencies assumed in our model based on the regression in (A) and the experiments-based relationship between pore space pH and seafloor basalt dissolution.

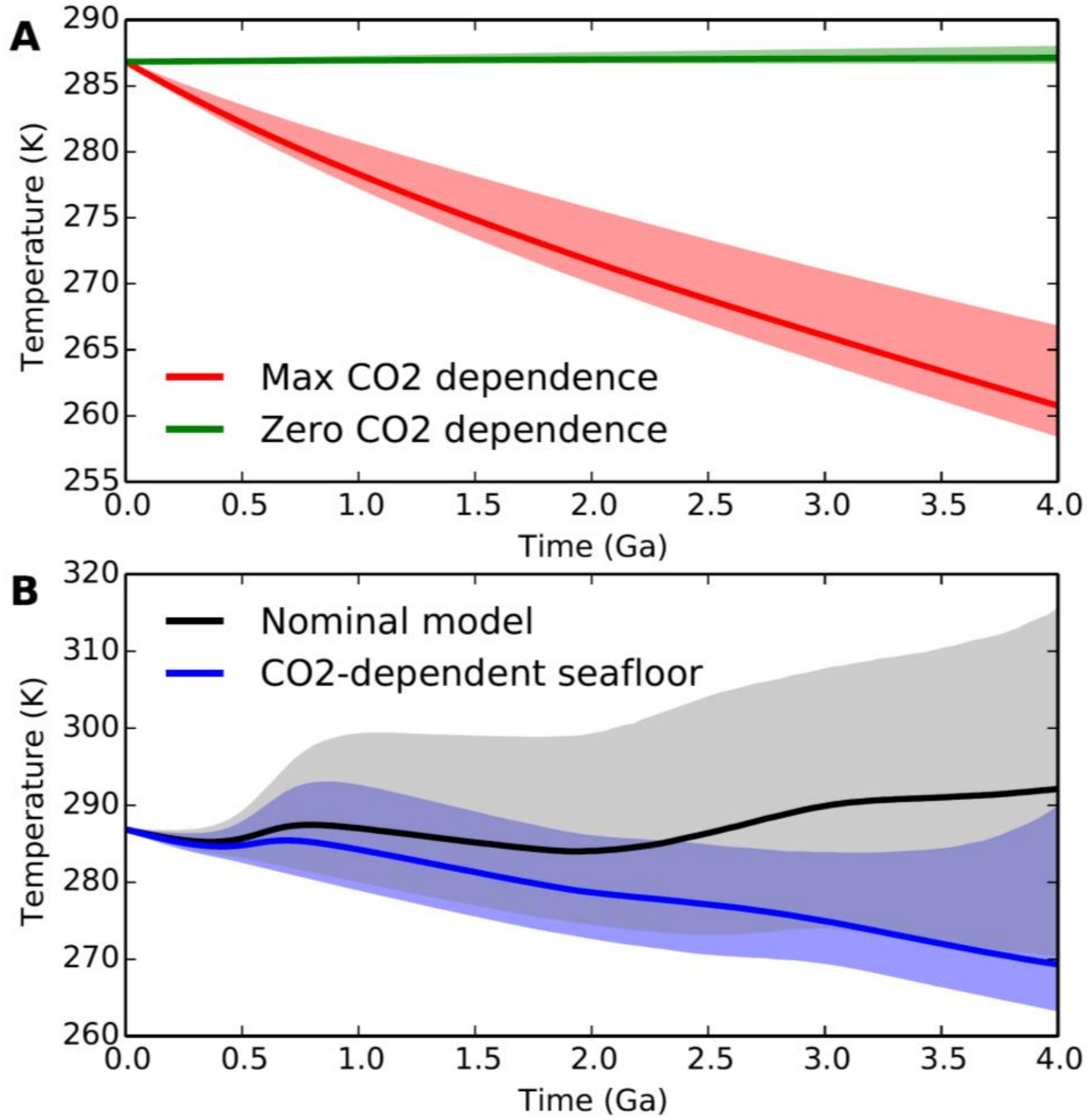

Fig. S6: The effect of temperature-dependent and $CO_2$-dependent weathering feedbacks on buffering surface temperature. (A) In this case evolving solar luminosity is the only major variable driving carbon cycle evolution (land fraction, biological enhancement of weathering, and internal heat flow held constant). Envelopes represent extreme cases where both continental and seafloor weathering feedbacks have zero $CO_2$-dependence (green) and $CO_2$-dependencies at the maximum end of our assumed parameter ranges (red). Temperature is not exactly constant in the green case because secondary variables (e.g. sediment thickness) affect temperature slightly. (B) Comparison between the nominal model and the model with our temperature- and pH-dependent seafloor weathering parameterization replaced by the purely $CO_2$-dependent parameterization from Sleep and Zahnle (exponent of 0.7).

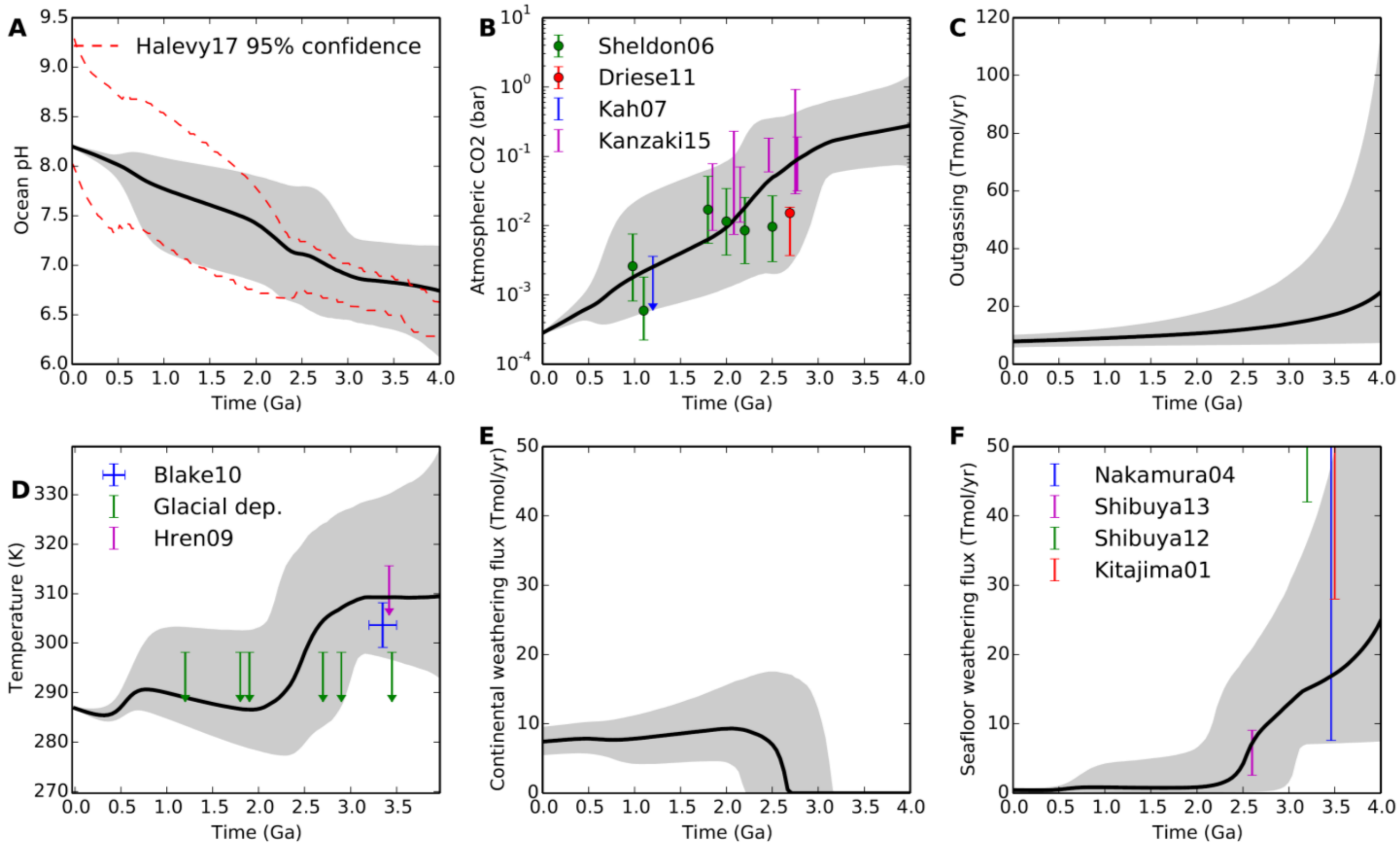

Fig. S7: Extreme endmember scenario with no Archean land and high Proterozoic and Archean methane abundances (100 ppm and 1% mixing ratios, respectively). Subfigures (a)-(f), lines, and, shadings are the same as in Fig. 3. (d) Even under this extreme scenario, Archean surface temperatures are likely temperate, with only a small probability of exceeding 50°C. This is because the temperature-dependent seafloor weathering flux (f) increases to balance the carbon cycle. The evolution of (a) ocean pH and (b) atmospheric $pCO_2$ is comparable to the nominal model because the $CO_2$ decrease from adding methane is roughly canceled by the $CO_2$ increase from removing continents.

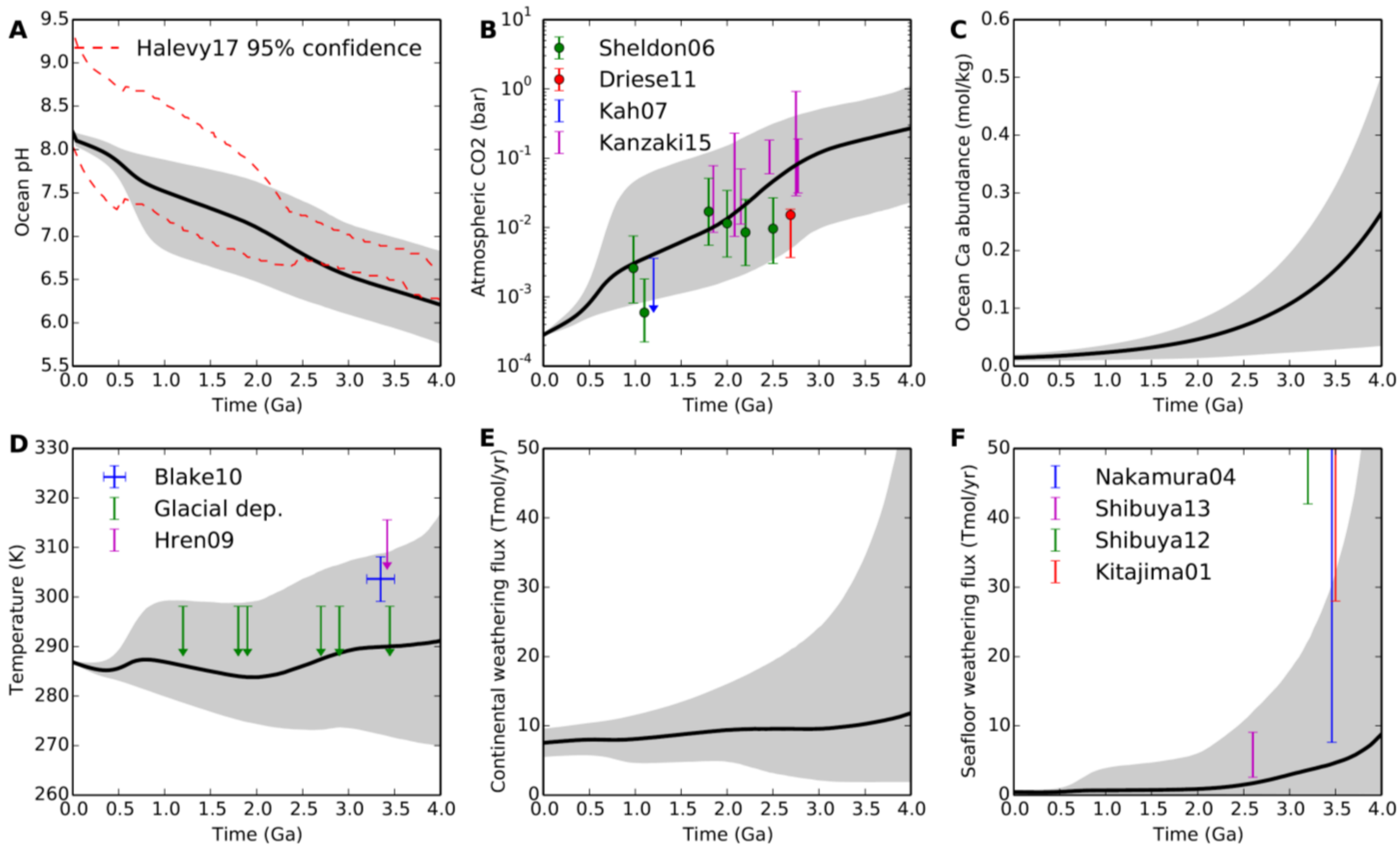

Fig. S8: Large range of ocean calcium abundance evolutions imposed. Subfigures (a)-(f), lines, and, shadings are the same as in Fig. 3 except the (c) now shows imposed ocean Ca evolution. The evolution of most variables is the same as in the nominal model, except that (a) the ocean pH envelope is shifted downward. This is because at high calcium abundances, carbonate must be lower for the same saturation state, thereby shifting the carbonate equilibria to more acidic values. However, this model output is incomplete because it ignores changes in activity coefficients (see main text).

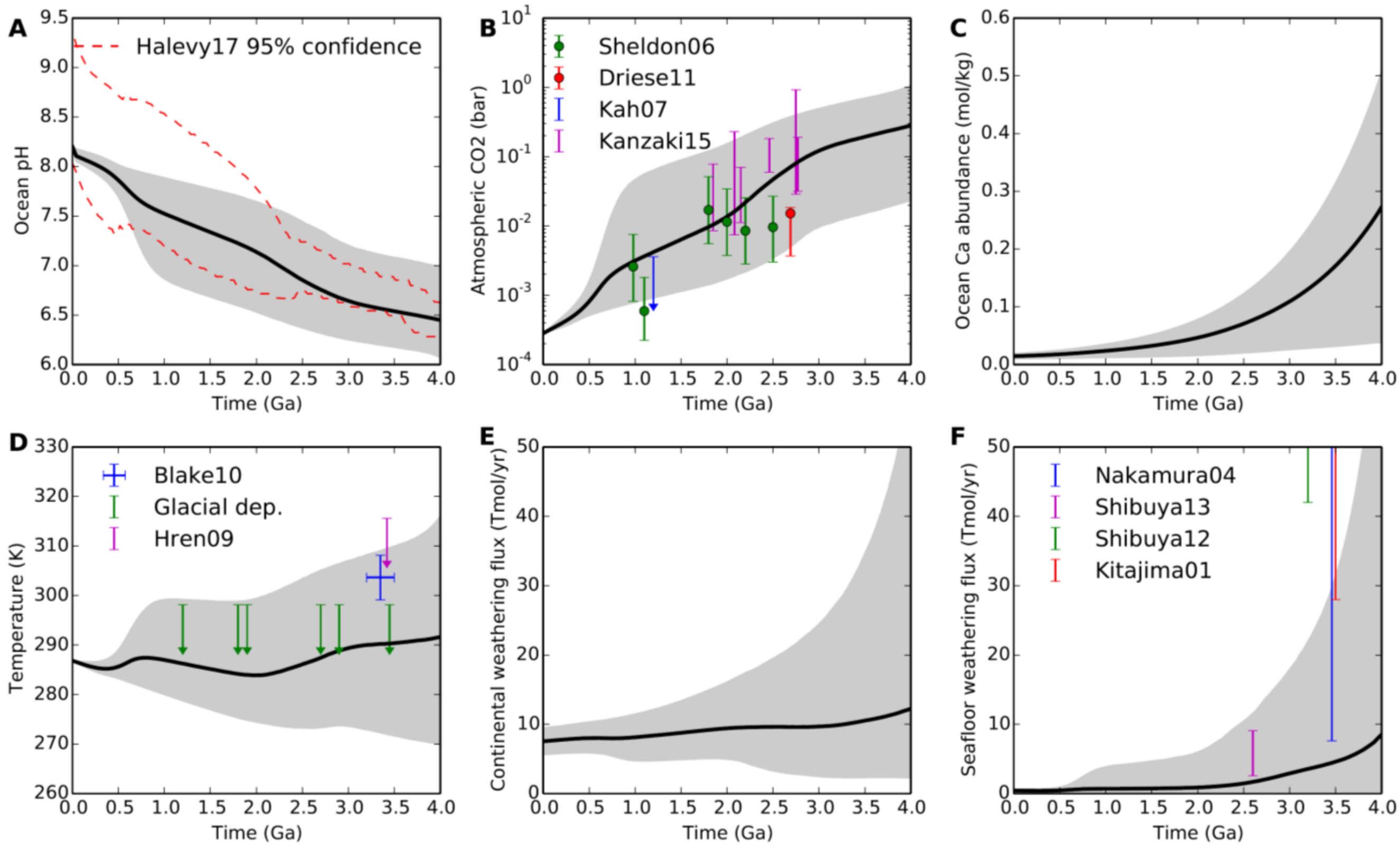

Fig. S9: Large range of ocean calcium abundances imposed and activity coefficients for calcium and carbonate calculated using the Pitzer equations. Subfigures (a)-(f), lines, and, shadings are the same as in Fig. 3 except that (c) now shows imposed ocean Ca evolution. The evolution of all variables is very similar to the nominal model because the effects of high calcium abundance are offset by lowering activity coefficients (see main text).

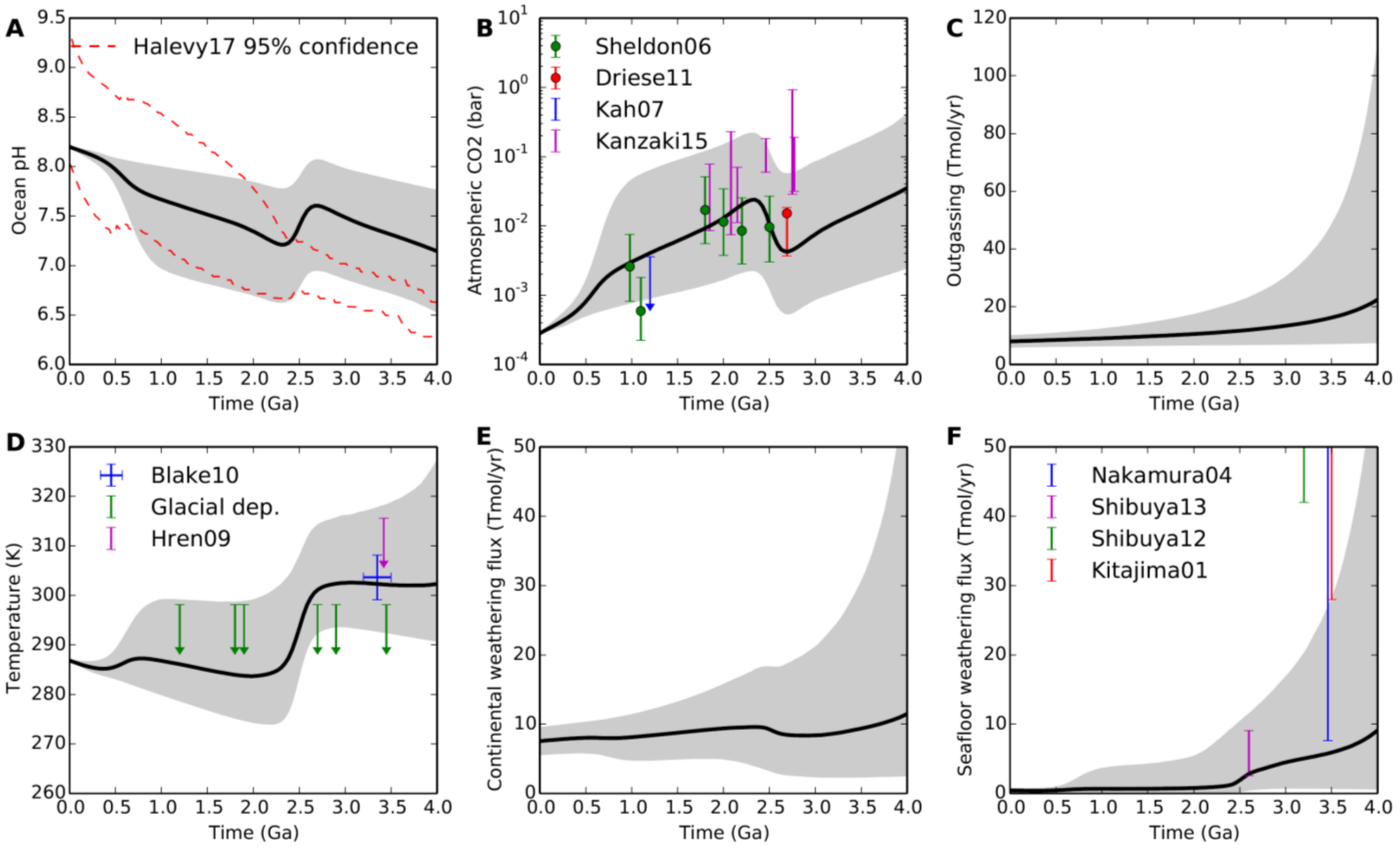

Fig. S10: Extreme warming from albedo changes where we have imposed +30 K surface warming in the Archean. Subfigures (a)-(f), lines, and, shadings are the same as in Fig. 3. (d) Surface temperature increase by less than 30 K in the Archean because there is a compensating decrease in (b) atmospheric $pCO_2$ is similar to that described for imposing high atmospheric methane. (a) Ocean pH is much higher than our nominal model because of the lower $pCO_2$. However, the mechanisms required to generate such large warmings from albedo changes are speculative, and so this scenario may not be realistic.

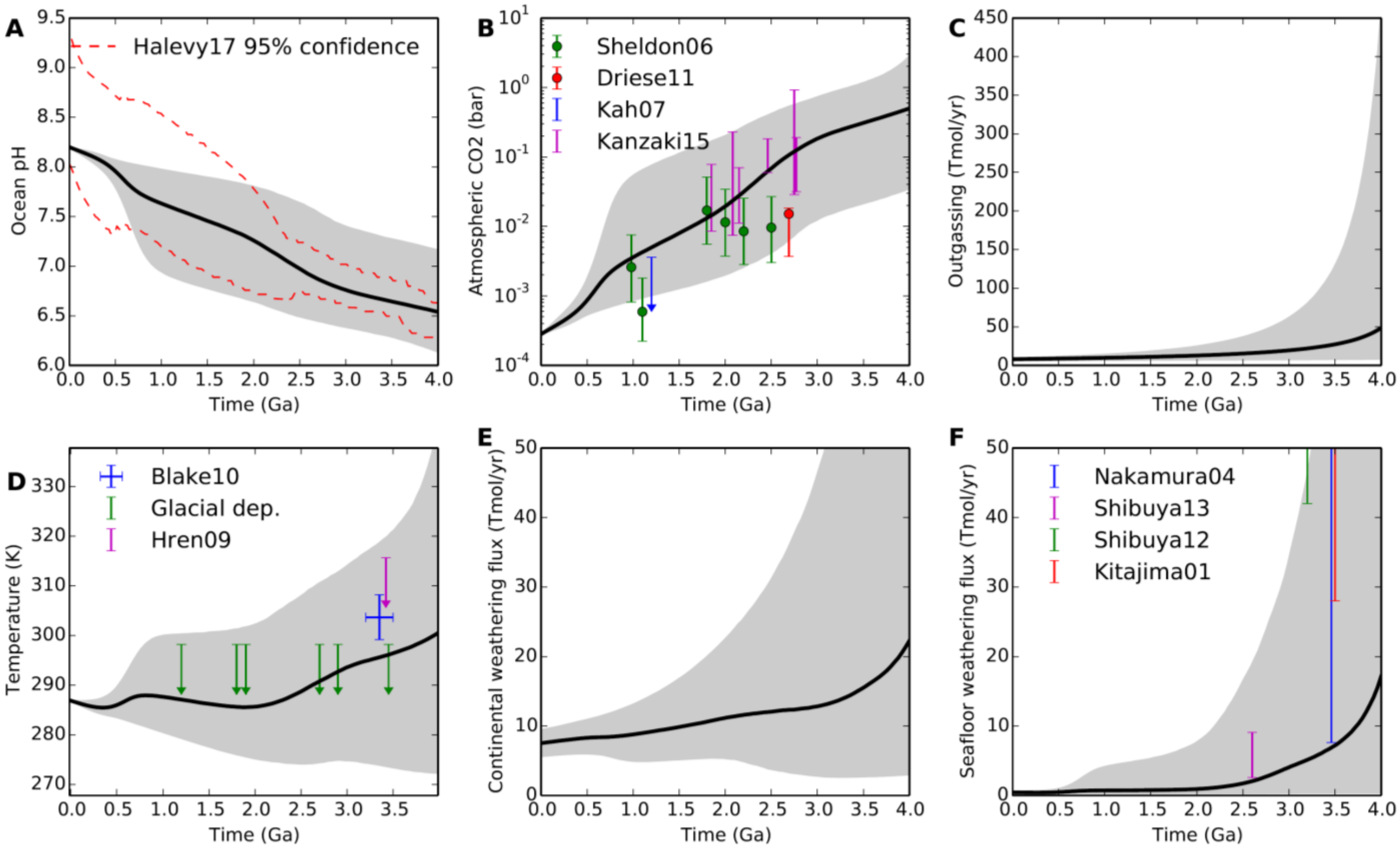

Fig. S11: Extreme Archean outgassing sensitivity test where we have set $m$ =2-3, thereby ensuring outgassing at 4.0 Ga extends up to 60x greater than the modern flux. Subfigures (a)-(f), lines, and, shadings are the same as in Fig. 3. Even in this extreme scenario, Archean surface temperatures were most likely temperate, with only a small probability of temperatures exceeding 50°C.

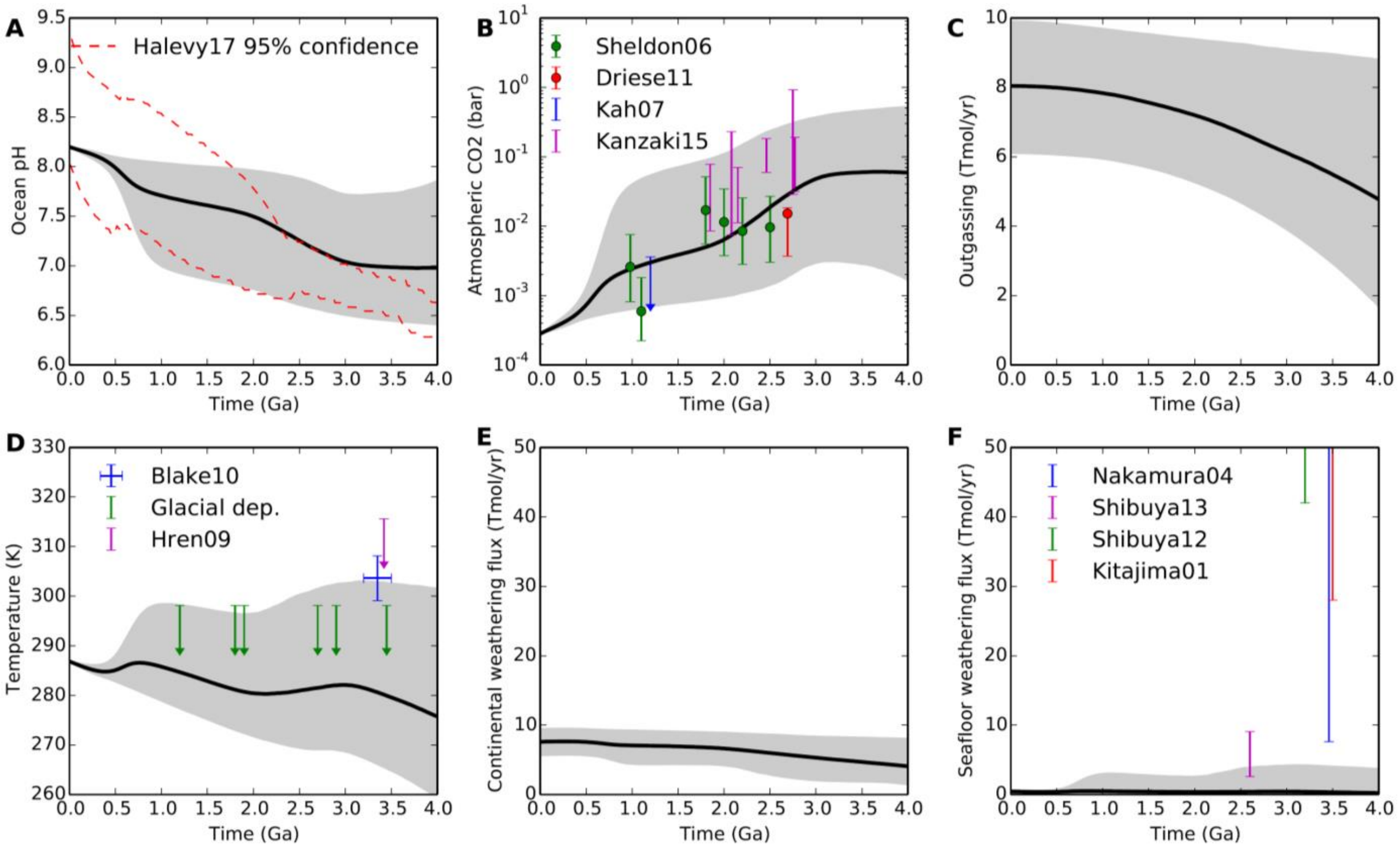

Fig. S12: Extremely low Archean outgassing sensitivity test where we have assumed outgassing at 4.0 Ga is anywhere between 1x and 5x lower than the modern flux. Subfigures (a)-(f), lines, and, shadings are the same as in Fig. 3. In this extreme scenario, frozen climates are possible (~50% likelihood). But as we argue in Appendix C and the main text, dramatically lower Archean outgassing is unlikely.

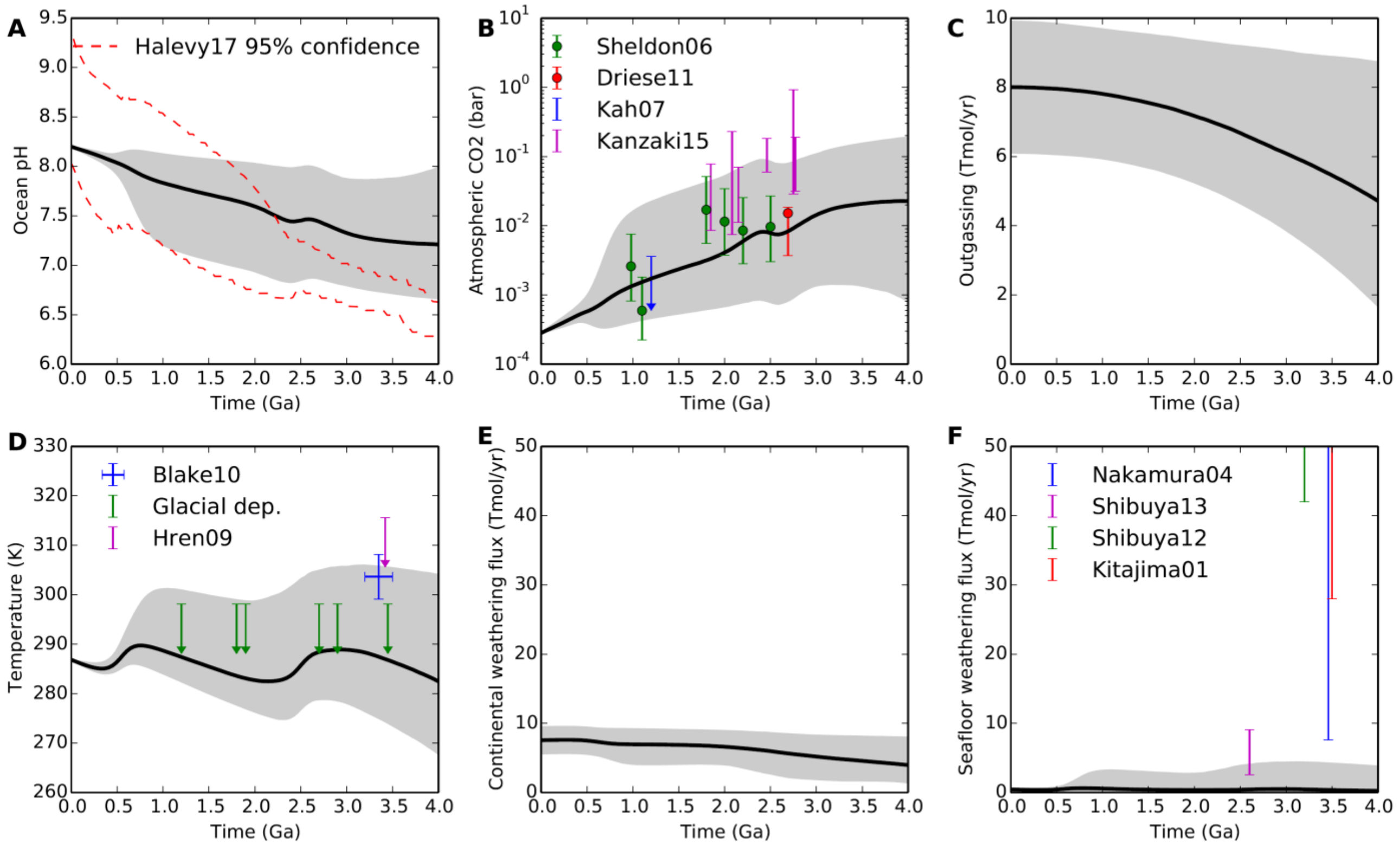

Fig. S13: Same as Fig. S12 except that we have imposed high Proterozoic (0.01%) and high Archean (1%) methane. This avoids freezing the Archean climate with a high likelihood.

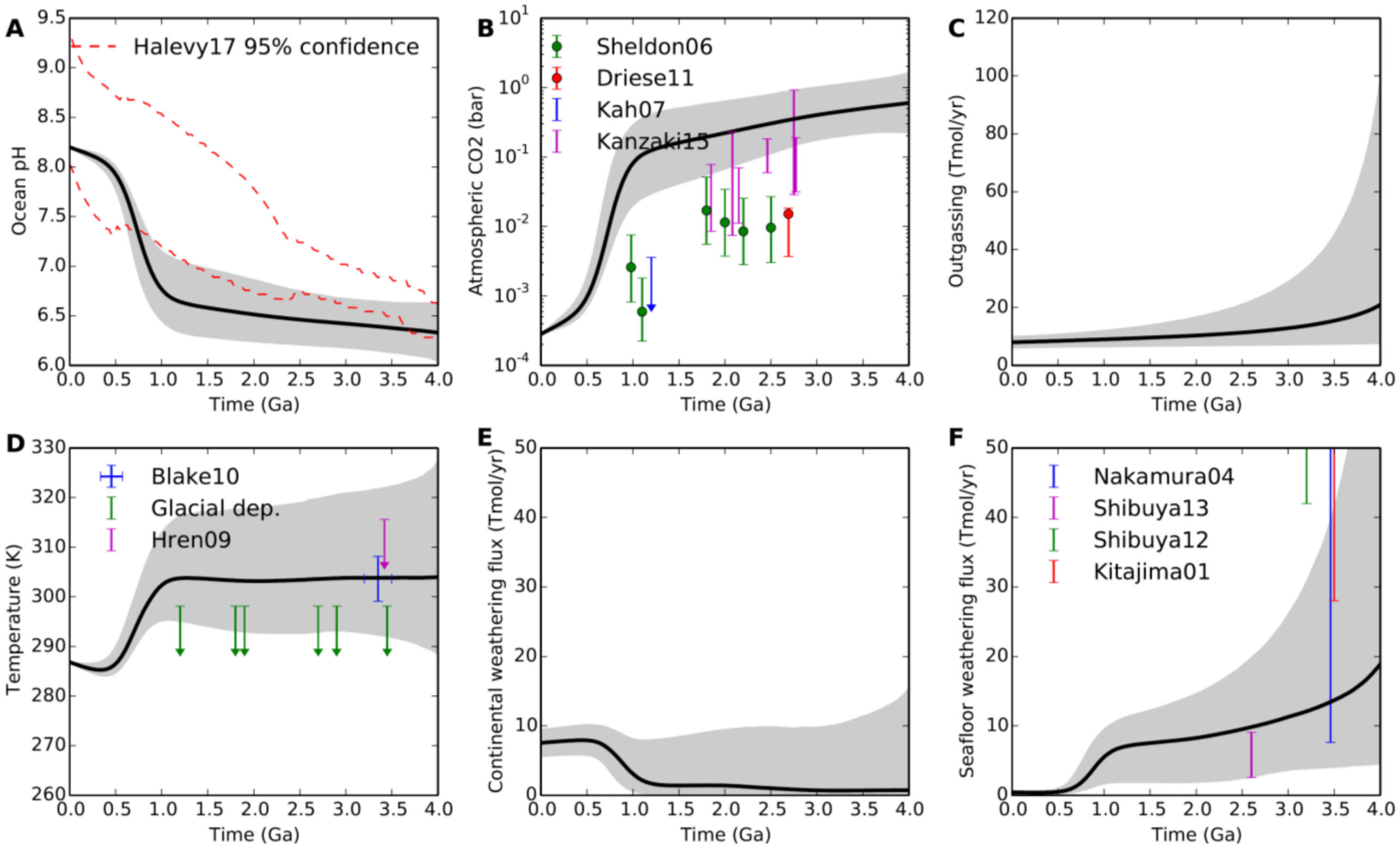

Fig. S14: Extreme biotic enhancement of weathering. Subfigures (a)-(f), lines, and, shadings are the same as in Fig. 3. Even with $f_{bio}$ ranging from 0.001 to 0.1 in the Archean (uniformly sampled in logspace), (d) surface temperatures are likely temperate. In this scenario (a) Ocean pH is acidic for most of Earth history, and (f) seafloor weathering is likely the dominant carbon sink. However, (b) atmospheric $pCO_2$ significantly overshoots almost all proxies, which suggests this extreme biotic enhancement of continental weathering is unlikely.

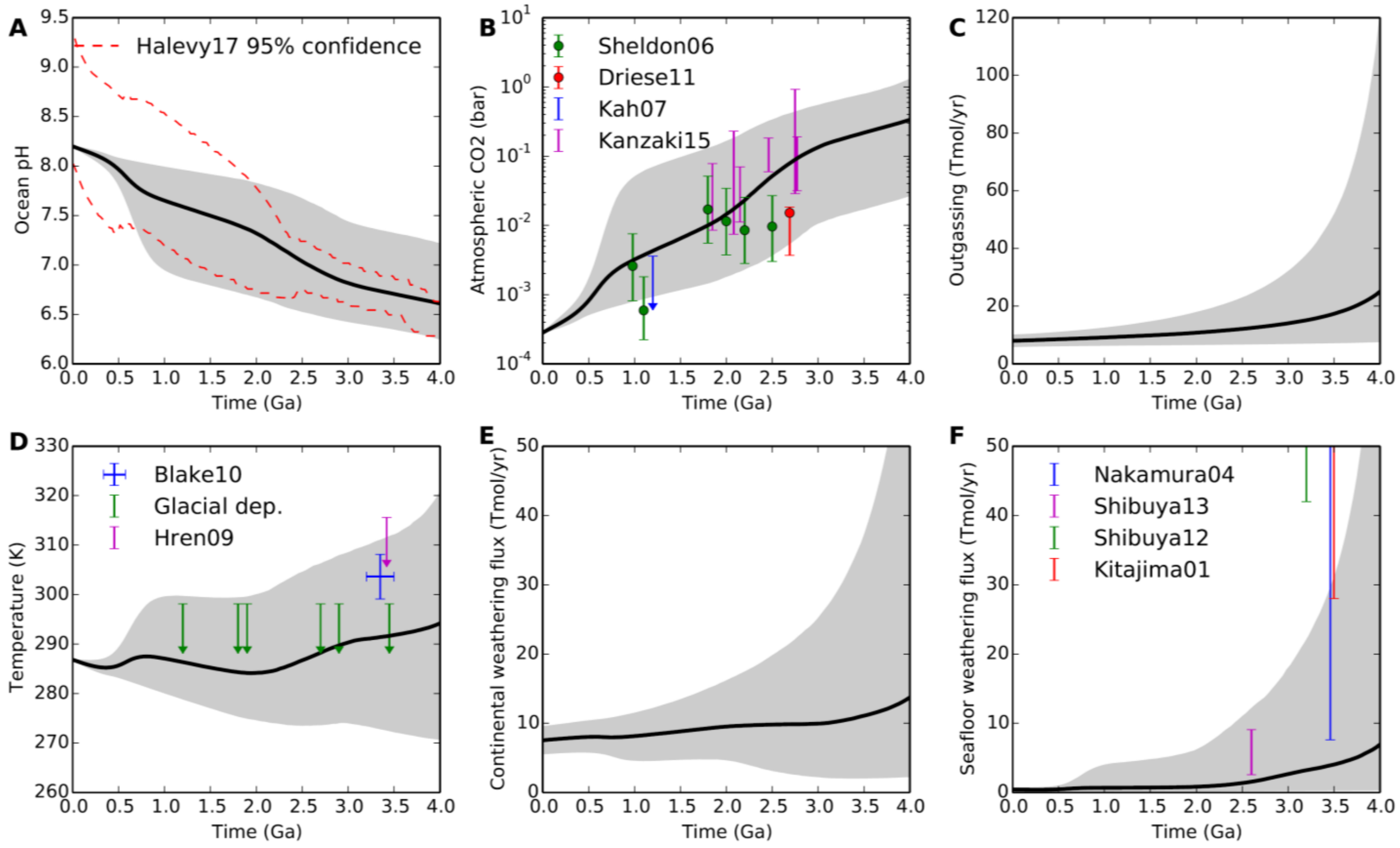

Fig. S15: Deep ocean temperature is not affected by changes in internal heat flow. Subfigures (a)-(f), lines, and, shadings are the same as in Fig. 3. Model outputs are very similar to the nominal model except (f) Archean seafloor weathering is somewhat lower due to the lack of a contribution from a larger internal heat flow.

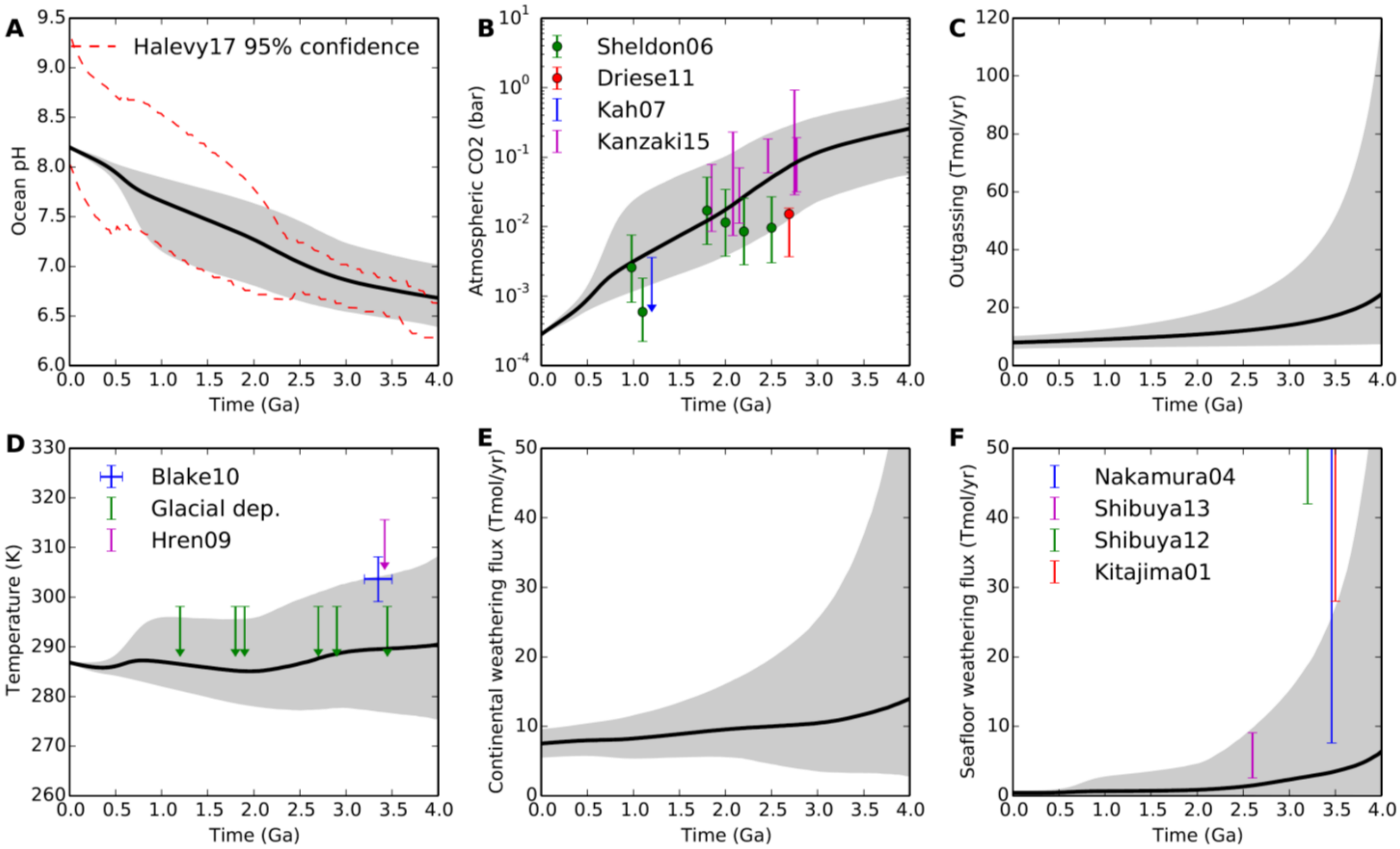

Fig. S16: More conventional range for the temperature sensitivity of continental weathering, $T_e$=5-15 K, as opposed to $T_e$=10-40 K as in our nominal model. Subfigures (a)-(f), lines, and, shadings are the same as in Fig. 3. Model outputs are very similar to the nominal model, except that the spread in Archean temperature and pH values are narrower.

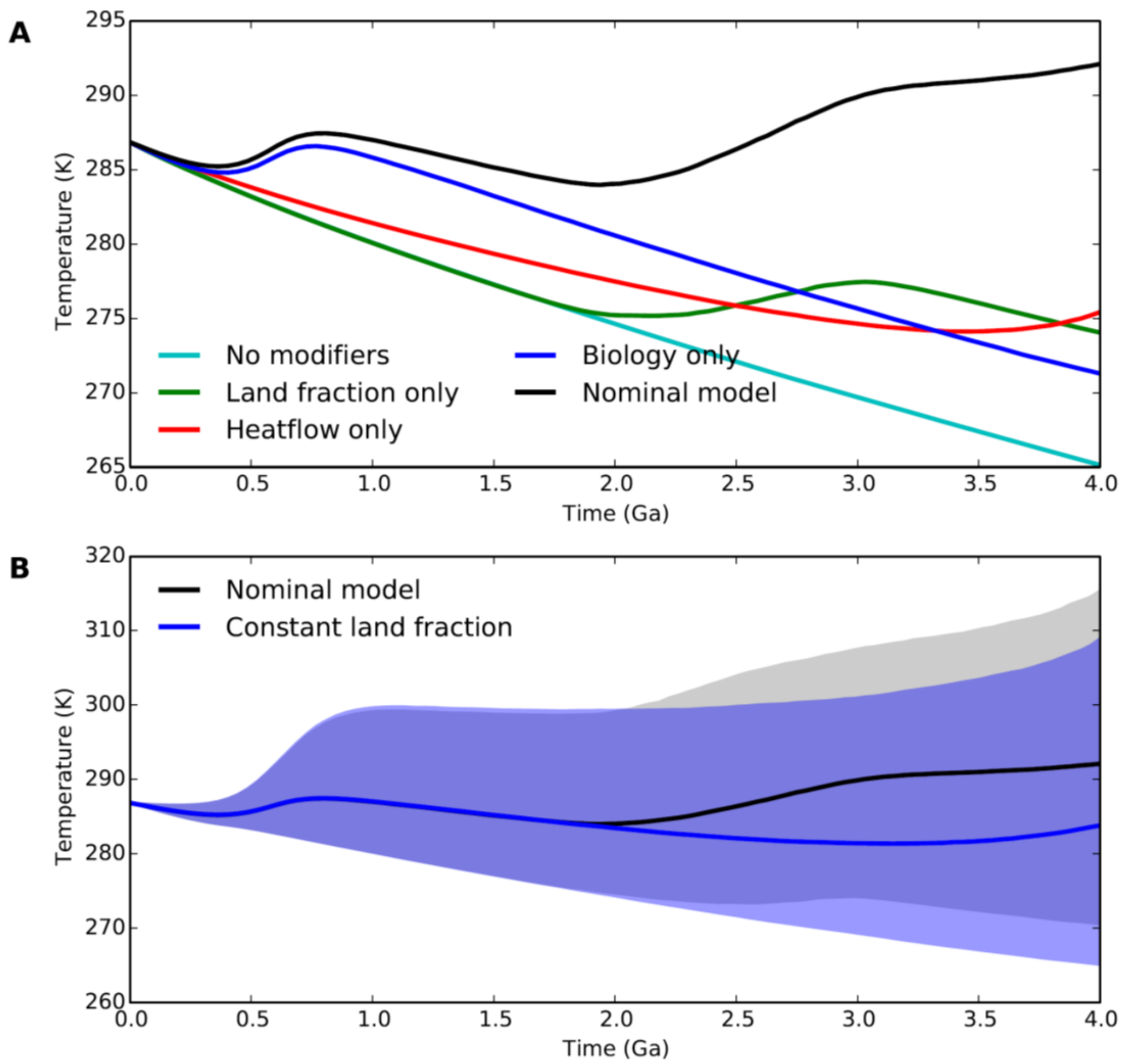

Fig. S17: Contributions to Archean warmth. (A) shows median model outputs for the nominal model (black), and the model when only solar evolution is driving the carbon cycle and continental fraction, biological enhancement of weathering, and heat flow evolution are held constant (cyan). The other three cases are median outputs where continental growth (green), biological enhancement of weathering (blue), and internal heat flow evolution (red) are the only factors modulating solar evolution and weathering feedbacks. We observe that all three factors contribute a similar amount to Archean warmth, on average. (B) Comparison between the nominal model 95% confidence envelope (grey) and the same model where land fraction is held constant throughout Earth history (blue). The difference between the two envelopes is small, implying that temperate Archean temperatures are not solely attributable to reduced Archean land fraction.

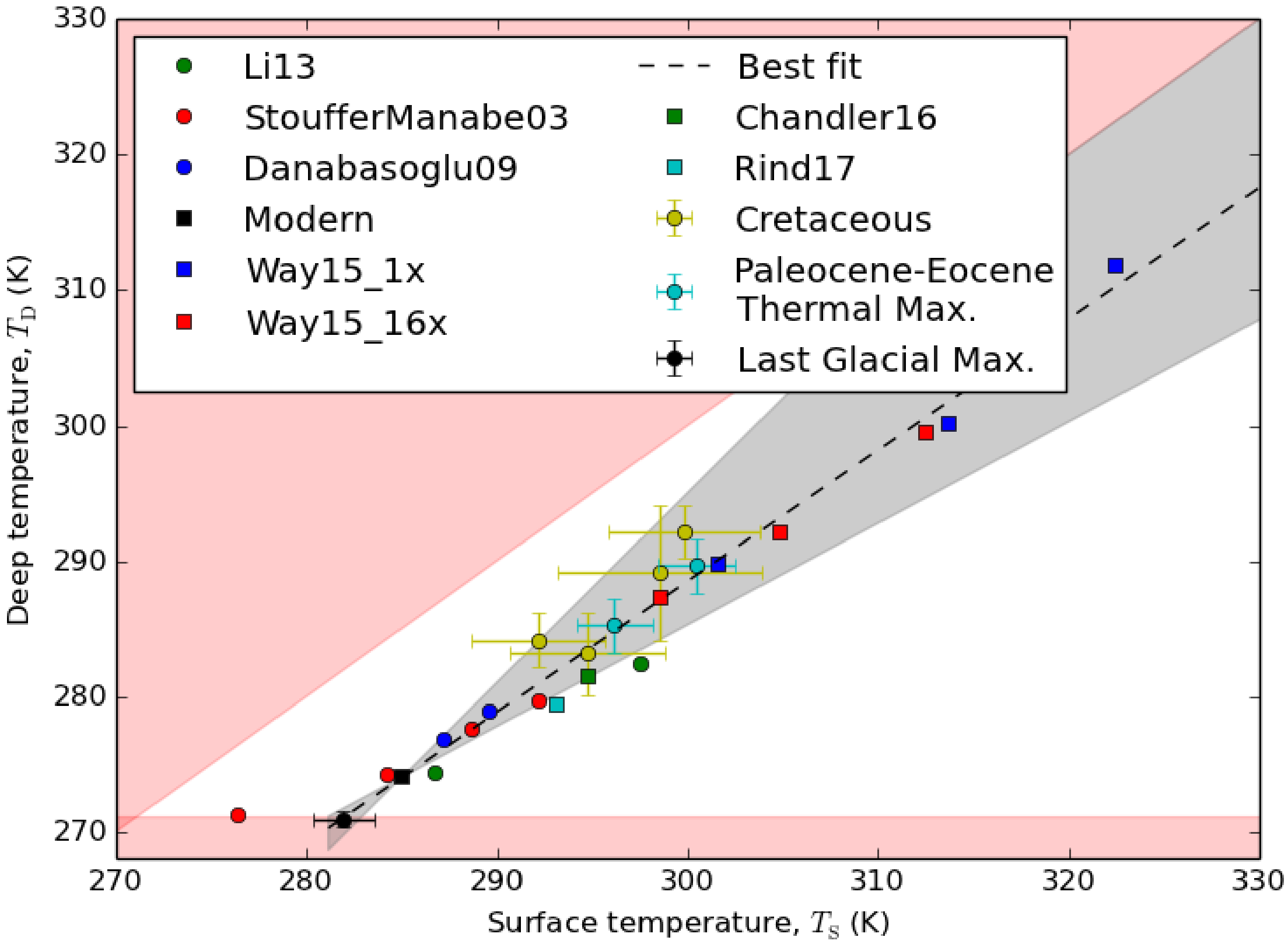

Fig. S18: Linear relationship between global mean surface temperature and deep ocean temperature, as determined by Global Circulation Model (GCM) outputs from the literature, and proxy data, updated from Krissansen-Totton and Catling (1). Full references for the plotted temperatures are provided in Appendix A. The Way, *et al.* (17) 1x and 16x points refer to different day lengths relative to the modern Earth. The deep ocean temperature for GCM outputs was taken to be the mean temperature of the deepest ocean layer (for bathtub oceans) or the mean temperature at the ocean-seafloor interface (for realistic bathymetry). The red regions are unphysical because the ocean is frozen or the deep ocean is warmer than the surface. The dotted line is the best fit to the proxies (gradient 0.96), and the grey shaded region is the range of $T_D - T_S$ relationships considered in our model.

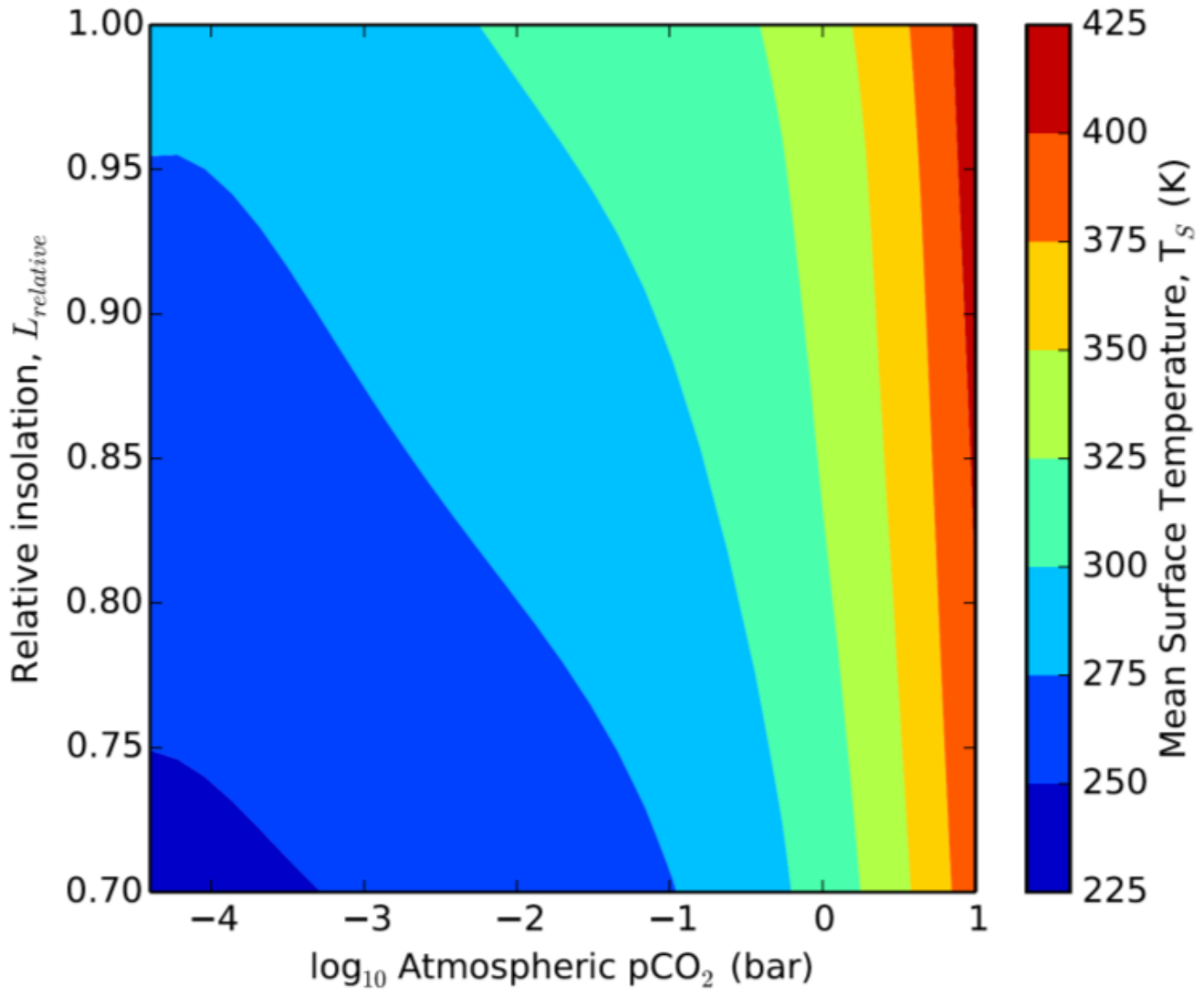

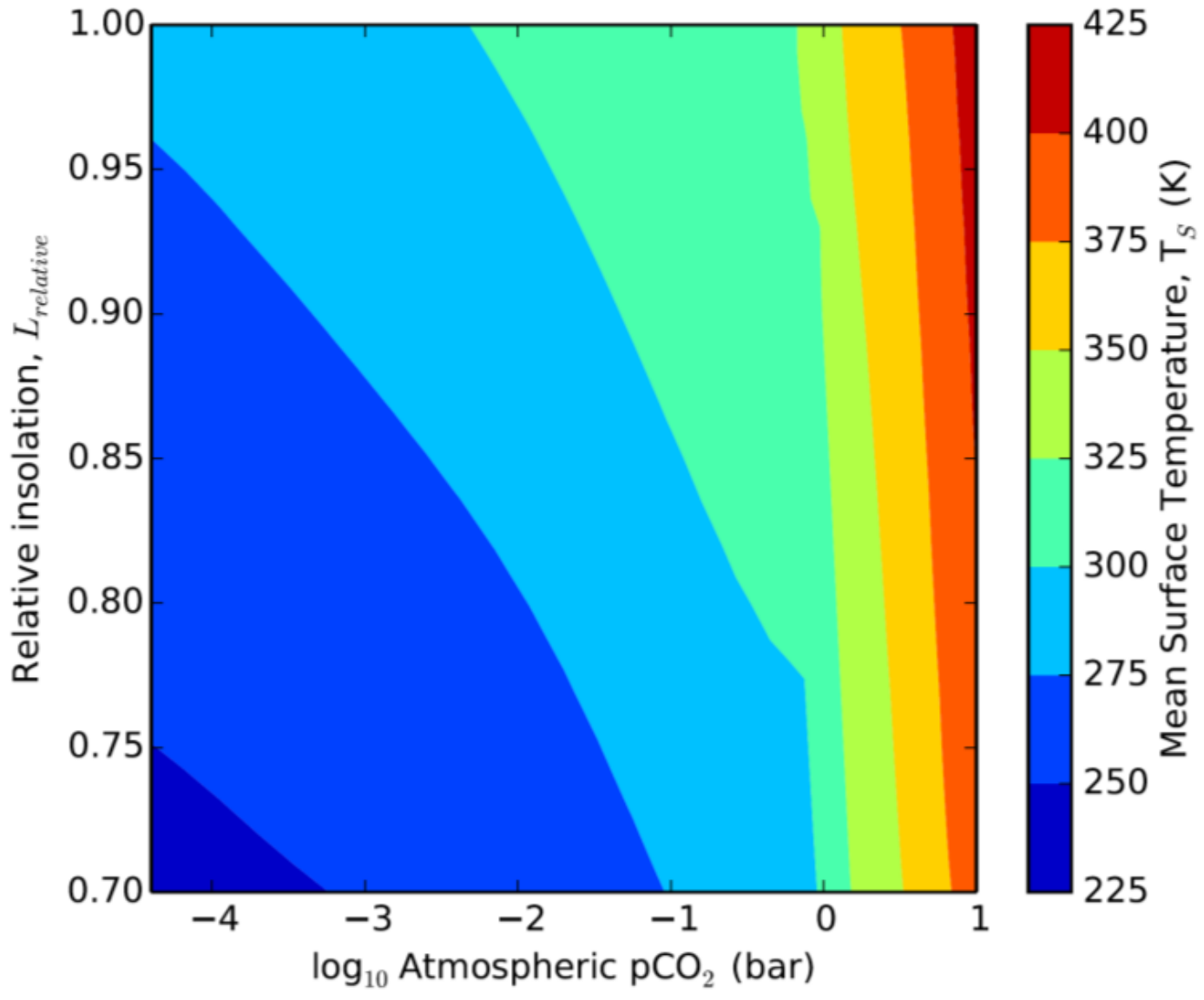

Fig. S19: (a) Parameterized climate model used in this study, described by the polynomial in Appendix E. (b) Outputs of radiative convective model. The polynomial in (a) was created by fitting these outputs.

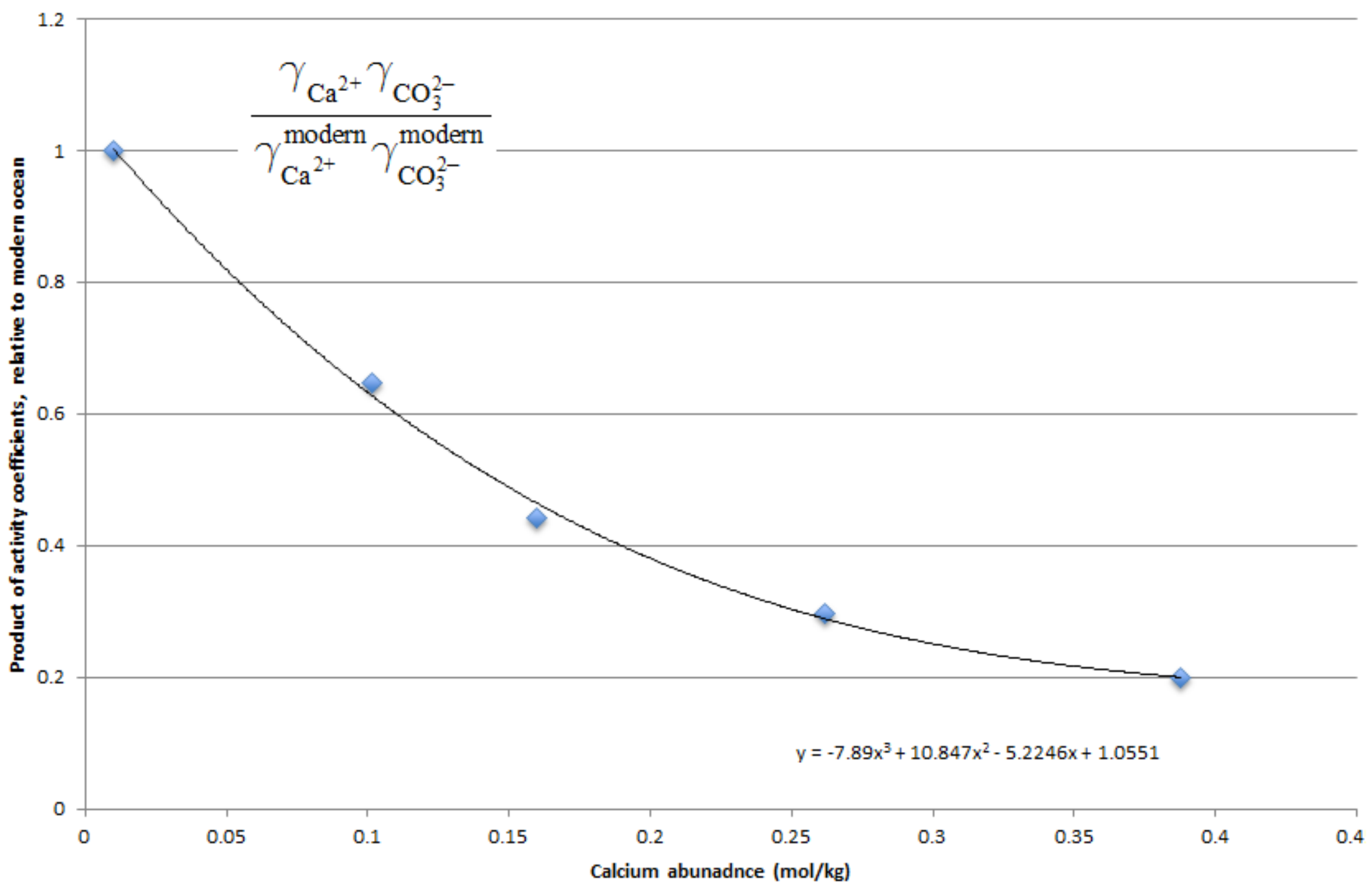

Fig. S20: Parameterization of activity coefficients used in sensitivity in analysis. The product of calcium and carbonate activity coefficients, relative to their modern values is plotted as a function of ocean calcium molality. The data points were calculated using the commercial aqueous chemistry package Aspen Plus (see Appendix C for details). These data points were then fitted with a 3rd order polynomial, and this polynomial was used to calculate relative activity coefficients went computing saturation states in our carbon cycle model.

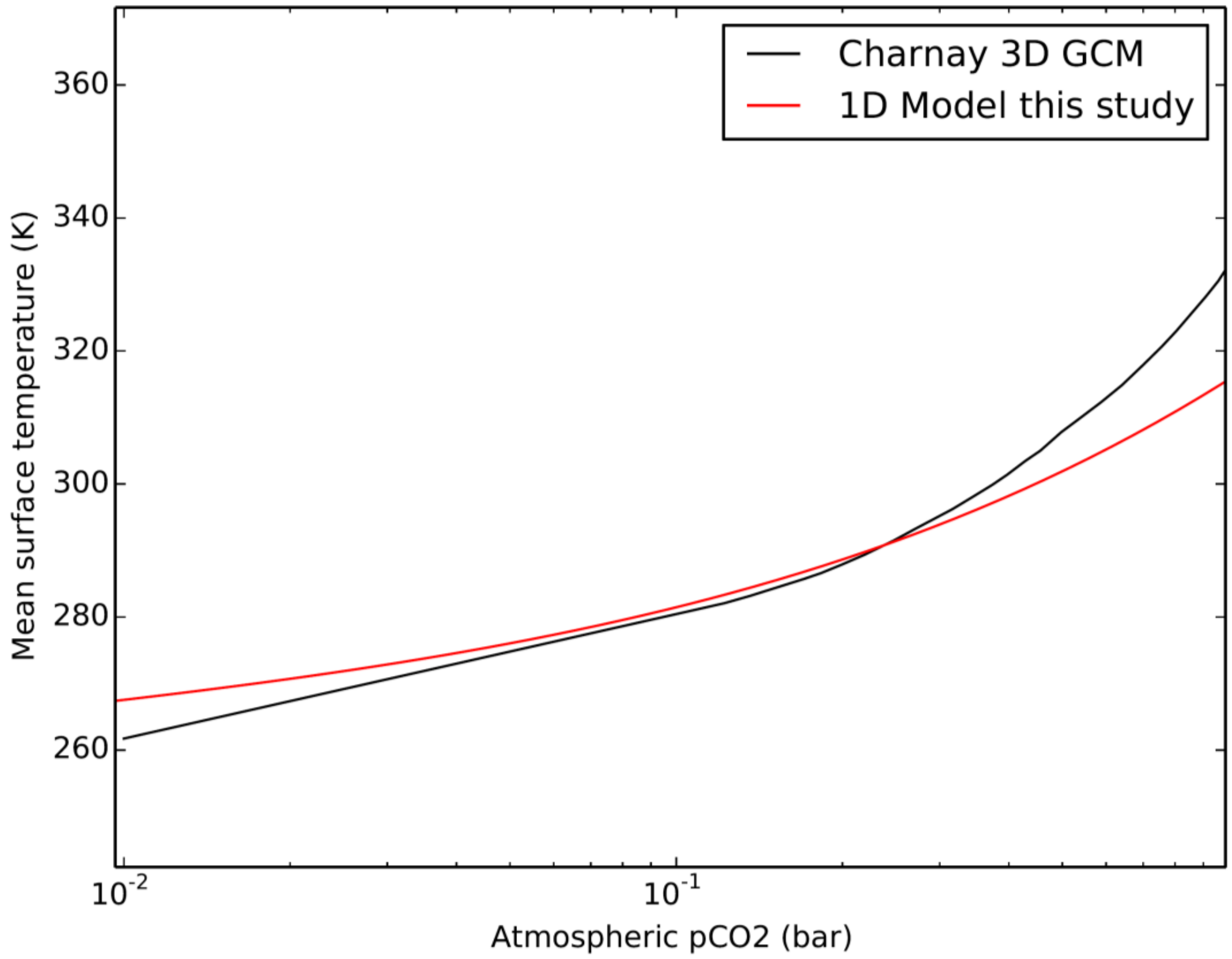

Fig. S21: Comparison between the climate model used in this study (red) and 3D GCM simulating the Archean Earth (3.8 Ga) from Charnay, *et al.* (73) (black). We observe the climate models are very similar across the range of atmospheric $CO_2$ values relevant to the Archean. The GCM has somewhat higher climate sensitivity than our model in this range, which suggests our model $pCO_2$ values may be slightly overestimated.

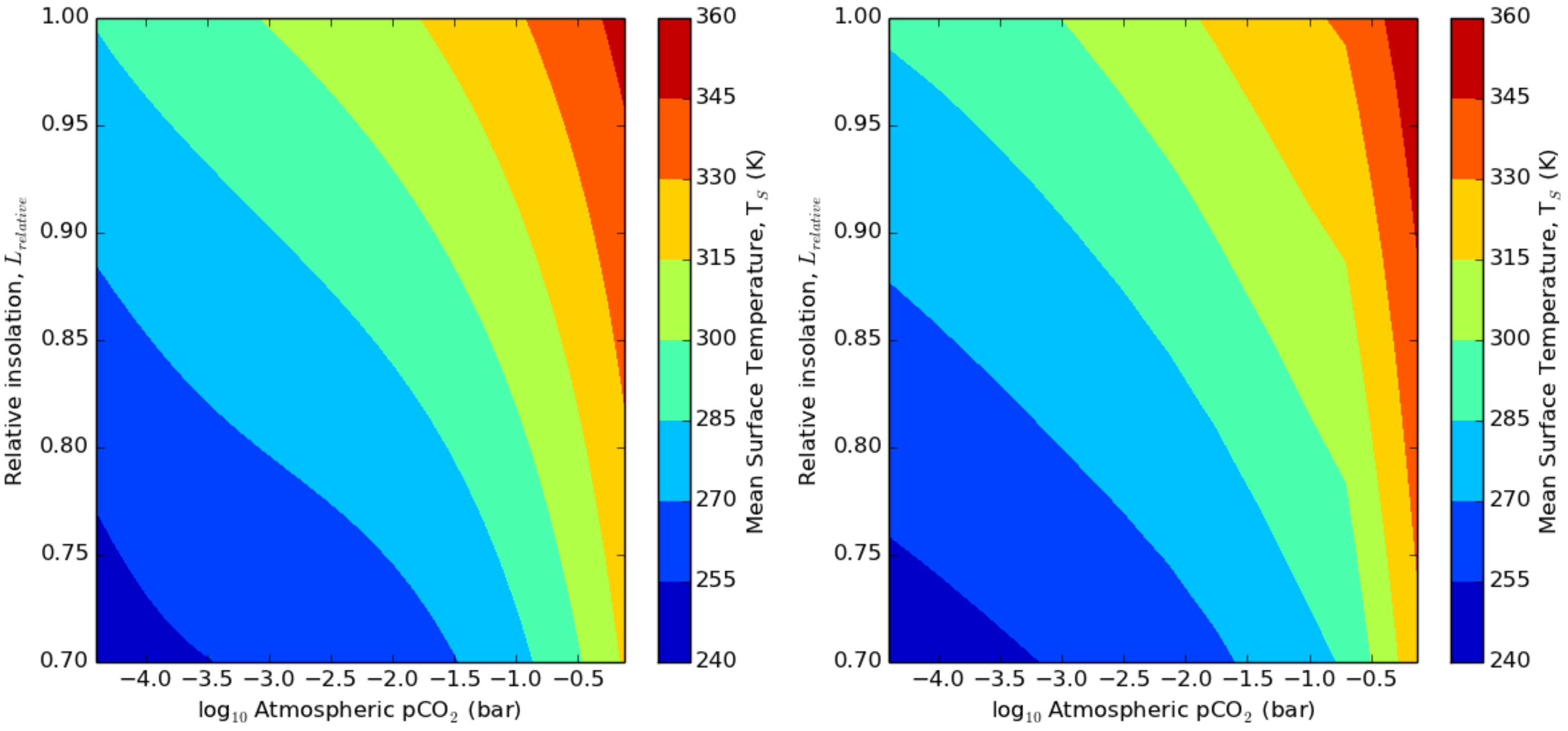

Fig. S22: (a) Parameterized climate model with 100 ppm methane, described by the polynomial in Appendix E. (b) Outputs of radiative convective model. The polynomial in (a) was created by fitting these outputs. Note that we do not attempt to extend the grid beyond 1 bar $pCO_2$ because the carbon cycle model never explores this region of parameter space.

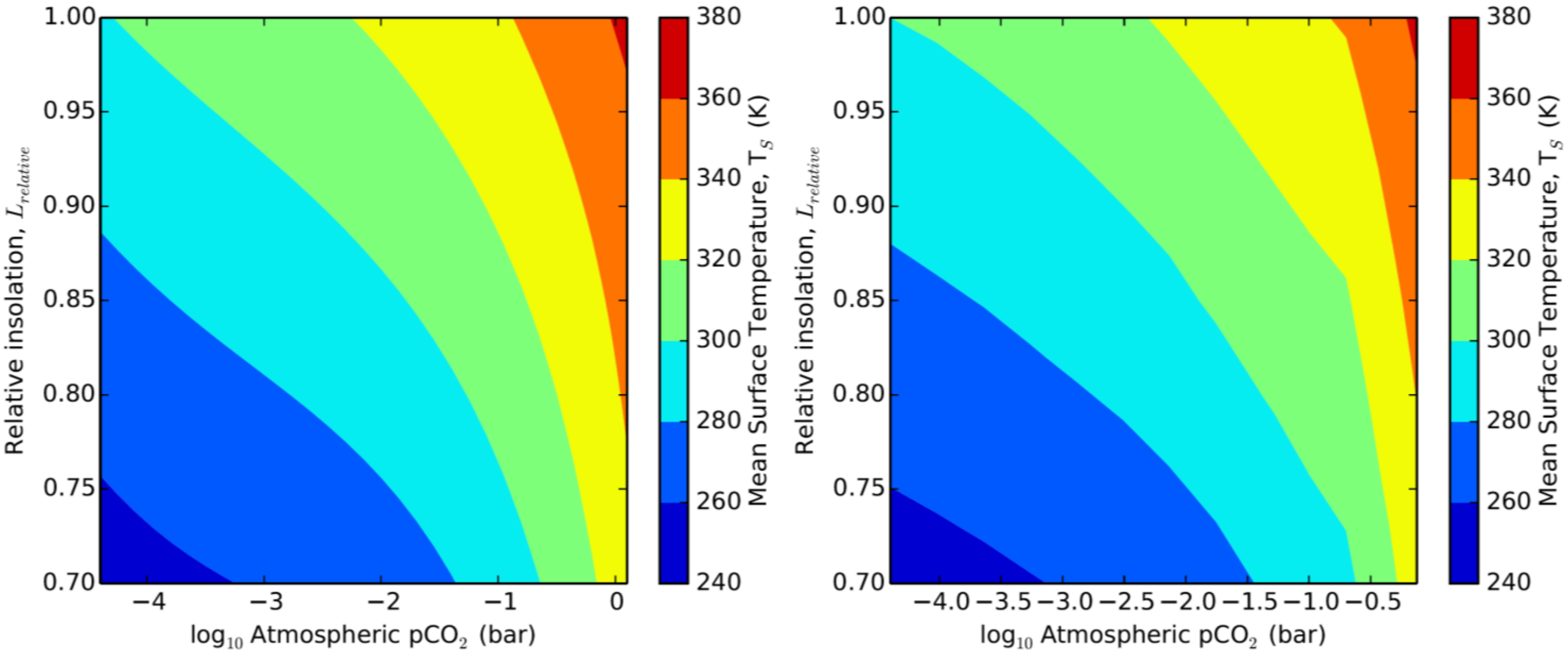

Fig. S23: (a) Parameterized climate model with 1% methane, described by the polynomial in Appendix E. (b) Outputs of radiative convective model. The polynomial in (a) was created by fitting these outputs. Note that we do not attempt to extend the grid beyond 1 bar $pCO_2$ because the carbon cycle model never explores this region of parameter space.

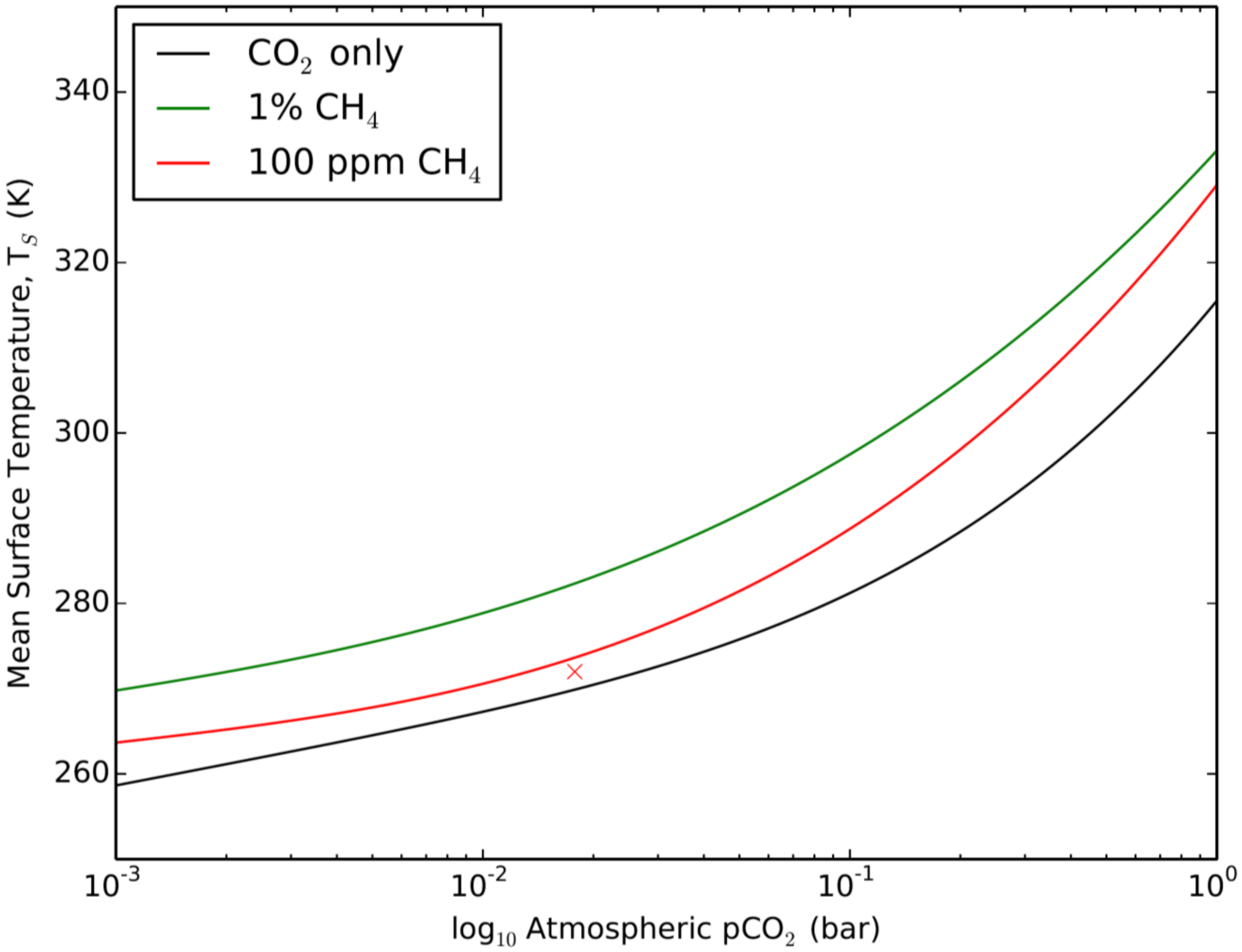

Fig. S24: Comparison of three climate model parameterizations (no methane, 1% methane, and 100 ppm) for relative insolation of 0.82. Our model outputs are comparable to those in Haqq-Misra, *et al.* (68), although with some minor differences because we are fitting a broad range of parameters. The red cross is a spot check with a coupled photochemical-climate model (69) for the 100 ppm methane case. It is slightly cooler than the nominal model without photochemistry because stratospheric water content is lower.

**Supplementary Tables**

Table S1: Parameter ranges

| Variable | Nominal model range (uniform distribution) | Reference/explanation |
|---|---|---|
| $CO_2$-dependence continental silicate weathering, $\alpha$ | 0.1-0.5 | (13, 39, 81) |
| $CO_2$-dependence continental silicate weathering, $\xi$ | 0.1-0.5 | (82) |
| e-folding temp. dep. of cont. weathering, $T_e$ (K) | 10-40 | Krissansen-Totton and Catling (1) and references therein. But see Appendix C for sensitivity tests using a more conventional range. |
| Relative Archean land fraction, $L_{\text{Archean}}$ * | 0.1-0.75 | Fig. 2a |
| Time of continental growth, $t_{\text{grow}}$ (Ga) | 2.0-3.0 | Fig. 2a |
| Precambrian relative weatherability, $B_{\text{Precambrian}}$ | 0.1-0.999 | Fig. 2c |
| Modern outgassing, $F_{\text{out}}^{\text{mod}}$ (Tmol C $yr^{-1}$) | 6-10 | (82, 83) |
| Modern carb. weathering, $F_{\text{carb}}^{\text{mod}}$ (Tmol C $yr^{-1}$) | 7-14 | (84, their table 2) |
| Pore-space circulation time, $\tau$ (kyr) | 20 - 1000 | (2, 85) |
| Carbonate precip. Coefficient, $n$ | 1.0-2.5 | (86) |
| Modern seafloor dissolution relative to precipitation, $x$ | 0.5-1.5 | (87). See Krissansen-Totton and Catling (1) for further explanation and justification |
| Surface-deep temp. gradient, $a_{\text{grad}}$ | 0.75-1.4 | Fig. S18 and accompanying discussion. |
| pH dependence seafloor, $\gamma$ | 0-0.5 | Gudbrandsson*, et al.* (88) but see discussion in Krissansen-Totton and Catling (1). |
| Temp. dependence seafloor, $E_{\text{bas}}$ (kJ $mol^{-1}$) | 60-100 | (1, 3) |
| Spreading rate dep., $\beta$ | 0.0-2.0 | See methods |
| Internal heat flow exponent, $n_{\text{out}}$ ** | 0-0.73 | Fig. 2d |
| Outgassing exponent, $m$ | 1.0-2.0 | See methods |
| Relative Archean sediment | 0.2-1.0 | See methods |
| thickness, $f_{sed}$ | | |

* For the model runs with no Archean land, we assume $L_{\text{Archean}} = -0.2$.

** For the no Archean land sensitivity test, model failure is slightly biased toward low heatflow exponent values. This was corrected by sampling the heatflow exponent to ensure the final distribution was uniform.

Table S2: Initial values or initial value ranges assumed in our model.

| Variable | Initial value or initial range | Reference |
|---|---|---|
| Modern pore-space carbonate precipitation, $P_{\text{pore}}^{\text{mod}}$ (Tmol C yr$^{-1}$) | 0.45* | (89) |
| Modern seafloor dissolution, $F_{\text{diss}}^{\text{mod}}$ (Tmol C yr$^{-1}$) | 0.225-0.675** | (87) |
| Modern outgassing, $F_{\text{out}}^{\text{mod}}$ (Tmol C yr$^{-1}$) | 6-10 | (82, 83) |
| Modern carb. Weathering, $F_{\text{carb}}^{\text{mod}}$ (Tmol C yr$^{-1}$) | 7-14 | (84, their table 2) |
| Preindustrial mean surface temperature, $T_S$ (K) | 285 | (39) |
| Modern ocean pH | 8.2 | (90) |
| Ocean Ca abundance (mMol kg$^{-1}$) | 10 | (90) |
| Preindustrial atmospheric $pCO_2$ (ppm) | 280 | - |

*Because we are adopting wide ranges for $F_{\text{out}}^{\text{mod}}$ and $F_{\text{carb}}^{\text{mod}}$, it is unnecessary to include a range for $P_{\text{pore}}^{\text{mod}}$ because its size relative to outgassing and weathering fluxes already encompasses a wide range (only the relative sizes of carbon cycle fluxes matter for predicting observable variables).

**Here was assume $F_{\text{diss}}^{\text{mod}} = xP_{\text{pore}}^{\text{mod}}$ where $x = 0.5$ to $1.5$. Coogan and Gillis (87) used a geochemical model of pore-space precipitation to show that at least 70% of pore-space precipitation is attributable to alkalinity release from basalt dissolution. Here we conservatively assume a lower limit of 50% instead. The upper limit is 150% to allow for the possibility that pore-space dissolution exceeds pore-space precipitation, and that the excess alkalinity is mixed into the ocean to form marine carbonates.

**Supplementary References**

1. Krissansen-Totton J & Catling DC (2017) Constraining climate sensitivity and continental versus seafloor weathering using an inverse geological carbon cycle model. *Nat Commun* 8:15423.
2. Caldeira K (1995) Long-term control of atmospheric carbon dioxide; low-temperature seafloor alteration or terrestrial silicate-rock weathering? *Am J Sci* 295(9):1077-1114.
3. Coogan LA & Dosso SE (2015) Alteration of ocean crust provides a strong temperature dependent feedback on the geological carbon cycle and is a primary driver of the Sr-isotopic composition of seawater. *Earth Planet Sci Lett* 415:38-46.
4. Divins D (2003) Total sediment thickness of the world's oceans and marginal seas. NOAA National Geophysical Data Center, Boulder.
5. Armstrong R & Harmon R (1981) Radiogenic isotopes: the case for crustal recycling on a near-steady-state no-continental-growth Earth [and discussion]. *Philosophical Transactions of the Royal Society of London A: Mathematical, Physical and Engineering Sciences* 301(1461):443-472.
6. Veizer J & Jansen SL (1979) Basement and sedimentary recycling and continental evolution. *J Geol* 87(4):341-370.
7. Schubert G & Reymer A (1985) Continental volume and freeboard through geological time.
8. McLennan SM & Taylor S (1982) Geochemical constraints on the growth of the continental crust. *J Geol* 90(4):347-361.
9. Reymer A & Schubert G (1984) Phanerozoic addition rates to the continental crust and crustal growth. *Tectonics* 3(1):63-77.
10. Flament N, Coltice N, & Rey PF (2008) A case for late-Archaean continental emergence from thermal evolution models and hypsometry. *Earth Planet Sci Lett* 275(3):326-336.
11. Viehmann S, Hoffmann JE, Münker C, & Bau M (2014) Decoupled Hf-Nd isotopes in Neoarchean seawater reveal weathering of emerged continents. *Geology* 42(2):115-118.
12. Davies GF (2009) Effect of plate bending on the Urey ratio and the thermal evolution of the mantle. *Earth Planet Sci Lett* 287(3):513-518.
13. Sleep NH & Zahnle K (2001) Carbon dioxide cycling and implications for climate on ancient Earth. *J Geophys Res: Planets* 106(E1):1373-1399.
14. Korenaga J (2008) Plate tectonics, flood basalts and the evolution of Earth's oceans. *Terra Nova* 20(6):419-439.
15. Godderis Y & Veizer J (2000) Tectonic control of chemical and isotopic composition of ancient oceans; the impact of continental growth. *Am J Sci* 300(5):434-461.
16. Schubert G, Turcotte DL, & Olson P (2001) *Mantle convection in the Earth and planets* (Cambridge University Press, Cambridge, UK).
17. Way M, Del Genio A, Kelley M, Aleinov I, & Clune T (2015) Exploring the Inner Edge of the Habitable Zone with Fully Coupled Oceans. *arXiv preprint arXiv:1511.07283*.

18. Chandler MA, Sohl LE, & Carter D (2016) ROCKE-3D: A Dynamical Modeling Approach to Exploring Rocky Planet Habitability. *Geological Society of America Annual Meeting*, pp 79-11.
19. Li C, von Storch J-S, & Marotzke J (2013) Deep-ocean heat uptake and equilibrium climate response. *Climate Dynamics* 40(5-6):1071-1086.
20. Stouffer R & Manabe S (2003) Equilibrium response of thermohaline circulation to large changes in atmospheric CO2 concentration. *Climate Dynamics* 20(7-8):759-773.
21. Danabasoglu G & Gent PR (2009) Equilibrium climate sensitivity: Is it accurate to use a slab ocean model? *Journal of Climate* 22(9):2494-2499.
22. Jones TD*, et al.* (2013) Climate model and proxy data constraints on ocean warming across the Paleocene–Eocene Thermal Maximum. *Earth-Science Reviews* 125:123-145.
23. Clark PU*, et al.* (2009) The last glacial maximum. *Science* 325(5941):710-714.
24. IPCC (2014) *Climate change 2013: the physical science basis: Working Group I contribution to the Fifth assessment report of the Intergovernmental Panel on Climate Change* (Cambridge University Press).
25. Franck S, Kossacki KJ, von Bloh W, & Bounama C (2002) Long‐term evolution of the global carbon cycle: historic minimum of global surface temperature at present. *Tellus B* 54(4):325-343.
26. Foriel J*, et al.* (2004) Biological control of Cl/Br and low sulfate concentration in a 3.5-Gyr-old seawater from North Pole, Western Australia. *Earth Planet Sci Lett* 228(3):451-463.
27. Halevy I & Bachan A (2017) The geologic history of seawater pH. *Science* 355(6329):1069-1071.
28. De Ronde CE, deR Channer DM, Faure K, Bray CJ, & Spooner ET (1997) Fluid chemistry of Archean seafloor hydrothermal vents: Implications for the composition of circa 3.2 Ga seawater. *Geochim Cosmochim Acta* 61(19):4025-4042.
29. Lowe DR & Byerly GR (2003) Ironstone pods in the Archean Barberton greenstone belt, South Africa: Earth's oldest seafloor hydrothermal vents reinterpreted as Quaternary subaerial springs. *Geology* 31(10):909-912.
30. Krissansen-Totton J, Bergsman DS, & Catling DC (2016) On detecting biospheres from chemical thermodynamic disequilibrium in planetary atmospheres. *Astrobiology* 16(1):39-67.
31. Schwartzman D (2002) *Life, temperature, and the Earth: the self-organizing biosphere* (Columbia University Press).
32. Avice G, Marty B, & Burgess R (2017) The origin and degassing history of the Earth's atmosphere revealed by Archean xenon. *Nat Commun* 8.
33. Debaille V*, et al.* (2013) Stagnant-lid tectonics in early Earth revealed by 142 Nd variations in late Archean rocks. *Earth Planet Sci Lett* 373:83-92.
34. Tosi N*, et al.* (2017) The habitability of a stagnant-lid Earth. *Astronomy & Astrophysics* 605:A71.

35. Trail D, Watson EB, & Tailby ND (2011) The oxidation state of Hadean magmas and implications for early Earth/'s atmosphere. *Nature* 480(7375):79-82.
36. O'Neill C, Lenardic A, Moresi L, Torsvik T, & Lee C-T (2007) Episodic precambrian subduction. *Earth Planet Sci Lett* 262(3):552-562.
37. Moyen J-F & Van Hunen J (2012) Short-term episodicity of Archaean plate tectonics. *Geology* 40(5):451-454.
38. West AJ, Galy A, & Bickle M (2005) Tectonic and climatic controls on silicate weathering. *Earth Planet Sci Lett* 235(1):211-228.
39. Walker JC, Hays P, & Kasting JF (1981) A negative feedback mechanism for the long - term stabilization of Earth's surface temperature. *Journal of Geophysical Research: Oceans (1978-2012)* 86(C10):9776-9782.
40. Berner RA & Kothavala Z (2001) GEOCARB III: a revised model of atmospheric CO2 over Phanerozoic time. *Am J Sci* 301(2):182-204.
41. White AF & Buss HL (2014) Natural Weathering Rates of Silicate Materials. *Surface and Ground Water, Weathering and Soils, Treatise on Geochemistry 2nd Edition*, ed Drever JI (Elsevier), pp 115-155.
42. Geboy NJ*, et al.* (2013) Re–Os age constraints and new observations of Proterozoic glacial deposits in the Vazante Group, Brazil. *Precambrian Research* 238:199-213.
43. Kuipers G, Beunk FF, & van der Wateren FM (2013) Periglacial evidence for a 1.91–1.89 Ga old glacial period at low latitude, Central Sweden. *Geology Today* 29(6):218-221.
44. Williams GE (2005) Subglacial meltwater channels and glaciofluvial deposits in the Kimberley Basin, Western Australia: 1.8 Ga low-latitude glaciation coeval with continental assembly. *Journal of the Geological Society* 162(1):111-124.
45. Ojakangas RW, Srinivasan R, Hegde V, Chandrakant S, & Srikantia S (2014) The Talya Conglomerate: an Archean (~ 2.7 Ga) Glaciomarine Formation, Western Dharwar Craton, Southern India. *Curr. Sci* 106(3):387-396.
46. Young GM, Brunn VV, Gold DJ, & Minter W (1998) Earth's oldest reported glaciation: physical and chemical evidence from the Archean Mozaan Group (~ 2.9 Ga) of South Africa. *J Geol* 106(5):523-538.
47. de Wit MJ & Furnes H (2016) 3.5-Ga hydrothermal fields and diamictites in the Barberton Greenstone Belt—Paleoarchean crust in cold environments. *Sci Adv* 2(2):e1500368.
48. Zachos JC, Dickens GR, & Zeebe RE (2008) An early Cenozoic perspective on greenhouse warming and carbon-cycle dynamics. *Nature* 451(7176):279-283.
49. Hren M, Tice M, & Chamberlain C (2009) Oxygen and hydrogen isotope evidence for a temperate climate 3.42 billion years ago. *Nature* 462(7270):205-208.
50. Blake RE, Chang SJ, & Lepland A (2010) Phosphate oxygen isotopic evidence for a temperate and biologically active Archaean ocean. *Nature* 464(7291):1029-1032.

51. Sheldon ND (2006) Precambrian paleosols and atmospheric CO 2 levels. *Precambrian Research* 147(1):148-155.
52. Driese SG*, et al.* (2011) Neoarchean paleoweathering of tonalite and metabasalt: Implications for reconstructions of 2.69 Ga early terrestrial ecosystems and paleoatmospheric chemistry. *Precambrian Research* 189(1):1-17.
53. Kanzaki Y & Murakami T (2015) Estimates of atmospheric CO 2 in the Neoarchean–Paleoproterozoic from paleosols. *Geochim Cosmochim Acta* 159:190-219.
54. Kah LC & Riding R (2007) Mesoproterozoic carbon dioxide levels inferred from calcified cyanobacteria. *Geology* 35(9):799-802.
55. Nakamura K & Kato Y (2004) Carbonatization of oceanic crust by the seafloor hydrothermal activity and its significance as a CO 2 sink in the Early Archean. *Geochim Cosmochim Acta* 68(22):4595-4618.
56. Shibuya T*, et al.* (2013) Decrease of seawater CO 2 concentration in the Late Archean: an implication from 2.6 Ga seafloor hydrothermal alteration. *Precambrian Research* 236:59-64.
57. Shibuya T*, et al.* (2012) Depth variation of carbon and oxygen isotopes of calcites in Archean altered upperoceanic crust: Implications for the CO 2 flux from ocean to oceanic crust in the Archean. *Earth Planet Sci Lett* 321:64-73.
58. Kitajima K, Maruyama S, Utsunomiya S, & Liou J (2001) Seafloor hydrothermal alteration at an Archaean mid ‐ ocean ridge. *J Metamorph Geol* 19(5):583-599.
59. Kasting JF & Ackerman TP (1986) Climatic consequences of very high carbon dioxide levels in the Earth's early atmosphere. *Science* 234(4782):1383-1385.
60. Kopparapu RK*, et al.* (2013) Habitable zones around main-sequence stars: new estimates. *Astrophys J* 765(2):131.
61. Segura A*, et al.* (2003) Ozone concentrations and ultraviolet fluxes on Earth-like planets around other stars. *Astrobiology* 3(4):689-708.
62. Rugheimer S, Kaltenegger L, Zsom A, Segura A, & Sasselov D (2013) Spectral fingerprints of Earth-like planets around FGK stars. *Astrobiology* 13(3):251-269.
63. Arney GN*, et al.* (2017) Pale orange dots: the impact of organic haze on the habitability and detectability of Earthlike exoplanets. *Astrophys J* 836(1):49.
64. Kitzmann D*, et al.* (2010) Clouds in the atmospheres of extrasolar planets-I. Climatic effects of multi-layered clouds for Earth-like planets and implications for habitable zones. *Astronomy & Astrophysics* 511:A66.
65. Kitzmann D, Patzer A, von Paris P, Godolt M, & Rauer H (2011) Clouds in the atmospheres of extrasolar planets-II. Thermal emission spectra of Earth-like planets influenced by low and high-level clouds. *Astronomy & Astrophysics* 531:A62.
66. Schwieterman EW*, et al.* (2016) Identifying planetary biosignature impostors: spectral features of CO and O4 resulting from abiotic O2/O3 production. *The Astrophysical Journal Letters* 819(1):L13.

67. Meadows VS, *et al.* (2016) The habitability of Proxima Centauri b: II: environmental states and observational discriminants. *arXiv preprint arXiv:1608.08620*.
68. Haqq-Misra JD, Domagal-Goldman SD, Kasting PJ, & Kasting JF (2008) A revised, hazy methane greenhouse for the Archean Earth. *Astrobiology* 8(6):1127-1137.
69. Arney G, *et al.* (2016) The pale orange dot: the spectrum and habitability of hazy Archean Earth. *Astrobiology* 16(11):873-899.
70. Toon OB, McKay CP, Ackerman TP, & Santhanam K (1989) Rapid calculation of radiative heating rates and photodissociation rates in inhomogeneous multiple scattering atmospheres. *Journal of Geophysical Research: Atmospheres* 94(D13):16287–16301.
71. Manabe S & Wetherald RT (1967) Thermal equilibrium of the atmosphere with a given distribution of relative humidity. *Journal of the Atmospheric Sciences* 24(3):241-259.
72. Pavlov AA, Kasting JF, Brown LL, Rages KA, & Freedman R (2000) Greenhouse warming by CH4 in the atmosphere of early Earth. *J Geophys Res: Planets* 105(E5):11981-11990.
73. Charnay B, Hir GL, Fluteau F, Forget F, & Catling DC (2017) A warm or a cold early Earth? New insights from a 3-D climate-carbon model. *arXiv preprint arXiv:1706.06842*.
74. Olson SL, Reinhard CT, & Lyons TW (2016) Limited role for methane in the mid-Proterozoic greenhouse. *Proc Natl Acad Sci USA* 113(41):11447-11452.
75. Kasting JF & Brown LL (1998) The early atmosphere as a source of biogenic compounds. *The Molecular Origins of Life*, ed Brack A (Cambridge University Press, Cambridge, UK), pp 35-56.
76. Kasting JF, Liu SC, & Donahue TM (1979) Oxygen levels in the prebiological atmosphere. *J Geophys Res Oceans* 84(C6):3097-3107.
77. Zahnle K, Claire M, & Catling D (2006) The loss of mass‐independent fractionation in sulfur due to a Palaeoproterozoic collapse of atmospheric methane. *Geobiology* 4(4):271-283.
78. Domagal-Goldman SD, Segura A, Claire MW, Robinson TD, & Meadows VS (2014) Abiotic ozone and oxygen in atmospheres similar to prebiotic Earth. *Astrophys J* 792(2):90.
79. Harman C, Schwieterman E, Schottelkotte J, & Kasting J (2015) Abiotic $O_2$ Levels on Planets around F, G, K, and M Stars: Possible False Positives for Life? *Astrophys J* 812(2):137.
80. Trainer MG, *et al.* (2006) Organic haze on Titan and the early Earth. *Proc Natl Acad Sci USA* 103(48):18035-18042.
81. Volk T (1987) Feedbacks between weathering and atmospheric CO2 over the last 100 million years. *Am. J. Sci* 287(8):763-779.
82. Berner RA (2004) *The Phanerozoic carbon cycle: CO2 and O2* (Oxford University Press).
83. Lee C-TA & Lackey JS (2015) Global continental arc flare-ups and their relation to long-term greenhouse conditions. *Elements* 11(2):125-130.

84. Hartmann J, Jansen N, Dürr HH, Kempe S, & Köhler P (2009) Global CO 2-consumption by chemical weathering: What is the contribution of highly active weathering regions? *Global and Planetary Change* 69(4):185-194.
85. Johnson HP & Pruis MJ (2003) Fluxes of fluid and heat from the oceanic crustal reservoir. *Earth Planet Sci Lett* 216(4):565-574.
86. Opdyke BN & Wilkinson BH (1993) Carbonate mineral saturation state and cratonic limestone accumulation. *Am J Sci* 293:217-217.
87. Coogan LA & Gillis KM (2013) Evidence that low‐temperature oceanic hydrothermal systems play an important role in the silicate‐carbonate weathering cycle and long‐term climate regulation. *Geochemistry, Geophysics, Geosystems* 14(6):1771-1786.
88. Gudbrandsson S, Wolff-Boenisch D, Gislason SR, & Oelkers EH (2011) An experimental study of crystalline basalt dissolution from 2⩽ pH⩽ 11 and temperatures from 5 to 75° C. *Geochim Cosmochim Acta* 75(19):5496-5509.
89. Gillis K & Coogan L (2011) Secular variation in carbon uptake into the ocean crust. *Earth Planet Sci Lett* 302(3):385-392.
90. Pilson ME (1998) *An Introduction to the Chemistry of the Sea* (Prentice-Hall, Inc.).